\documentclass{article}

\usepackage{amssymb}
\usepackage{amsmath}
\usepackage{graphics}
\usepackage{epsfig}
\def\ra{\rangle}

\def\1{1 \!\! 1}
\newcommand{\ket}[1]{|{#1}\ra}

\newcommand{\Ref}[1]{(\ref{#1})}
\def\spose#1{\hbox to 0pt{#1\hss}}

\def\lta{\mathrel{\spose{\lower 3pt\hbox{$\mathchar"218$}}
     \raise 2.0pt\hbox{$\mathchar"13C$}}}
\def\gta{\mathrel{\spose{\lower 3pt\hbox{$\mathchar"218$}}
     \raise 2.0pt\hbox{$\mathchar"13E$}}}

\newcommand{\be}{\begin{equation}}
\newcommand{\en}{\end{equation}}
\newcommand{\bea}{\begin{eqnarray}}
\newcommand{\ena}{\end{eqnarray}}

\newcommand{\dd}{\mbox{d}}

\def\setR{\mathbb{R}}

\newcommand{\dee}{\mathrel{\mathop:}=}
\newcommand{\no}{\nonumber\\}

\begin{document}

\begin{center}
\large{\bf QUANTUM COSMOLOGY FROM THE\\ DE BROGLIE-BOHM PERSPECTIVE}\\
\end{center}
\begin{center}
\begin{center}
{N. Pinto-Neto}\footnote[1]{email: nelson.pinto@pq.cnpq.br} \\
{\it ICRA - CBPF, Rua Dr. Xavier Sigaud, 150, Urca, Rio de Janeiro, RJ, CEP: 22290-180, Brazil}
\end{center}
J.C. Fabris\footnote[2]{email: fabris@pq.cnpq.br}\\
{\it Departamento de F\'{i}sica, CCE, UFES, Av. Fernando Ferrari, 514, Vit\'oria, ES, CEP:29060-910, Brazil}
\end{center}

\hspace{1cm}
\begin{abstract}
We review the main results that have been obtained in quantum cosmology from
the perspective of the de Broglie-Bohm quantum theory. As it is a dynamical theory
of assumed objectively real trajectories in the configuration space of the physical
system under investigation, this quantum theory is not essentially probabilistic
and dispenses the collapse postulate, turning it suitable to be applied to cosmology.
In the framework of minisuperspace
models, we show how quantum cosmological
effects in the de-Broglie-Bohm's approach can avoid the initial singularity, and isotropize
the Universe. We then extend minisuperspace in order to include linear cosmological
perturbations. We present the main equations which govern the dynamics of quantum cosmological
perturbations evolving in non-singular quantum cosmological backgrounds, and calculate
some of their observational consequences. These results are not known how to be obtained
in other approaches to quantum theory. In the general case of full superspace,
we enumerate the possible structures of quantum space and time that emerge from the de Broglie-Bohm
picture. Finally, we compare some of the results coming from the de Broglie-Bohm theory
with other approaches, and discuss the physical reasons for some discrepancies that occur.
\end{abstract}
\newpage

\section{Introduction}

The great majority of the physics community believes that quantum mechanics is a universal and
fundamental theory, applicable to any physical system, from which
classical physics can be recovered.  The Universe is, of course, a
valid physical system: there is a theoretical model, the standard cosmological model, which is
able to describe it in physical terms, and make predictions which can
be confirmed or refuted by observations. In fact, the observations
until now confirm the standard cosmological scenario (even though important tensions exist,
like the unexpected results that first came from supernova data in 1998 \cite{supernova}).
Hence, supposing
the universality of quantum mechanics, the Universe itself must be
described by quantum theory, from which we could recover classical
cosmology. However, the Copenhagen interpretation of quantum mechanics
\cite{bohr,hei,von}, which is the one
taught in undergraduate courses and employed by the majority of
physicists in all areas, cannot
be used in a quantum theory of cosmology. This is because it imposes
the existence of a classical domain or of an external agent outside the
physical system. In von Neumann's view, for
instance, the necessity of an external agent, comes from the way it
solves the measurement problem (see Ref. \cite{omn} for a good
discussion).  In an impulsive measurement of some observable, the wave
function of the observed system plus macroscopic apparatus splits into
many branches which almost do not overlap (in order to be a good
measurement), each one containing the observed system in an eigenstate
of the measured observable, and the pointer of the apparatus pointing
to the respective eigenvalue. However, in the end of the measurement,
we observe only one of these eigenvalues, and the measurement is robust
in the sense that if we repeat it immediately after, we obtain the same
result. So it seems that the wave function collapses into one of the branches, the others
disappear. The Copenhagen interpretation assumes that this
collapse is real.  However, a real collapse cannot be described by the
unitary Schr\"{o}dinger  evolution. Hence, the Copenhagen
interpretation must assume that there is a fundamental process in a
measurement which must occur outside the quantum world, by an external agent in a classical
domain.  Of course, if we want to quantize the whole Universe, there is
no place for a classical domain or any external agent outside it, and the Copenhagen
interpretation cannot be applied.

In order to save the situation, one could try to evoke the
phenomenon of decoherence \cite{deco}. In fact, the interaction of
the observed quantum
system with its environment yields an effective diagonalization of the
reduced density matrix, obtained by tracing out the irrelevant degrees of
freedom. Decoherence can
explain why the splitting of the wave function is given in terms of the
pointer basis states, and why we do not see superpositions of
macroscopic objects. In this way, classical properties emerge from
quantum theory without the need of being assumed.  In the framework of
quantum gravity, it can also explain how a classical background
geometry can emerge in a quantum universe \cite{2kie}. In fact, it is the first
quantity to become classical. However, decoherence is not yet a
complete answer to the measurement problem \cite{omn,muk,zeh}. It does not
explain the apparent collapse after the measurement is completed, or
why all but one of the diagonal elements of the density matrix become
null when the measurement is finished. The theory is unable to give an
account of the existence of facts, their uniqueness as opposed to the
multiplicity of possible phenomena. Further developments are still in progress, like the consistent
histories approach \cite{omn,har}, which are however incomplete until now.

Hence, if someone insists with the
Copenhagen interpretation, at least in its present form, she or he must assume that quantum theory
is not universal, that quantum cosmology does not make any sense at all,
and we are stuck. This is a perfect example of the adequacy of Albert Einstein sentence:
"Contemporary quantum theory � constitutes an optimum formulation
of  [certain] connections �  [but] offers no useful point of departure
for future developments."

Fortunately, there are
some alternative solutions to this quantum cosmological dilemma which
can solve the measurement problem maintaining the universality of quantum
theory. One can say that the Schr\"{o}dinger
evolution is an approximation of a more fundamental non-linear
theory which can accomplish the collapse \cite{rim,pen}, or that the
collapse is effective but not real, in the sense that the other
branches disappear from the observer but do not disappear from
existence. In this second category we can cite the Many-Worlds
Interpretation (MW) \cite{eve} and the de Broglie-Bohm Theory (dBB)
\cite{dBB,boh,hol}. In the former, all the possibilities in the splitting
are actually realized. In each branch there is an observer with the
knowledge of the corresponding eigenvalue of this branch, but she or he
is not aware of the other observers and the other possibilities because
the branches do not interfere. In the latter, a point-particle in
configuration space describing the observed system and apparatus is
supposed to exist, independently on any observations.  In the
splitting, this point particle will enter into one of the branches
(depending on the initial position of the point particle before
the measurement, which is unknown), and the other branches will be
empty. It can be shown \cite{hol} that the empty waves can neither
interact with other particles, nor with the point particle containing
the apparatus.  Hence, no observer can be aware of the other branches
which are empty.  Again we have an effective but not real collapse (the
empty waves continue to exist), but now with no multiplication of
observers. Of course these theories can be used in quantum
cosmology. Schr\"{o}dinger  evolution is always valid, and there is no need
of a classical domain or any external agent outside the physical system.

We call quantum cosmology \cite{mis1,mis2,dew3}
as this attempt to apply quantum theory
to the Universe as a whole.
In the following, we will apply the de Broglie-Bohm (dBB) theory to quantum cosmology.
One of the important results is the elimination of cosmological singularities
which has been proved at least for some particular but relevant cases. In general,
it may depend on the chosen initial state, according to the cosmological model which has been taken. As we know,
the singularity theorems \cite{he} show that, under reasonable physical
assumptions, the universe has developed an initial singularity,
which is called the big bang,
and will develop future singularities in the form of black holes
and, perhaps, of a big crunch. Until now, singularities are out of the
scope of any physical theory, hence one should assume that
the `reasonable physical assumptions'  of the theorems are
not valid under extreme situations of very high energy density and curvature,
which is very plausible.
We may say that general relativity or any other matter field theory
must be changed under these extreme conditions. One good point of
view (which is not the only one)
is to think that quantum gravitational effects become
important.
We should then construct a quantum theory of gravitation
and apply it to cosmology. This is a good strategy because, besides
the possibility of obtaining from quantum gravity a solution
to the singularity problem,
we gain from quantum
theory the possibility of constructing a theory of initial
conditions for the Universe. This theory could then
explain why the universe is remarkably
homogeneous and isotropic, and even why the constants of nature
have the values we observe they have.
Moreover, it could give the spectrum
of quantum fluctuations of geometry and matter of primordial origin
and provide us with a complete theory of structure formation.

However, there is no complete and well founded theory of quantum gravity.
In the covariant perturbative approach, the best candidate is string theory.
However, string theory is not yet predictive
about the field theory that should be valid in the physical domain
of 4-dimensional space-time dimensions, and their standard particles states.
Nonperturbative quantum gravity, like loop quantum gravity,
has also a lot of unresolved problems (complicate constraint
equations, unclear classical limit, etc).
The application of
quantum gravity to cosmology adds new problems, as we mentioned above. How do we
interpret and extract results from the wave function of the Universe?
How can we recover the standard cosmological model?
Finally, in quantum mechanics, time, in spite of
seeming to be a measurable physical quantity, is not treated
as an observable (hermitian operator) but as an external evolution
parameter ($c$-number). In the quantum cosmology of a closed
universe, there is no place for an external parameter. So, what
happens with time; does it become an operator?
These are some of the difficult issues which the subject of
quantum cosmology has to give an answer in order to have a meaning.

In this review we will try to explain the ideas on
how one can deal with some of these deep issues concerning quantum cosmology within the
framework of the de Broglie-Bohm quantum theory. In fact, the simple hypothesis
of the objective physical reality of trajectories in the configuration
space of fields and particles governed by a suitable dynamical
law, from which probabilistic concepts become secondary, yields
a framework where many of these problems become less contrived and capable
to be solved. In the next section we will review the standard canonical
quantization of general relativity, which leads to the so called
Wheeler-DeWitt equation. We know that there are much better mathematical
implementations of nonperturbative quantum general relativity, like
loop quantum gravity \cite{ash4} or path integral quantization \cite{tri},
but our aim here is to apply our ideas to quantum cosmology, where the
resulting quantum equations of these approaches should be quite similar
(or even coincide, if one is not
so close to the Planck scale) under the restricted domain of
degrees of freedom of homogeneous spaces with linear perturbations on it, characteristic
of the standard classical cosmological model.
In section 3, we will present three of the most used quantum theories
applied to quantum cosmology, the many worlds theory (MW), the consistent histories
approach (CH), and the de Broglie-Bohm theory (dBB). In section 4, we will show how some
basic problems of quantum cosmology
can be solved in the light of the de Broglie-Bohm theory. In section
5, we will compare the de Broglie-Bohm results with other approaches, and discuss
the reasons for some unexpected discrepancies. In section 6, we will
extend the cosmological models from which we have obtained the results of the preceding sections, and introduce
perturbations on them in order to obtain results concerning structure
formation and the anisotropies of the cosmic background radiation, which
could be tested by the new cosmological observations which will be implemented
in the near future. We end up with the conclusions in section 7.

We end this introduction by refereeing the reader to other investigations concerning
the de dBB approach in cosmology and quantum black holes \cite{dBBbis}

\newpage

\section{Foundations of quantum cosmology}

\subsection{Canonical quantization of general relativity}

As we have explained in the introduction, quantum cosmology
is the application of a theory of quantum gravity to the Universe
as a whole. However, until now, there is no consistent and
widely accepted theory of quantum gravity.

As we have seen, there are two basic
ways to quantize gravity: the covariant and the canonical
approaches. We will focus here on the canonical one, which
is essentially non-perturbative. But can quantum general relativity make
sense if its perturbation expansion is not renormalizable?
The answer is affirmative. There are examples of theories
that are exactly solvable non-perturbatively but which perturbation
expansion is not renormalizable \cite{ed1}.
Non-perturbative technics are also being developed in superstring theory,
but we will stay in the framework of general relativity itself and
present the canonical quantization of this theory.

The canonical approach is based on the hamiltonian of
general relativity. The idea is to obtain a quantum
functional equation for a wave functional, which is
analogous to the Schr\"{o}dinger equation. For historical reasons,
this approach is not very popular in other quantum field theories.
Some papers have been published with comparisons
of this approach with the more usual covariant approach
in quantum electrodynamics and other quantum field
theories \cite{kie1,fou}. To construct the hamiltonian
of general relativity we must assume that spacetime
can be splitted into a family of spacelike hypersurfaces
and a timelike direction. It means that we are restricting
the topology of the manifold to be of the type:
$M^{4} = R \bigotimes M^{3}$. Hence, we are discarding
spacetimes with rotation and with closed timelike curves.
Questions about the existence of closed timelike
curves cannot be answered within this formalism.

Let us now split the metric into the timelike direction
and the spacelike direction.

The spacelike hypersurfaces can be defined by the
equations $\phi(x^{\mu}) = {\rm const.}$. Their normals
are given by one-forms $\eta = {\eta}_{\mu} dx^{\mu} =
{\partial}_{\mu} \phi dx^{\mu}$. As they are spacelike,
there is always a timelike coordinate $x^{0} = t$ that
parametrizes the hypersurfaces yielding
${\eta}_{\mu} = - N {\delta}^0_{\mu}$. $N$ is a normalization
factor, $g^{\mu\nu}{\eta}_{\mu}{\eta}_{\nu} = -1$, which implies
that $g^{00} = -\frac{1}{N^2}$. The projector onto the hypersurfaces
is given by $h^{\mu\nu} \equiv g^{\mu\nu} + {\eta}^{\mu}
{\eta}^{\nu}$ whose components are $h^{00}=0$, $h^{0i}=0$
and $h^{ij} = g^{ij} + N^2 g^{i0} g^{j0}$. Defining
$N^{i} = g^{i0} N^2$, the components of the contravariant
metric are:
\begin{equation}
\label{31}
g^{00} = -\frac{1}{N^2}; \; g^{0i} = \frac{N^{i}}{N^2};
\; g^{ij} = h^{ij} - \frac{N{i} N^{j}}{N^2}
\end{equation}

We can calculate the inverse covariant metric $g_{\mu\nu}$
yielding the following line element:
\begin{eqnarray}
\label{32}
ds^2 &=& g_{\mu\nu}dx^{\mu}dx^{\nu} \nonumber \\
&=& (N_iN^i-N^2)dt^2 + 2 N_i dx^i dt + h_{ij} dx^i dx^j =
\nonumber \\ &=& N^2 dt^2 + h_{ij} (N^i dt + dx^i) (N^j dt + dx^j)
\end{eqnarray}
where $N_i = h_{ij} N^j$, $h_{ij}$ is the inverse of $h^{ij}$
and it is, by contruction,
the intrinsic covariant metric of the spacelike hypersurfaces. Examining
Eq. (\ref{32}) we can see that
$N(t,x^k)$ is the rate of change with respect to the coordinate
time $t$ of the proper time of an
observer with four-velocity ${\eta}^{\mu}(t,x^{k})$ at the
point$(t,x^k)$.
It is called the lapse function. Also, $N^i(t,x^k)$
is the rate of change with respect to coordinate time $t$
of the shift of the points with the same label $x^i$ when we
go from one hypersurface to another. It is called the shift
function. It can also be viewed as the projection onto the spacelike
hypersurface of the tangent vector $\frac{\partial}{\partial t}$
to the $t$-time coordinate curves. For more details, see Ref. \cite{mtw}.

Another useful quantity is the extrinsic curvature. It measures
how much the 3-dimensional hypersurfaces are curved with respect
to the 4-dimensional manifold in which it is embedded. It does that
by comparing the normal vector $\eta _{\mu}$ at one point with the
parallel transported
normal vector from a neighbor point
to this same point. Precisely, it is defined as follows:
\begin{equation}
\nonumber
K_{\mu\nu} \equiv - h^{\alpha}_{\mu} \, h^{\beta}_{\nu} \,
{\nabla}_{(\alpha} {\eta}_{\beta)}
\end{equation}

The relevant components of the extrinsic curvature are:
\begin{eqnarray}
\label{33}
K_{ij} &=& - N {\Gamma}^{0}_{ij} \nonumber \\
&=& \frac{1}{2N}(2D_{(i}N_{j)}-{\partial}_th_{ij}) ,
\end{eqnarray}
where $D_i$ is the 3-dimensional covariant derivative.

Using Eqs. (\ref{31}), (\ref{32}) and (\ref{33}), we obtain for
the four-dimensional Ricci scalar:
\begin{equation}
\label{34}
R=R^{(3)} +K^{ki}K_{ki}+K^2-\frac{2}{N}{\partial}_{t}K+\frac{2N^i}{N}
{\partial}_iK-\frac{2}{N}D_k({\partial}^kN),
\end{equation}
where $R^{(3)}$ is the 3-dimensional Ricci scalar,
and hence, for the Einstein-Hilbert lagrangian density, after discarding surface terms,

\begin{equation}
\label{36}
{\cal L}[N,N^i,h_{ij}]=Nh^{1/2}(R^{(3)}+K^{ij}K_{ij}-K^2) .
\end{equation}

Let us now construct the hamiltonian of general relativity.
As the lagrangian density (\ref{36}) does not depend on
${\partial}_{t} N$ and on ${\partial}_{t} N^{i}$, their
canonical conjugate momenta are zero. These are the so
called primary constraints.
Therefore, general
relativity is a theory with constraints and it will be treated
with the Dirac formalism \cite{constr1,constr2,sun}.

The canonical momenta conjugate to $h^{ij}$ are given by:
\begin{equation}
\label{37}
\Pi _{ij} = \frac{\delta L}{\delta (\partial _t h^{ij})} =
- h^{1/2}(K_{ij}-h_{ij}K).
\end{equation}

For consistency, the primary constraints must be conserved in
time. This implies the following weak equations:
\begin{equation}
\label{h01}
{\cal H} = G_{ijkl}\Pi ^{ij}\Pi ^{kl}-h^{1/2}R^{(3)} \approx 0
\end{equation}
\begin{equation}
\label{hi}
{\cal H}^j = -2D_i\Pi ^{ij} \approx 0,
\end{equation}
where
\begin{equation}
\label{301}
G_{ijkl}=\frac{1}{2}h^{-1/2}(h_{ik}h_{jl}+h_{il}h_{jk}-h_{ij}h_{kl}),
\end{equation}
which is called the DeWitt metric.
They are secondary constraints and are called super-hamiltonian
and super-momentum constraints, respectively. Their conservations in time do
not lead to any new constraints. An straightforward calculation then shows
that the hamiltonian of general
relativity is simply given by:
\begin{equation}
\label{303}
H_{GR} = \int d^3x(N{\cal H}+N_j{\cal H}^j)
\end{equation}

As $N$ and $N^{i}$ have no dynamics
and they multiply secondary constraints in the total hamiltonian,
they can be viewed as lagrangian
multipliers of these constraints, and they can be eliminated from the
phase space of the theory \cite{sun}.

It can be shown that the secondary constraints have weakly zero
Poisson brackets among each other. They are called first class
constraints and they are generators of gauge transformations.
In fact, it can be shown that:
\begin{equation}
\label{304}
\delta h_{ij} (x) = \biggl \{ h_{ij}(x), \int d^3 y \xi ^k(y) {\cal H}_k (y)\biggr \} =
D_j \xi _i(x) + D_i \xi _j(x) = {\bf \pounds _{\xi}} h_{ij}
\end{equation}
\begin{equation}
\label{305}
\delta h_{ij} (x) = \biggl \{ h_{ij}(x), \int d^3 y \zeta (y) {\cal H} (y)\biggr \} =
- 2 \zeta (x) K_{ij}(x) =  \zeta (x) {\bf \pounds _{\eta}} h_{ij}
\end{equation}
where ${\bf \pounds _{\xi}}$ is the Lie derivative along the infinitesimal
spacelike vector ${\bf \xi}$ and ${\bf \pounds _{\eta}}$ is the
Lie derivative along the direction orthogonal to the spacelike
hypersurfaces with metric $h_{ij}$. The function $\zeta (x)$ is
infinitesimal. Analogous results can be obtained for the momenta
$\Pi _{ij}$. Therefore, the first constraint is the generator of
spatial coordinate transformations while the second one is the generator
of time reparametrization, which are the gauge transformations
of the theory. As can be seen from Eq. (\ref{305}), the second
constraint is connected to time evolution.

Let us then quantize the
theory. We will work in the $h_{ij}$ representation.
The wave function $\Psi[h_{ij}, t]$ must satisfy the
Schr\"{o}dinger-like functional equation:
\begin{equation}
\label{sch}
i \hbar \frac{\partial \Psi (h_{ij}, t)}{\partial t} =
{\hat{H}}_{GR} \Psi (h_{ij}, t)
\end{equation}
where ${\hat{H}}_{GR}$ is the operator coming from the classical
hamiltonian (\ref{303}).

For the first
constraints (\ref{h01}) and (\ref{hi}), Dirac rules impose that
\begin{equation}
\label{320}
\hat{H} \Psi (h_{ij}, t) = 0
\end{equation}
\begin{equation}
\label{321}
{\hat{H}}^{k} \Psi (h_{ij}, t) = 0.
\end{equation}

If Eqs. (\ref{320}) and (\ref{321}) are correct, then the
right-hand-side of Eq. (\ref{sch}) is zero, and it implies that
$\Psi$ does not depend on $t$.

Equation (\ref{321}) has a simple meaning. It states that
the value of the wave function does not change
if the spacelike metric changes by a spatial coordinate transformation.
Therefore, Eq. (\ref{321}) implies that the wave function
is a functional of the equivalence class of metrics which describe
the same geometry, not of one particular metric. It is a functional
defined on the space of all spacelike geometries, not on the
space of all spacelike metrics. The space of all three-dimensional
spacelike geometries is called superspace. This is the
quantum version of the meaning of the constraint (\ref{hi}), which classically
is interpreted as the generator of spacelike coordinate transformations.

\subsection{The Wheeler-DeWitt equation and the issue of time}

Let us turn now to the Eq. (\ref{320}), which is called the
Wheeler-DeWitt equation \cite{dew3}. We
should expect that the dynamics of the wave function be
contained in it. There should
exist one momentum which is canonically conjugate to the time
in which the quantum dynamics takes place. In the reparametrization
invariant formulation of non-relativistic classical mechanics of
point particles, this particular momentum is easily
distinguishable from the others because it appears linearly
in this equation, while the others appear quadratically \cite{sun}.
However, in Eq. (\ref{320}), even when one adds non-gravitational
degrees of freedom, there is no momentum which
appears linearly in general (with some exceptions which we will
discuss later on); all of them appear quadratically. Hence, where
is time?

There are some proposals of solution to this problem,
which is called the issue of time. We will now expose some of them:
\vspace{0.8cm}

i) The DeWitt metric (\ref{301}) is a $6 \, X \, 6$ matrix per
space point and
it can be shown that it has signature $(-,+,+,+,+,+)$ \cite{dew3}.
Adding conventional non-gravitational degrees of freedom
(satisfying the null energy condition) will just add other dimensions
to the DeWitt metric with positive signature, not altering its lorentzian nature.
The minus sign is related to the square root
of the determinant of the spacelike metric \cite{dew3,nen1,nen2},
$\sqrt{h}$. Thus, it
seems that we should identify this quantity with time. However,
$\sqrt{h}$ is the volume of the spacelike hypersurfaces. Does it
mean that in a contracting universe time goes backwards?
Quite unplausible. Furthermore, because of the Lorentzian signature
of the DeWitt metric, the Wheeler-DeWitt
Eq. (\ref{320}) is like a Klein-Gordon equation with a variable
`mass' term, $R^{(3)}(h^{ij})$, which depends on the `time'
$\sqrt{h}$. Consequently, if we want to give some kind of probabilistic
interpretation to $\Psi$, we will have to face
all the problems with negative probabilities which are characteristic
of this type of equation. The presence of the
variable `mass' term turns this problem very difficult to solve \cite{kuc1},
and one cannot work with the single frequency approach. One possibility
is to define a new inner product, as it was done in Ref.~\cite{halli2}.
This is something yet to be explored in full superspace.

In quantum field theory, this
problem is solved by second quantizing the Klein-Gordon field.
This field operator is expanded into creation and annihilation operators
of spin zero particles. The vacuum state is the state with no particles.
If this quantum field is submitted to a time variable potential energy,
or if it is embedded in a time variable curved background, then
spin zero particles are created out of the vacuum.

For the Wheeler-DeWitt equation, this procedure would lead us
to a third quantization
of gravity by quantizing the wave function itself \cite{ter1,ter2}.
The particles are now universes that can be created by the
action of creation operators which are obtained by an expansion of
the wave function, which is now an operator.
The vacuum state is the real nothing, the absence
of matter {\it and} spacetime. Since the DeWitt metric (\ref{301}),
as well as $R^{(3)}(h^{ij})$, depends on
$\sqrt{h}$, which is considered here as `time', this quantum
wave function is like a quantum scalar field propagating in a
time variable curved background and submitted to a time variable
potential energy. Thus, universes can be spontaneously created
from nothing! This is a rather exotic and interesting picture,
but with no rigorous mathematical foundations.
\vspace{1.0cm}

ii) We could try to identify some degrees of freedom which, playing the
role of time, put the Wheeler-DeWitt
Eq. (\ref{320}) in a Schr\"odinger form. This variable can
come from matter degrees of freedom, and we will show in section
IV that this is indeed possible if matter is described by a hydrodynamical
perfect fluid. It can also come from gravitational degrees of freedom, but
this is possible only implicitly \cite{kuc1,nen2}. The variable that
plays the role of time is the trace of $\Pi _{ij}$, which is
a monotonically increasing function
of time whenever the dominant energy condition is
satisfied.
\vspace{1.0cm}

\subsection{Minisuperspace models}

The Wheeler-DeWitt
Eq. (\ref{320}) is a very complicate functional differential
equation, which is equivalent to an intricate system of partial
differential equations, one for each space point $x^i$. Such a system
is not well defined mathematically. However, one would like to investigate
the problem of time and other issues related to the quantization
of the Universe, like the singularity problem in classical cosmology
and the interpretation of the wave function of the Universe, more deeply.
Hence, it should be a good strategy to get rid of the very difficult
technical problems characteristic of the Wheeler-DeWitt
equation in full superspace, and work in a more restricted
framework. Furthermore, the great degree of space homogeneity of the
primordial Universe suggests that this simplification can be physically
reasonable when dealing with quantum cosmology.

In order to do that, one
usually simplifies the Wheeler-DeWitt
equation by freezing out the degrees of freedom of
gravity and matter, reducing the superspace to a minisuperspace where
only a finite amount of degrees of freedom are still available.

More precisely, we expand the spacelike metric, non-gravitational fields and their conjugate
momenta in some complete set $f_n$:
\begin{equation}
h^{ij}(x,t) = h^{ij}_{(0)}(t) + {\sum}_{n=1}^{\infty} h^{ij}_{(n)}(t) f_n(x)
\end{equation}
\begin{equation}
\Pi _{ij}(x,t) = \Pi _{ij}^{(0)}(t) + {\sum}_{n=1}^{\infty}
\Pi _{ij}^{(n)}(t) f_n(x)
\end{equation}
\begin{equation}
\Phi^{A}(x,t) = \Phi^{A}_{(0)}(t) + {\sum}_{n=1}^{\infty} \Phi^{A}_{(n)}(t) f_n(x)
\end{equation}
\begin{equation}
\Pi _{A}(x,t) = \Pi _{A}^{(0)}(t) + {\sum}_{n=1}^{\infty}
\Pi _{A}^{(n)}(t) f_n(x)
\end{equation}

A minisuperspace is the set of spacelike geometries and matter fields where all but
a finite set of the $h^{ij}_{(n)}(t),\Phi^{A}_{(n)}(t)$ and their corresponding
$\Pi _{ij}^{(n)}(t),\Pi _{A}^{(n)}(t)$
are put identically to zero.

Evidently, this procedure violate the uncertainty principle. However,
we expect that the quantization of these minisuperspace models retains
many of the qualitative features of the full quantum theory, which
are easier to study in this simplified model. For more details
on minisuperspace models, see Refs.  \cite{rya,kucmin,hal0}.

In the case of homogeneous models, which describe quite accurately
the primordial Universe, the
supermomentum constraint ${\cal H}^i$ is identically zero, and the shift
function $N_i$ can be set to zero in Eq. (\ref{303}) without loosing
any of the Einstein's equations. The hamiltonian (\ref{303}) is
reduced to:
\begin{equation}
\label{homham}
H_{GR} = N(t) {\cal H}[p^{\alpha}(t), q_{\alpha}(t)],
\end{equation}
where $p^{\alpha}(t)$ and $q_{\alpha}(t)$ represent the homogeneous
degrees of freedom coming from $\Pi _{ij}(x,t)$, $\Pi _{A}(x,t)$, $h^{ij}(x,t)$ and
$\Phi ^{A}(x,t)$.
The Weeler-DeWitt equation then reads
\begin{equation}
\label{hoqg1}
\frac{1}{2}f_{\alpha\beta}(q_{\mu})\frac{\partial ^2\Psi(q_{\mu})}{\partial q_{\alpha}\partial q_{\beta}}
+ U(q_{\mu}) \Psi(q_{\mu}) = 0,
\end{equation}
where
\begin{equation}
\label{h}
p^{\alpha} =
f^{\alpha\beta}\frac{1}{N}\frac{\partial q_{\beta}}{\partial t},
\end{equation}
and $f_{\alpha\beta}(q_{\mu})$ and $U(q_{\mu})$ are the minisuperspace
particularizations of $G_{ijkl}$ and $-h^{1/2}R^{(3)}(h_{ij})+V(\Phi^A)$, respectively,
where $V$ represents the potential terms coming from the matter degrees of freedom.

Equations of the type of Eq.~(\ref{hoqg1}) are well defined mathematically and solvable.
We will discuss and solve some minisuperspace examples in the sequence, but first we
must agree on how to extract information from these solutions.
\newpage

\section{Interpreting the wave function of the Universe}

\subsection{The many-worlds theory}

As we have emphasized in the Introduction, we need a new
interpretation of quantum mechanics that can be applied
consistently to the wave function of the Universe,
which should be a solution of the Wheeler-DeWitt Eq. (\ref{320}).

In 1957, Hugh Everett III presented his Ph.D. thesis {\emph{On the foundations of quantum mechanics}}, after
publishing it under the title {\emph{``Relative State" Formulation of Quantum Mechanics}} \cite{mwi}.
In this article he asserts that his motivation was to construct a formulation of quantum theory which should
be adequate to general relativity and cosmology. He was aware of the incompatibility of the collapse postulate
of the Copenhaguen interpretation with a quantum theory of everything. He then proposed that the
collapse postulate is not necessary, that it is in fact a foreign and artificial concept which is not imposed
by the mathematical structure of quantum mechanics. In fact, in a good measurement,
the different branches in which the total wave function bifurcates,
describing the quantum system and experimental apparatus pointing to one particular possible result of
the measurement, do not interfere among each other. This bifurcation, with such properties
happen for whatever complicate experimental apparatus the branch is describing
(apparata measuring other apparata, yielding consistent results among them in the branch, and
so on), including observers. Each branch looks like a world where a particular outcome of
the measurement has been obtained and every happening afterwards is compatible with this
fact. Also, as the branches do not interfere, they cannot be aware of what is happening
in the other branches. Hence it seems that each branch describes a sensible world where
facts are present. These observations led Everett to the following argumentation:
{\it From the view point of the theory, all elements of a superposition
(all ``branches") are ``actual", none anymore ``real"
than another. It is completely unnecessary to suppose that
after an observation somehow one element of the final
superposition is selected to be awarded with a mysterious
quality called ``reality" and the others condemned to oblivion.
We can be more charitable and allow the others to coexist
- they won't cause any trouble anyway because all the
separate elements of the superposition (``branches") individually obey
the wave equation with complete indifference to the
presence or absence (``actuality" or not) of the other elements.}

The question concerning the meaning of the reality of these branches, whether they
represent minds of the observer, worlds or universes is still under
debate, but a crucial point in this theory is the following:
if all possibilities are realized in Nature, what is the sense
to assign probabilities to one of the outcomes of a measurement?
Or in other words, how to recover the Born rule in this framework?
This question is also under debate, with no final answer yet. Hence
the Born measure of the worlds is a postulate of the MW theory. Note, however, that this theory does
not need an external agent or classical world amended to the
quantum system: the Schr\"odinger evolution is always valid, facts
are present in each world, no collapse is needed, and
the theory can be applied to any physical system.

\subsection{The consistent histories approach}

In the consistent histories interpretation, quantum mechanics is not viewed as a theory
of many worlds, but as a theory of many histories. It was developed
by Griffiths and Omn\`es in order to get a consistent interpretation
of quantum mechanics without the problems mentioned above.

The first basic assumption of this scheme is that,
according to Omn\'es \cite{omn},
`every physical system, whether an atom or a star, is assumed to be
described by a universal kind of mechanics, which is quantum mechanics'.
There are two immediate important consequences of this assumption:
first that the theory deals with individual systems (there is no
sense in dealing with an ensemble of planets Mars in order to
study this planet), and second that classical mechanics must be
derived from quantum mechanics in the situations where it is a
good approximation. Here, classical mechanics means not only classical
dynamics (Newton's laws, in the non-relativistic case)
but also classical logic (common sense), determinism, and
everything characteristic of the classical world. Therefore, the classical
world must be derived from the quantum world.

Evidently, this kind of interpretation is better suited for quantum
cosmology than the Copenhaguen one. That is why Hartle and Gell-Mann
have developed an analogous framework in order to apply it to quantum cosmology.

In the history interpretation, probabilities\footnote{Here, probability
has only a formal meaning, a mathematical object which must satisfy
some mathematical requirements, as will see later on. Its connection
with the relative frequences of measurement data is something
to be established when a theory of measurements is formulated.
It is argued in Ref. \cite{omn} that there are some probabilities which
cannot be tested by measurements while there are others which may have an
empirical sense.} are not assigned
to events as in usual quantum mechanics, but to whole histories.
However, as we know, we cannot assign probabilities
to every history in quantum mechanics. The interference figure
obtained from the two slit experiment is an evidence of this
fact. Hence, we must establish
what are the conditions on families of histories in order to be
possible to assign probabilities to all members of such families.
Once we obtain these conditions, we will have the possibility
of saying, for instance, that a history of the universe with
inflation is more probable than another one without inflation,
without mentioning observers or measurements.
Let us give more details on how this interpretation works.

A history of an isolated physical system is a succession of properties
of this system occurring at different times.
An example of a property of a system is the sentence `the eigenvalue of
the observable $\hat{B}$ is in the set $D$'. To each property is associated
a projector operator. In the above example, it would be the projector $P$
onto the subspace of the Hilbert space containing all eigenvectors
with eigenvalues in the set $D$. Another way to say the above property
is `the value of P is 1'.

The probability of a property, designed by its projector $P$,
must satisfy the following conditions:
\begin{eqnarray}
0 \leq & p \, (P)& \leq 1 \\
 p \, (I) &=& 1 \\
 p \, (P + P') &=&  p \, (P) +  p \, (P')
\end{eqnarray}
where $P$ and $P'$ are projectors into disjoint sets $D$ and
$D'$.

There is a theorem due to Gleason \cite{gle}, which shows
that there exists a trace-class (with unit trace) positive operator $\rho$
(the density operator), where a $ p \, (P)$ satisfying the above conditions
can be written as:
\begin{equation}
\label{pro}
 p \, (P)={\rm Tr}(\rho P) .
\end{equation}

The probability of a history can also be obtained from some
logical conditions (for details, see Ref.  \cite{omn}).
The unique\footnote{The uniqueness is only proved for histories with
two instants of time or histories where the projectors refers either
to position or momentum operators.} probability is given by:
\begin{equation}
\label{hpr}
p = {\rm Tr}\{P_n(t_n) ... P_k(t_k) ... P_1(t_1) \rho
P_1(t_1) ... P_k(t_k) ... P_n(t_n)\}
\end{equation}
where $\rho$ is the density matrix of the initial
state of the system. One of the projectors $P_n(t_n)$ can be
omitted due to the cyclic property of the trace and the fact
that $P_n(t_n)$ is a projector. Note that for $n=1$ this probability
reduces to Eq. (\ref{pro}). Also, if $\rho$ represents a pure
state, $\rho = |\Psi \! > < \!\Psi|$, this probability reduces to the
reasonable equation:
\begin{equation}
p = |P_n(t_n) ... P_k(t_k) ... P_1(t_1) |\Psi \! >|^2
\end{equation}

If we have more than one history, constituting what will be called
a family of histories, then the additivity condition on probabilities
must be checked. Let us make some definitions.

Two histories $h$ and $h'$ are said to be disjoint if
$P_k'(t_k)P_k(t_k)=0$ for some $k$. The union of two histories
is defined if the histories have $P_i'(t_i)=P_i(t_i)$ for all $i$
except for one $i=k$ where $P_k'(t_k)P_k(t_k)=0$ (they must
be disjoint). The union is the history given by the sequence
$\{P_1(t_1) ... P_i(t_i) ... P_k(t_k)+P_k'(t_k) ... P_n(t_n)\}$

A  consistent family of histories is one
where each probability of each possible union of two disjoint histories
is the sum of the probabilities of each disjoint history:
\begin{equation}
\label{add}
 p \, (h + h')= p \, (h)+ p \, (h')
\end{equation}

Therefore, in a consistent family of histories, a probability can
be assigned to each history of the family. Equation (\ref{add}) implies
some consistency conditions. Let us examine a simple example.
Take a family constituted of two histories and two instants of time.
The history $h$ is $\{P_1(t_1),P_2(t_2)\}$ and $h'$ is
$\{P_1'(t_1),P_2(t_2)\}$, with $P_1(t_1)P_1'(t_1)=0$. The initial
state is given by the density matrix $\rho$. Then we have:
\begin{eqnarray}
\label{sum}
 p \, (h + h') &=& {\rm Tr}\{P_2(t_2)(P_1(t_1)+P_1'(t_1))\rho (P_1(t_1)+
P_1'(t_1))\}
\nonumber \\
&=&  p \, (h) +  p \, (h') + \nonumber \\
&+& {\rm Tr}\{P_2(t_2)P_1'(t_1)\rho P_1(t_1)\} +
{\rm Tr}\{P_2(t_2)P_1(t_1)\rho P_1'(t_1)\}
\end{eqnarray}

Hence, probabilities can be assigned to this family of histories if:
\begin{equation}
\label{con}
{\rm Tr}\{P_2(t_2)P_1'(t_1)\rho P_1(t_1)\} +
{\rm Tr}\{P_2(t_2)P_1(t_1)\rho P_1'(t_1)\}=0
\end{equation}

Using that the projectors are hermitean operators,
Eq. (\ref{con}) is equivalent to:
\begin{equation}
\label{re}
{\rm Re Tr}\{P_2(t_2)P_1(t_1)\rho P_1'(t_1)\}=0
\end{equation}

This is the consistency condition for this family of histories.

For more complicate families of histories, the necessary and sufficient
consistency conditions are more involved \cite{omn}. That is why Hartle and
Gell-Mann \cite{har} prefer to use a simpler sufficient, but not
necessary, condition. They defined the `decoherence functional' as:
\begin{equation}
\label{df}
D(\{P_{\alpha '}\}\{P_{\alpha}\}) =
{\rm Tr}\{P^n_{\alpha _{n}'}(t_n) ... P^1_{\alpha _{1}'}(t_1) \rho
P^1_{\alpha _{1}}(t_1) ... P^n_{\alpha _{n}}(t_n)\}
\end{equation}
(the indices $\alpha _{n}$ are to emphasize that we may have many projectors
at each instant $t_n$).

Their sufficient condition is:
\begin{equation}
\label{hgc}
D(\{P_{\alpha '}\}\{P_{\alpha}\}) = 0 \: ; \;
\alpha _{k'} \neq \alpha _{k}
\end{equation}
This implies that the decoherence functional can be written as:
\begin{equation}
\label{dfd}
D(\{P_{\alpha '}\}\{P_{\alpha}\})  =
\delta _{\alpha _{1}' \alpha _{1}} ...
\delta _{\alpha _{n}' \alpha _{n}}  p \, (h=\{P_{\alpha}\})
\end{equation}
for each history $h=\{P_{\alpha}\}$ of the given family of
histories.

Families of fine grained histories (for instance, precise values of
the position operator at every instant of time) are not consistent.
Usually we have to deal with coarse grained histories (for instance,
values of the position operator belonging to some set of values
at some instants of time). These coarse grained histories may satisfy,
at least approximately, Eq. (\ref{hgc}) (recall that our observations
in cosmology are very coarse grained). In this case, we may
assign probabilities to them. There must exist some families of coarse
grained
histories which satisfy Eq. (\ref{hgc}) with no finer-grained
family which satisfies it. These families are called maximal sets.
The time evolution contained in some histories belonging to consistent
families may be approximately equal to the time evolution obtained
from the classical equations of motion. These are quasi-classical histories.
They involve quasi-classical projectors associated with collective
observables (e.g., the center of mass position of a collection
of atoms).

In quantum cosmology, the goal would be to find collective
observables (related with concrete observations), and their
connections with fundamental quantum gravity operators,
identify consistent family of histories, impose as initial
condition some solution of the Wheeler-DeWitt equation obtained
from some suitable boundary conditions, and finally calculate probabilities of histories.
This was subject of intense research \cite{his1,2kie,his3,his4,his5,halli1,halli2,har2}.

\subsection{The de Broglie-Bohm theory}

The dBB quantum theory works as follows:
take the Schr\"{o}dinger equation for a single non-relativistic particle
in the coordinate representation with the hamiltonian
$H = \frac{P^2}{2m} + V(x)$
\begin{equation}
\label{bsc}
i \hbar \frac{d \Psi (x,t)}{d t} = \biggl[-\frac{\hbar ^2}{2m} \nabla ^2 +
V(x)\biggr] \Psi (x,t)
\end{equation}

Write $\Psi = R \exp (iS/\hbar)$ and substitute it in (\ref{bsc}).
We obtain the following equations:
\begin{equation}
\label{bqp}
\frac{\partial S}{\partial t}+\frac{(\nabla S)^2}{2m} + V
-\frac{\hbar ^2}{2m}\frac{\nabla ^2 R}{R} = 0
\end{equation}
\begin{equation}
\label{bpr}
\frac{\partial R^2}{\partial t}+\nabla .(R^2 \frac{\nabla S}{m}) = 0
\end{equation}

The main features of the dBB theory, based on these two equations, are \cite{dBB,hol}:

i) The quantum particles follow trajectories $x(t)$, {\it independent
of observations}.

ii) The particles are never separated from a quantum field
$\Psi$ which acts on them and satisfies the Schr\"{o}dinger
Eq. (\ref{bsc}).

iii) As one is assuming the ontology of position, one must postulate
an equation for it. De Broglie proposed that the momentum of the particle
should be given by
\begin{equation}
p=m\dot{x}=\nabla S.
\label{guidance}
\end{equation}
This is the so called guidance equation. If one solves the Schr\"odinger
equation and obtains $S$, then one can integrate (\ref{guidance}) to
obtain $x(t)$. The precise trajectory depends on an integration constant,
the initial position, which is however unknown and for many physicists they are the hidden
variable of the theory.

iv) Equation (\ref{bqp}) is a Hamilton-Jacobi type equation
for a particle submitted to an external potential which is the classical
potential plus a new quantum potential
\begin{equation}
\label{qp}
Q \equiv -\frac{\hbar ^2}{2m}\frac{\nabla ^2 R}{R}.
\end{equation}
Hence, the particle trajectory $x(t)$ satisfies the equation of motion
\begin{equation}
\label{beqm}
m \frac{d^2 x}{d t^2} = -\nabla V - \nabla Q.
\end{equation}

v) In a statistical ensemble of particles in the same quantum
field $\Psi$, if the probability density for the unknown initial
position is given by $P(x_0)=R^2(x=x_0,t=t_0)$, Eq. (\ref{bpr})
guarantees that $R^2(x,t)$ will give the distribution of positions
at any time, and all statistical predictions of quantum mechanics are recovered.

Let us make some comments:

a) Even in the regions where $\Psi$ is very small, the quantum potential
can be very high, as we can see from Eq. (\ref{qp}). It depends
only on the form of $\Psi$, not on its absolute value. This fact,
when one is dealing with many particles in an entangled state,
brings home the non-local character of the quantum potential. This is
very important because the Bell's inequalities together
with Aspect's experiments show that, in general, a
quantum theory must be either non-local or non-ontological. As
the dBB theory is ontological, it must be non-local, as it is.
The quantum potential is responsible for the quantum effects.

b) An image proposed by Bohm and Hiley is that the wave function
$\Psi$ acts like a radio wave emitted to an automatic pilot in
a ship and guide it. It has not the energy to pull the ship
but it gives the information for its engine to do so. This information
is given through the guidance relation (\ref{guidance}). For instance,
in the two-slit experiment, the wave function informs the particle
about the other slit (its size, the separation from the other slit, etc), affecting its motion between the slit and the
screen, which is no longer a straight line. In terms of the quantum potential,
the quantum force (the gradient of the quantum potential)  is infinite
exactly on the points of destructive interference; particles cannot be there.

c) It is not always true that one can write the probability density
of an statistical ensemble of quantum particles as $P=R^2$. In fact,
there is not in the theory any logical connection between
the distribution of the unknown initial positions with $R^2$.
However, whenever $P\neq R^2$, equation $p=m\dot{x}=\nabla S$
and (\ref{bpr}) makes $P$ rapidly relax to $R^2$, at least in a coarse
grained level. This is an analog of the $H$-theorem of statistical mechanics
applied to quantum mechanics (see Ref.~\cite{valentini} for details).
In fact, almost all known physical systems have relaxed to $P=R^2$,
hence all statistical predictions of quantum mechanics are recovered
in the dBB theory. Note that if one can find physical systems which
have not relaxed to $P=R^2$, then their statistical predictions will
not agree with conventional quantum mechanics, and the dBB theory could be tested.
The possibility of existence of such systems, like relic gravitational
ways, are now under investigation \cite{valentini2}.
Hence, probabilities are not fundamental
in this theory, and the Born rule is not postulated: it is obtained
through the dynamics.

d) The classical limit is very simple: we have only to find
the conditions for having $Q\approx0$ when compared with the classical kinetic and
potential energy terms.

e) As we have discussed above, in a measurement
the wave function is a superposition of non-overlapping wave
functions. The particle will enter in only
one region, a branch of the wave function, and it will be influenced by the quantum potential
obtained from the non-zero wave function of this region only. The other branches are
not eliminated but they do not have any physical consequence to the evolution of the
physical system. There is an apparent but not real collapse.

Note that, although assuming the ontology of position of particles in space
through the new proposed guidance relation (\ref{guidance}),
the dBB theory has, in the end, less postulates than the Copenhagen
interpretation as long as it dispenses the Born rule and the collapse postulate.

A detailed analysis of the dBB theory of
quantum field theory is given in Ref. \cite{oqed} for the
case of quantum electrodynamics. See also Ref. \cite{str}
and references therein for a discussion on the general case.
\newpage

\section{Solving quantum cosmology with the de Broglie-Bohm theory}

\subsection{The issue of time}

For minisuperspace models, there is no issue of time in
the framework of the de Broglie-Bohm theory. Substituting $\Psi = Re^{iS/\hbar}$
in Eq.~(\ref{hoqg1}), the time evolution of the trajectories in minisuperspace
is then obtained from the Hamilton-Jacobi like equation,
\begin{equation}
\label{hoqg}
\frac{1}{2}f_{\alpha\beta}(q_{\mu})\frac{\partial S}{\partial q_{\alpha}}
\frac{\partial S}{\partial q_{\beta}}+ U(q_{\mu}) +
Q(q_{\mu}) = 0,
\end{equation}
with
\begin{equation}
\label{hqgqp}
Q(q_{\mu}) = -\frac{1}{R} f_{\alpha\beta}\frac{\partial ^2 R}
{\partial q_{\alpha} \partial q_{\beta}},
\end{equation}
from which one can obtain the $S$ function, and from the guidance relation
\begin{equation}
\label{hm}
p^{\alpha} = \frac{\partial S}{\partial q_{\alpha}} =
f^{\alpha\beta}\frac{1}{N}\frac{\partial q_{\beta}}{\partial t} = 0.
\end{equation}

Of course Eq. (\ref{hm}) will lead to a different evolution from
the classical one if $Q$ is not negligible. However,
Eq. (\ref{hm}) is invariant under time reparametrization. Hence,
even at the quantum level, different choices of $N(t)$ yield the same
spacetime geometry for a given non-classical solution $q_{\alpha}(x,t)$.

In the case of full superspace, the situation is rather different.
Let us take as an example the case of a non-gravitational field described
by a canonical scalar field with potential $V(\phi)$ with the following lagrangian:

\begin{equation}
\label{lg1}
{\it L_m} = \sqrt{-g}\biggr(- \frac{1}{2}\phi_{,\rho}\phi^{,\rho} - V(\phi)\biggl)
\quad.
\end{equation}

Substitution of $\Psi = A \exp (iS/\hbar)$
into the Wheeler-DeWitt Eq. (\ref{320}) for the full system yields the
Hamilton-Jacobi like equation

\begin{equation}
\label{hj27}
\kappa G_{ijkl}\frac{\delta S}{\delta h_{ij}}
\frac{\delta S}{\delta h_{kl}}
 + \frac{1}{2}h^{-1/2} \biggr(\frac{\delta S}{\delta \phi}\biggl)^2
+U+Q=0 ,
\end{equation}
where $U$ is the classical potential given by
\begin{equation}
\label{v27}
U = h^{1/2}\biggr[-{\kappa}^{-1}(R^{(3)} - 2\Lambda)+
\frac{1}{2}h^{ij}\partial _i \phi\partial _j \phi+
V(\phi)\biggl] .
\end{equation}

We will investigate the following important problem. From
the guidance relations in full superspace
\begin{equation}
\label{grh}
\Pi ^{ij} = \frac{\delta S(h_{ab},\phi)}{\delta h_{ij}} ,
\end{equation}
\begin{equation}
\label{grf}
\Pi _{\phi} = \frac{\delta S(h_{ij},\phi)}{\delta \phi} ,
\end{equation}
we obtain
the following first order partial differential equations:
\begin{equation}
\label{hdot}
{\dot{h}}_{ij} =
2NG_{ijkl}\frac{\delta S}{\delta h_{kl}} + D _i N_j + D _j N_i
\end{equation}
and
\begin{equation}
\label{fdot}
\dot{\phi}=Nh^{-1/2}\frac{\delta S}{\delta \phi} + N^i \partial _i \phi .
\end{equation}
The question is, given some initial 3-metric and scalar field,
what kind of structure do we obtain when we integrate these equations
in the parameter $t$? Does this structure form a 4-dimensional
geometry with a scalar field for any choice of the lapse and shift
functions? Note that if the functional $S$ were a solution of the
classical Hamilton-Jacobi equation, which does not contain the quantum
potential term,
then the answer would be in the affirmative because we would be in the
scope of GR. But $S$ is a solution of the {\it modified} Hamilton-Jacobi
Eq. (\ref{hj27}), and we cannot guarantee that this will continue
to be true. We may obtain a complete different structure due to
the quantum effects driven by the quantum potential term
in Eq. (\ref{hj27}). To answer this question we will move from
this Hamilton-Jacobi picture of quantum geometrodynamics to a
hamiltonian picture. This is because many strong results concerning
geometrodynamics were obtained in this later picture \cite{kuc2,tei1}.
We will construct a hamiltonian formalism which is consistent with
the guidance relations (\ref{grh}) and (\ref{grf}). It yields the bohmian
trajectories (\ref{hdot}) and (\ref{fdot}) if the guidance relations
are satisfied initially. Once we have this hamiltonian, we can use
well known results in the literature to obtain strong results about
the de Broglie-Bohm view of quantum geometrodynamics.

Taking the super-momentum constraint
\begin{equation}
\label{smos}
-2 h_{li}D_j\frac{\delta S(h_{ij},\phi)}{\delta h_{lj}} +
\frac{\delta S(h_{ij},\phi)}{\delta \phi} \partial _i \phi = 0 ,
\end{equation}
\begin{equation}
\label{smoa}
-2 h_{li}D_j\frac{\delta A(h_{ij},\phi)}{\delta h_{lj}} +
\frac{\delta A(h_{ij},\phi)}{\delta \phi} \partial _i \phi = 0 .
\end{equation}
and Eq.~(\ref{hj27}), we can easily guess that
the hamiltonian which generates the bohmian trajectories, once the
guidance relations (\ref{grh}) and (\ref{grf}) are satisfied initially,
should be given by:
\begin{equation}
\label{hq}
H_Q = \int d^3x\biggr[N({\cal H} + Q) + N^i{\cal H}_i\biggl]
\end{equation}
where we define
\begin{equation}
\label{hq0}
{\cal H}_Q \equiv {\cal H} + Q .
\end{equation}
The quantities ${\cal H}$ and ${\cal H}_i$ are the usual
GR super-hamiltonian and
super-momentum constraints.
In fact, the guidance relations (\ref{grh}) and (\ref{grf}) are consistent
with the constraints ${\cal H}_Q \approx 0$ and ${\cal H}_i \approx 0$
because $S$ satisfies (\ref{smos}) (which means that $S$ is invariant by
spatial coordinate transformations, and ${\cal H}_i$ is the unique generator
of such transformations) and (\ref{hj27}). Furthermore, they are
conserved by the hamiltonian evolution given by (\ref{hq0}).
Then we can show that indeed Eqs.(\ref{hdot},\ref{fdot}) can be obtained from $H_Q$
with the guidance relations (\ref{grh}) and (\ref{grf}) viewed as additional
constraints. For details, see Ref.\cite{euc}.

We have a hamiltonian, $H_Q$, which generates the bohmian trajectories
once the guidance relations (\ref{grh}) and (\ref{grf}) are imposed
initially. In the following, we can investigate whether the evolution of the fields
driven by $H_Q$ forms a four-geometry like in classical geometrodynamics.
First we recall a result obtained by Claudio Teitelboim \cite{tei1}.
In this paper, he shows that if the 3-geometries and field configurations
defined on hypersurfaces are evolved by some hamiltonian with the form
\begin{equation}
\label{hg}
\bar{H} = \int d^3x(N\bar{{\cal H}} + N^i\bar{{\cal H}}_i) ,
\end{equation}
and if this evolution can be viewed as the ``motion" of a 3-dimensional
cut in a 4-dimensional spacetime (the 3-geometries can be embedded in
a four-geometry with invertible four-metric), then the constraints
$\bar{{\cal H}} \approx 0$ and $\bar{{\cal H}}_i
\approx 0$ must satisfy the following algebra

\begin{eqnarray}
\{ \bar{{\cal H}} (x), \bar{{\cal H}} (x')\}&=&-\epsilon[\bar{{\cal
H}}^i(x) {\partial}_i \delta^3(x',x)]
-  \bar{{\cal H}}^i(x') {\partial}_i \delta^3(x,x')
\label{algebra1} \\
\{\bar{{\cal H}}_i(x),\bar{{\cal H}}(x')\} &=& \bar{{\cal H}}(x)
{\partial}_i \delta^3(x,x')
\label{algebra2} \\
\{\bar{{\cal H}}_i(x),\bar{{\cal H}}_j(x')\} &=& \bar{{\cal H}}_i(x)
{\partial}_j \delta^3(x,x')-
\bar{{\cal H}}_j(x') {\partial}_i \delta^3(x,x')
\label{algebra3}
\end{eqnarray}
The constant $\epsilon$ in (\ref{algebra1}) can be $\pm 1$
depending if the four-geometry
in which the 3-geometries are embedded is euclidian, with signature $(++++)$,
($\epsilon = 1$) or hyperbolic, with signature $(-+++)$, ($\epsilon = -1$),
because the stacking of 3-geometries which are evolved by the hamiltonian (\ref{hq})
can either yield an euclidian or a hyperbolic 4-geometry.
Because the 4-metric is invertible, there is no extra preferred vector field that appears besides the metric itself. If that would not be the case, there would come out absolute vector fields, which
would be the null eigen-vectors of the metric, which could be associated with absolute time and/or
absolute space, breaking spacetime into space+time, as in Newtonian physics.
Hence, we would have violations of the necessary conditions
for the existence of the single entity which we call "spacetime".

The above algebra is the same as the algebra of GR
if we choose $\epsilon = -1$. Note that the hamiltonian (\ref{hq}) is
different from the hamiltonian of GR only by the presence of
the quantum potential term $Q$ in ${\cal H}_Q$. The Poisson bracket
$\{{\cal H}_i (x),{\cal H}_j (x')\}$ satisfies Eq.
(\ref{algebra3}) because the ${\cal H}_i$ of $H_Q$ defined in Eq.
(\ref{hq}) is the same as in GR. Also
$\{{\cal H}_i (x),{\cal H}_Q (x')\}$ satisfies Eq. (\ref{algebra2})
because ${\cal H}_i$ is the generator of spatial coordinate tranformations,
and as ${\cal H}_Q$ is a scalar density of weight one ($Q$ must be a scalar
density of weight one because of Eq. (\ref{hj27})), then it must satisfies the
Poisson bracket relation (\ref{algebra2}) with ${\cal H}_i$. What remains to be verified
is whether the Poisson bracket
$\{{\cal H}_Q (x),{\cal H}_Q (x')\}$ closes as in Eq. (\ref{algebra1}).
We now recall the result of Ref. \cite{kuc2}. There it is shown that a
general super-hamiltonian $\bar{{\cal H}}$ which satisfies Eq.
(\ref{algebra1}), is a scalar density of weight one, whose geometrical
degrees of freedom are given only by the three-metric $h_{ij}$ and its
canonical momentum, and
contains only even powers and no
non-local term in the momenta (together with the other requirements,
these last two conditions are also satisfied
by ${\cal H}_Q$ because it is quadratic in the momenta and the quantum
potential does not contain any non-local term in the momenta), then
$\bar{{\cal H}}$ must have the following form:

\begin{equation}
\label{h0g}
\bar{{\cal H}} = \kappa G_{ijkl}\Pi ^{ij}\Pi ^{kl} +
\frac{1}{2}h^{-1/2}\pi ^2 _{\phi} + U_G ,
\end{equation}
where

\begin{equation}
\label{vg}
U_G \equiv -\epsilon h^{1/2}\biggl[-{\kappa}^{-1}(R^{(3)} - 2\bar{\Lambda})+
\frac{1}{2}h^{ij}\partial _i \phi\partial _j \phi+\bar{V}(\phi)\biggr] .
\end{equation}
With this result we can now establish two possible scenarios for the
Bohm-de Broglie quantum geometrodynamics, depending on the form of the quantum
potential:
\vspace{1.0cm}

\subsubsection{Quantum geometrodynamics evolution creates a non degenerate four-geometry}

In this case, the Poisson bracket $\{{\cal H}_Q (x),{\cal H}_Q (x')\}$
must satisfy Eq. (\ref{algebra1}). Then $Q$ must be such that
$U+Q=U_G$ with $U$ given by (\ref{v27}) yielding:
\begin{equation}
\label{q4}
Q = -h^{1/2}\biggr[(\epsilon + 1)\biggr(-{\kappa}^{-1} R^{(3)}+
\frac{1}{2}h^{ij}\partial _i \phi\partial _j \phi\biggl)+
\frac{2}{\kappa}(\epsilon\bar{\Lambda} + \Lambda)+
\epsilon\bar{V}(\phi) + V(\phi)\biggl] .
\end{equation}
Then we have two possibilities:

\begin{enumerate}

\item The spacetime is hyperbolic ($\epsilon = -1$).

In this case $Q$ is
\begin{equation}
\label{q4a}
Q = -h^{1/2}\biggr[\frac{2}{\kappa}(-\bar{\Lambda} + \Lambda)
-\bar{U}(\phi) + U(\phi)\biggl] .
\end{equation}
Hence $Q$ is like a classical potential. Its effect is to renormalize the
cosmological constant and the classical scalar field potential, nothing more.
The quantum geometrodynamics is indistinguishable from the classical one.
It is not necessary to require the classical limit $Q=0$ because $V_G=V+Q$
already may describe the classical universe we live in.

\item The spacetime is euclidean ($\epsilon = 1$).

In this case $Q$ is
\begin{equation}
\label{q4b}
Q = -h^{1/2}\biggr[2\biggr(-{\kappa}^{-1} R^{(3)}+
\frac{1}{2}h^{ij}\partial _i \phi\partial _j \phi\biggl)+
\frac{2}{\kappa}(\bar{\Lambda} + \Lambda)+
\bar{U}(\phi) + U(\phi)\biggl] .
\end{equation}
Now $Q$ not only renormalizes the cosmological constant and the
classical scalar field potential, but also changes the signature of spacetime.
The total potential $V_G=V+Q$ may describe some era of the early universe
when it had euclidean signature,
but not the present era, when it is hyperbolic. The transition between these
two phases must happen in a hypersurface where $Q=0$, which is the classical
limit.

\end{enumerate}

We can conclude from these considerations that if a quantum
spacetime exists with different features from the classical observed one,
then it must be euclidean. In other words, the unique relevant quantum effect
which maintains the non-degenerate nature of the four-geometry of spacetime is its
change of signature to a euclidean one. The other quantum effects are either
irrelevant or break completely the
spacetime structure.
\vspace{1.0cm}

\subsubsection{Quantum geometrodynamics evolution does
not create a non degenerate four-geometry}

In this case, the Poisson bracket $\{{\cal H}_Q (x),{\cal H}_Q (x')\}$
does not satisfy Eq. (\ref{algebra1}) but is weakly zero in some other
way. Let us examine some examples.

\begin{enumerate}

\item Real solutions of the Wheeler-DeWitt equation.

For real solutions of the Wheeler-DeWitt equation, which is a real
equation, the phase $S$ is null. Then, from Eq. (\ref{hj27}), we can
see that $Q=-U$. Hence, the quantum super-hamiltonian
(\ref{hq0}) will contain only the kinetic term, yielding
\begin{equation}
\label{car}
\{{\cal H}_Q (x),{\cal H}_Q (x')\} = 0.
\end{equation}
This is a strong equality. This case is connected with the strong gravity
limit of GR \cite{tei2,hen}. If we take the limit of large gravitational
constant $G$ (or small speed of light $c$, where we arrive at the Carroll group
\cite{poin}), then the potential in the super-hamiltonian constraint of GR
can be neglected and we arrive at a super-hamiltonian containing only
the kinetic term. The de Broglie-Bohm theory is telling us
that any real solution of the Wheeler-DeWitt equation yields a quantum
geometrodynamics satisfying precisely this strong gravity limit.
The classical limit $Q=0$ in this case implies also that $U=0$.
It should be interesting to investigate further the structure we obtain here.

\item Non-local quantum potentials.

Any non-local quantum potential breaks spacetime.
An explicit example is given in Ref.~\cite{euc}.

\end{enumerate}

Hence, for general wave functionals, the de Broglie-Bohm picture
of quantum geometrodynamics does not yield a non-degenerate 4-dimensional
geometry, indicating the presence of absolute quantities.

\subsection{Solving the singularity problem and the isotropization of the Universe}

In the framework of quantum cosmology in minisuperspace models,
non singular bouncing models have been obtained.
Let us examine concretely two examples: perfect fluids
and a free massless scalar field. In the second case, we will extend
mini-superspace in order to contain anisotropic models and show that not only singularities
are avoided but also quantum isotropization of the universe occurs.

\subsubsection{Perfect fluid: a Schr\"odinger equation}

One can describe fluids within many frameworks, but we will adopt here the Schutz method \cite{schutz1,schutz2}.
The Schutz's variables is an implementation of the degrees of freedom of a fluid,
which is described through some potentials. Applied to quantum cosmology in mini-superspace,
it leads to a Schr\"odinger-like equation, with a time variable associated to the fluid's potentials.
Let us revise this construction.

In general relativity, the matter lagrangian ${\cal L}_m$ yields
\begin{equation}
T_{\mu\nu} = \frac{2}{\sqrt{-g}}\frac{\partial {\cal L}_m}{\partial g^{\mu\nu}}.
\end{equation}
Defining ${\cal L}_m = \sqrt{-g}L_m$, we obtain
\begin{equation}
\label{fluido1}
T_{\mu\nu} = 2\frac{\partial L_m}{\partial g^{\mu\nu}} - g_{\mu\nu}L_m.
\end{equation}

Comparing the expression (\ref{fluido1}) with the canonical form of the energy-momentum tensor of a fluid
in a FLRW model,
which cannot contain heat conduction and anisotropic pressures, the following identifications can be made:
\begin{eqnarray}
\rho &=& 2\frac{\partial L_m}{\partial g_{\mu\nu}} - L_m,\\
p &=& L_m.
\end{eqnarray}

The description given above is quite general. For the case of hydrodynamical fluids with
equation of state $p=w\rho$, Schutz tried to give a more
fundamental description for the fluid Lagrangian, including dynamical
degrees of freedom \cite{schutz1,schutz2}. He defines the four-velocity of the fluid
in terms of velocity potentials which, in the case of no rotation, reads
\begin{equation}
u_\nu = \frac{1}{\mu}(\varphi_{,\nu} + \theta\,s_{,\nu}).
\end{equation}
The functions $\mu$ and $s$ describe the specific enthalpy and entropy of the system specifically, while $\varphi$ and $\theta$ are
auxiliary quantities.
The normalization condition $u^\mu u_\mu = -1$, implies that
\begin{equation}
\mu = \frac{\dot\varphi + \theta\dot s}{N}.
\end{equation}

The pressure as a function of $s$ and $\mu$ yields he first law of thermodynamics:
$$\dd\mu = T\dd s + \frac{1}{n}\dd p, \quad\Rightarrow\quad \frac{\partial p}{\partial s}\vert_{\mu} = -T n, \quad \frac{\partial p}{\partial \mu}\vert_{s} = n,$$ where $T$ denotes the temperature here.
Hence one can show that the pressure $p$
should have the form
\begin{equation}
p= w p_{0}\left( \frac{\mu}{1+w}
\right)^{\frac{1+w}{w}}
\exp{\left[-\frac{s}{s_{0}\left(w\right)} \right]}.
\end{equation}
The quantities $p_{0}$ and $s_{0}$ are arbitrary constants related to the
initial conditions of the fluid.

The canonical momenta $p_{\varphi}$,
$p_{s}$ and $p_{\theta}$ can be obtained in the usual way, yielding for the hamiltonian
density, as expected, ${\cal H}_{\rm fluid} = p/w$. One can then
perform the canonical transformation (for details see
Ref.~\cite{rubakov})
\begin{equation}
T=-p_s\exp \left( -\frac{s}{s_{0}}\right) p_\varphi^{-(1+w)}
p_{0}^{w}s_{0},
\label{can1}
\end{equation}
and
\begin{equation}
\varphi_N=\varphi + \lambda s_0 \frac{p_s}{p_\varphi},
\label{can3}
\end{equation}
leading to the momenta
\begin{equation}p_{_T}=\frac{p_\varphi^{1+w}}{p_{0}^{w}}
\exp\left(\frac{s}{s_{0}}\right),
\label{can2}
\end{equation}
and
\begin{equation}
p_{\varphi_N}=p_{\varphi}.
\label{can4}
\end{equation}
As it turns out, these variables are more suitable than the original
ones as the fluid Hamiltonian expressed in terms of them gets the
simple form
\begin{equation}
H_{\rm fluid} = N\frac{P_{_T}}{a^{3w}}.
\label{h1}
\end{equation}

Combining this perfect fluid hamiltonian with the gravitational hamiltonian
for a FLRW geometry, one gets for the total minisuperspace hamiltonian

\begin{equation}
H=N\left\{-\frac{P_a^{2}}{4a}-ka+ \frac{P_{_T}}{a^{3(w)}}\right\}\!\!,
\label{h2}
\end{equation}
Hence, because one momentum appears linearly in the hamiltonian,
the Wheeler-DeWitt equation assumes the Schr\"odinger form \cite{lemos-ini,pinto,fabris}
\begin{equation}
 i \frac{\partial}{\partial T}\Psi =\frac{1}{4} \left\{
a^{(3\omega-1)/2}\frac{\partial}{\partial a} \left[
a^{(3\omega-1)/2}\frac{\partial}{\partial a}\right]
\right\}\Psi
\end{equation}
where we have chosen the factor ordering in $a$ in order to yield a
covariant Schr\"odinger equation under field redefinitions.

We now
change variables to
$$\chi=\frac{2}{3} (1-\omega)^{-1} a^{3(1-\omega)/2},$$ obtaining the
simple equation
\begin{equation}
i\frac{\partial\Psi_{(0)}(a,T)}{\partial T}= \frac{1}{4}
\frac{\partial^2\Psi_{(0)}(a,T)}{\partial \chi^2}. \label{es202}
\end{equation}
This is just the time reversed Schr\"odinger equation for a one
dimensional free particle constrained to the positive axis. As $a$
and $\chi$ are positive, solutions which have unitary evolution must
satisfy the condition
\begin{equation}
\label{cond27} \biggl[\Psi_{(0)}^{\star}\frac{\partial\Psi_{(0)}}{\partial
\chi} -\Psi_{(0)}\frac{\partial\Psi_{(0)}^{\star}}{\partial
  \chi}\biggr]\Biggl|_{\chi=0}=0
\end{equation}
(see Ref.~\cite{fabris} for details).

We will choose the initial
normalized wave function
\begin{equation}
\label{initial}
\Psi_{(0)}^{(\mathrm{init})}(\chi)=\biggl(\frac{8}{T_0\pi}\biggr)^{1/4}
\exp\left(-\frac{\chi^2}{T_0}\right) ,
\end{equation}
where $T_0$ is an arbitrary constant. The Gaussian
$\Psi_{(0)}^{(\mathrm{init})}$ satisfies condition (\ref{cond27}).

Using the propagator procedure of Refs.~\cite{pinto,fabris}, we
obtain the wave solution for all times in terms of $a$:

\begin{eqnarray}\label{psi1t}
\Psi_{(0)}(a,T)=\left[\frac{8 T_0}{\pi\left(T^2+T_0^2\right)}
\right]^{1/4}
\exp\biggl[\frac{-4T_0a^{3(1-\omega)}}{9(T^2+T_0^2)(1-\omega)^2}\biggr]
\nonumber\\
\times\exp\left\{-i\left[\frac{4Ta^{3(1-\omega)}}{9(T^2+T_0^2)(1-\omega)^2}
+\frac{1}{2}\arctan\biggl(\frac{T_0}{T}\biggr)-\frac{\pi}{4}\right]\right\}.
\end{eqnarray}

Due to the chosen factor ordering, the probability density
$\rho(a,T)$ has a non trivial measure and it is given by
$\rho(a,T)=a^{(1-3\omega)/2}\left|\Psi_{(0)}(a,T)\right|^2$.  Its
continuity equation coming from Eq.~(\ref{es202}) reads
\begin{equation}
\label{cont1} \frac{\partial\rho}{\partial T}
-\frac{\partial}{\partial a}\biggl[\frac{a^{(3\omega-1)}}{2}
\frac{\partial S}{\partial a}\rho\biggr]=0 ,
\end{equation}
which implies in the dBB theory that
\begin{equation}
\label{guidancec} \dot{a}=-\frac{a^{(3\omega-1)}}{2} \frac{\partial
S}{\partial a} ,
\end{equation}
in accordance with the classical relations $\dot{a}=\{a,H\}=
-a^{(3\omega-1)}P_a/2$ and $P_a=\partial S/\partial a$.

Note that $S$ satisfies the modified Hamilton-Jacobi equation,
\begin{equation}
\label{hamilton-jacobi}
\frac{\partial S}{\partial T}-\frac{a^{(3\omega-1)}}{4}
\biggl(\frac{\partial S}{\partial a}\biggr)^2 +
\frac{a^{(3\omega-1)/2}}{4R}
\frac{\partial}{\partial a} \left[
a^{(3\omega-1)/2}\frac{\partial R}{\partial a}\right]=0 ,
\end{equation}
with the quantum potential, given by
\begin{equation}
\label{quantumpotential}
Q\equiv -\frac{a^{(3\omega-1)/2}}{4R}
\frac{\partial}{\partial a} \left[
a^{(3\omega-1)/2}\frac{\partial R}{\partial a}\right] .
\end{equation}
Hence, the trajectory (\ref{guidancec}) will not coincide with the classical trajectory
whenever $Q$ is comparable with the other terms present in Eq.~(\ref{hamilton-jacobi})
because $S$ will be different from the classical Hamilton-Jacobi function.

Inserting the phase of (\ref{psi1t}) into Eq.~(\ref{guidancec}), we
obtain the bohmian quantum trajectory for the scale factor:
\begin{equation}
\label{at} a(T) = a_0
\left[1+\left(\frac{T}{T_0}\right)^2\right]^\frac{1}{3(1-\omega)} .
\end{equation}
Note that this solution has no singularities and tends to the
classical solution when $T\rightarrow\pm\infty$. Remember that we
are in the gauge $N=a^{3\omega}$, and $T$ is related to conformal
time through
\begin{equation}
\label{jauge} Nd T = a d \eta \quad \Longrightarrow d\eta =
\left[a(T)\right]^{3\omega-1} d T.
\end{equation}
The solution (\ref{at}) can be obtained for other initial wave
functions (see Ref.~\cite{fabris}).

An important aspect of this approach to recover a time variable, using the degrees of freedom of the
fluid through the Schutz's variable, is because a Schr\"odinger-like equation emerges naturally. There are other ways to obtain the
time evolution from the Wheeler-de Witt equation in mini-superspace. One example is to use the $WKB$ approach. This has been done, for example, in
Ref. \cite{gg} for gravity coupled to a scalar field system. In order to compare both cases, we
must add a radiative fluid to gravity/scalar field system of Ref. \cite{gaussian}. Moreover, since the $WKB$ method is an approximative approach, it may lead to different predictions. However, a complete study about the possible relations between different ways of obtaining a time variable, and consequently a dynamics (Schutz against $WKB$, for example), is still lacking.

\subsubsection{The scalar field: a Klein-Gordon equation with quantum isotropization}

One of the fluids which may represent the matter content of the
very early Universe is a massless free scalar field, which is
equivalent to stiff matter ($p=\rho$, sound
velocity equal to the speed of light). In a FLRW model,
the energy density of such a field
(which is an excellent approximation for the very early Universe, see Ref. \cite{zel})
evolves as $\rho\propto a^{-6}(t)$ and, in the very early Universe when $a(t)$ becomes small,
it dominates over radiation and dust, whose energy densities
depend on $a(t)$ as $a^{-4}(t)$ and $a^{-3}(t)$, respectively.
We will concentrate on this
model now. If this scalar field is not present, radiation will be the dominant
term in the early Universe. Note that a massive or self-interacting scalar field, obtained by adding a potential term, would not satisfy the stiff matter condition
and the general features of the evolution of the universe would depend on the form of the potential.

Let us take the lagrangian
\begin{equation}
\label{lg2}
{\it L} = \sqrt{-g}\biggr(R - \frac{1}{2}\phi_{,\rho}\phi^{,\rho}\biggl)
\quad,
\end{equation}
where $R$ is the Ricci scalar of the metric $g_{\mu\nu}$ with determinant $g$, and
$\phi$ is the scalar field.

For the gravitational part we will take the more general minisuperspace model given by the
homogeneous and anisotropic Bianchi I line element
\begin{eqnarray} \label{bia}
ds^{2} &&=-N^{2}(t)dt^{2}+\exp [2\beta _{0}(t)+2\beta _{+}(t)+2\sqrt{3}\beta
_{-}(t)]\;dx^{2}+ \nonumber \\
&&\exp [2\beta _{0}(t)+2\beta _{+}(t)-2\sqrt{3}\beta _{-}(t)]\;dy^{2}+
\exp [2\beta _{0}(t)-4\beta _{+}(t)]\;dz^{2} \;.
\end{eqnarray}
This line element will be isotropic if and only if $\beta _{+}(t)$ and
$\beta _{-}(t)$ are constants \cite{he}. The aim here, besides setting a more general
framework for the investigation of the singularity problem, is to check whether quantum
effects can also make isotropic an anisotropic model which would classically never becomes isotropic.

Inserting Equation (\ref{bia}) into the
action $S=\int {{\it L\,}d^{4}x}$,
supposing that the scalar field $\phi $
depends only on time, discarding surface terms, and performing a Legendre
transformation, we obtain the following minisuperspace classical Hamiltonian
\begin{equation} \label{hbiaf}
H=\frac{N}{24\exp {(3\beta _{0})}}(-p_{0}^{2}+p_{+}^{2}+p_{-}^{2}+p_{\phi
}^{2}) \;,
\end{equation}
where $(p_{0},p_{+},p_{-},p_{\phi })$ are canonically conjugate to $(\beta
_{0},\beta _{+},\beta _{-},\phi )$, respectively.

We can write this Hamiltonian in a compact form by defining $y^{\mu} =
(\beta _0, \beta _+, \beta _-, \phi)$ and their canonical momenta $p_{\mu} =
(p_0, p_+, p_-, p_{\phi})$, obtaining
\begin{equation}  \label{ham}
H = \frac{N}{24 \exp{(3y^0)}}\eta ^{\mu\nu}p_{\mu}p_{\nu} \;,
\end{equation}
where $\eta ^{\mu\nu}$ is the Minkowski metric with signature $(-+++)$. The
equations of motion are the constraint equations obtained by varying the
Hamiltonian with respect to the lapse function $N$

\begin{equation}  \label{hbia1f}
{\cal H} \equiv \eta ^{\mu\nu}p_{\mu}p_{\nu} = 0 \;,
\end{equation}
and the Hamilton's equations

\begin{equation}  \label{hbia2f}
\dot{y}^{\mu} = \frac{\partial{\cal H}}{\partial p_{\mu}} = \frac{N}{12 \exp{
(3y_0)}}\eta ^{\mu\nu}p_{\nu} \;,
\end{equation}

\begin{equation}  \label{hbia3f}
\dot{p}_{\mu} = -\frac{\partial{\cal H}}{\partial y^{\mu}} = 0 \;.
\end{equation}
The solution to these equations in the gauge $N=12\exp (3 y_0)$ is

\begin{equation}  \label{solc}
y^{\mu} = \eta ^{\mu\nu}p_{\nu}t + C^{\mu} \;,
\end{equation}
where the momenta $p_{\nu}$ are constants due to the equations of motion and
the $C^{\mu}$ are integration constants. We can see that the only way to
obtain isotropy in these solutions is by making $p_{1}=p_{+}=0$ and $
p_{2}=p_{-}=0$, which yields solutions that are always isotropic, the usual
Friedmann-Robertson-Walker (FRW) solutions with a scalar field. Hence, there
is no anisotropic solution in this model which can classically become
isotropic during the course of its evolution. Once anisotropic, always
anisotropic. If we suppress the $\phi$ degree of freedom, the unique
isotropic solution is flat space-time because: in this case the constraint (\ref{hbia1f})
enforces $p_0 =0$.

To discuss the appearance of singularities, we need the Weyl square tensor $
W^{2}\equiv W^{\alpha \beta \mu \nu }W_{\alpha \beta \mu \nu }$. It reads
\begin{equation} \label{w2}
W^{2}=\frac{1}{432}e^{-12\beta
_{0}}(2p_{0}p_{+}^{3}-6p_{0}p_{-}^{2}p_{+}+p_{-}^{4}+2p_{+}^{2}p_{-}^{2}+
p_{+}^{4}+p_{0}^{2}p_{+}^{2}+p_{0}^{2}p_{-}^{2}) \;.
\end{equation}
Hence, the Weyl square tensor is proportional to $\exp {(-12\beta _{0})}$
because the $p$'s are constants (see Equations (\ref{hbia3f})), and the singularity
is at $t=-\infty $. The classical singularity can be avoided only if we set $
p_{0}=0$. But then, due to Equation (\ref{hbia1f}), we would also have $
p_{i}=0$, which corresponds to the trivial case of flat space-time. Therefore,
the unique classical solution which is non-singular is the trivial flat
space-time solution.

The Dirac quantization procedure yields the Wheeler-DeWitt equation through
the imposition of the condition
\begin{equation}  \label{nense}
\hat{{\cal H}} \Psi = 0 \;,
\end{equation}
on the quantum states, with $\hat{{\cal H}}$ defined as in Equation
(\ref{hbia1f}). Using
the substitutions
\begin{equation}
p_{\mu} \rightarrow - i\frac{\partial }{\partial y^{\mu}} \;,
\end{equation}
equation (\ref{nense}) reads
\begin{equation}  \label{wdw40}
\eta ^{\mu\nu}\frac{\partial ^2}{\partial y^{\mu} y^{\nu}} \Psi (y^{\mu}) =0
\;.
\end{equation}

For the minisuperspace we are investigating, the guidance relations in the
gauge $N=12\exp (3 y_0)$ are (see Equations (\ref{hbia2f}))
\begin{equation}  \label{gui}
p_{\mu} = \frac{\partial S}{\partial y^{\mu}} = \eta _{\mu\nu}{\dot{y}}
^{\nu} \;,
\end{equation}
where $S$ is the phase of the wave function.

Let us
investigate spherical-wave solutions of Equation (\ref{wdw40}). They read
\begin{equation}  \label{psi3}
\Psi_1 = \frac{1}{y}\biggl[ f(y^0 + y) + g(y^0 - y)\biggr] \;,
\end{equation}
where $y\equiv \sqrt{\sum _{i=1}^{3} (y^i)^2}$.

The guidance relations (\ref{gui}) are
\begin{equation}  \label{gui0'}
p_{0} = \partial _0 S = {\rm Im} \biggl(\frac{\partial _0 \Psi _1}{\Psi _1}
\biggr) = -{\dot{y}}^{0} \;,
\end{equation}
\begin{equation}  \label{guii'}
p_{i} = \partial _i S = {\rm Im} \biggl(\frac{\partial _i \Psi _1}{\Psi _1}
\biggr) = {\dot{y}}^{i} \;,
\end{equation}
where $S$ is the phase of the wave function. In terms of $f$ and $g$ the
above equations read
\begin{equation} \label{gui0}
{\dot{y}}^{0}=-{\rm Im}\biggl(\frac{f^{\prime }(y^{0}+y)+g^{\prime }(y^{0}-y)
}{f(y^{0}+y)+g(y^{0}-y)}\biggr) \;,
\end{equation}
\begin{equation} \label{guii}
{\dot{y}}^{i}=\frac{y^{i}}{y}{\rm Im}\biggl(\frac{f^{\prime
}(y^{0}+y)-g^{\prime }(y^{0}-y)}{f(y^{0}+y)+g(y^{0}-y)}\biggr) \;,
\end{equation}
where the prime means derivative with respect to the argument of the
functions $f$ and $g$, and $Im(z)$ is the imaginary part of the complex
number $z$.

From Eqs. (\ref{gui0}) and (\ref{guii}) we obtain that

\begin{equation}  \label{yi}
\frac{dy^{i}}{dy^{j}} = \frac{y^i}{y^j} \;,
\end{equation}
which implies that $y^{i}(t)=c_{j}^{i}y^{j}(t)$, with no sum in $j$, where
the $c_{j}^{i}$ are real constants, $c_{j}^{i}=1/c_{i}^{j}$ and $%
c_{1}^{1}=c_{2}^{2}=c_{3}^{3}=1$. Hence, apart some positive multiplicative
constant, knowing about one of the $y^{i}$ means knowing about all $y^{i}$.
Consequently, we can reduce the four Eqs. (\ref{gui0}) and (\ref{guii})
to a planar system by writing $y=C|y^{3}|$, with $C>1$, and working only
with $y^{0}$ and $y^{3}$, say. The planar system now reads
\begin{equation} \label{gui0p}
{\dot{y}}^{0}=-{\rm Im}\biggl(\frac{f^{\prime }(y^{0}+C|y^{3}|)+g^{\prime
}(y^{0}-C|y^{3}|)}{f(y^{0}+C|y^{3}|)+g(y^{0}-C|y^{3}|)}\biggr) \;,
\end{equation}
\begin{equation} \label{guiip}
{\dot{y}}^{3}=\frac{{\rm sign}(y^3)}{C}{\rm Im}\biggl(\frac{f^{\prime
}(y^{0}+C|y^{3}|)-g^{\prime }(y^{0}-C|y^{3}|)}{
f(y^{0}+C|y^{3}|)+g(y^{0}-C|y^{3}|)}\biggr) \;.
\end{equation}
Note that if $f=g$, $y^{3}$ stabilizes at $y^{3}=0$ because ${\dot{y}}^{3}$
as well as all other time derivatives of $y^{3}$ are zero at this line. As $%
y^{i}(t)=c_{j}^{i}y^{j}(t)$, all $y^{i}(t)$ become zero, and the
cosmological model isotropizes forever once $y^{3}$ reaches this line. Of
course one can find solutions where $y^{3}$ never reaches this line, but in
this case there must be some region where ${\dot{y}}^{3}=0$, which implies ${
\dot{y}}^{i}=0$, and this is an isotropic region. Consequently, quantum
anisotropic cosmological models with $f=g$ always have an isotropic phase,
which can become permanent in many cases.

As a concrete example, we have examined the case where $f=-g$
is a Gaussian in $\Psi _1$ given in
Equation (\ref{psi3}).
The bohmian trajectories have been obtained numerically in Ref.~ \cite{iso}, obtaining realistic cosmological
models without singularities (in fact, periodic Universes),
with expanding phases (increasing $\beta_0$) with
isotropic phases
which can be made arbitrarily large in the region $|\phi |>>|\beta _{0}|$.
Hence, what was classically forbidden
(a nonempty, nonsingular anisotropic model, which become isotropic in the
course of its evolution)
is possible within the bohmian quantum dynamics described above.
See Ref.~ \cite{iso} for details.

The simpler isotropic case with a scalar field was also studied, see Ref.~ \cite{gaussian},
and bohmian trajectories are also non-singular. We will return to this example in the next section.

\subsubsection{Perfect fluids with other degrees of freedom}

In the same framework of a time variable selected from a hydrodynamical perfect fluid,
we will now discuss some complications related to the addition
of new degrees of freedom, either gravitational or non-gravitational:
\begin{enumerate}
\item In the case of a multi-fluid model, how to select a unique time coordinate?
\item How to treat a scalar-tensor theory and anisotropic models?
\end{enumerate}
\vspace{0.5cm}

{\bf Multi-fluid models}

The multi-fluid model has been treated in Ref. \cite{nelson1}, by introducing radiation and dust matter in the Hamiltonian description.
Both the conjugate momenta associated to dust and radiation appear linearly in the Hamiltonian, which reads,
\begin{eqnarray}
{\cal H} = - \frac{p_a^2}{2\,a} + \frac{p_\eta}{a} + p_\phi,
\end{eqnarray}
where $p_\eta$ is related to the radiation fluid and $p_\phi$ to dust.
The Schr\"odinger-like equation reads:
\begin{eqnarray}
\biggr(\frac{1}{24\,a}\frac{\partial^2}{\partial a^2} - \frac{i}{a}\frac{\partial }{\partial\eta} - i\frac{\partial }{\partial\phi}\biggl)\Psi(a,\eta,\phi) = 0.
\end{eqnarray}
The radiation fluid variable was chosen as the time coordinate, for the simple reason that the classical solution takes a closed form using the conformal time coordinate, which is
naturally linked to the radiation fluid variable. Our choice thus refers
to the classical scenario, but of course any choice could be made, which would probably lead to inequivalent predictions. This is something yet to be checked.
\par
In Ref. \cite{nelson1}, two possibilities were explored: the wave function is an eigenstate of the dust matter operator, such that $\hat p_\phi|\Psi> = p_\phi|\Psi>$; or
the wave function is a superposition of the dust operator eigenstate, such that dust is not conserved separately. In both cases there are bohmian bouncing non singular solutions. However, when
the wave function is an eigenstate of the dust operator, the evolution is not unitary, while the superposition of matter eigenstate recovers unitarity and, at the same time, allows a conversion
of exotic matter into ordinary one, leading to a natural transition to the classical scenario.
\vspace{0.5cm}

{\bf Anisotropic models}

The problem of a non-unitary evolution of the wave function of the universe
when a perfect fluid is present appears in many other contexts in quantum cosmology.
One important situation is the case of anisotropic universes.
In that case, taking the Bianchi I model described above, the resulting Schr\"odinger like equation reads,
\begin{eqnarray}
\biggr(\frac{\partial^2 }{\partial\beta_0^2} - \frac{\partial^2 }{\partial\beta_+^2} - \frac{\partial^2 }{\partial\beta_-^2}\biggl)\Psi = - 24\frac{\partial\Psi}{\partial T}.
\end{eqnarray}
This Schr\"odinger-like equation admits a simple solution:
\begin{eqnarray}
\Psi = C_\pm e^{i(k_+\beta_+ + k_-\beta_-)}J_{\pm\nu}\biggr(\frac{\sqrt{24E}}{r}a^r\biggl),
\end{eqnarray}
where $C_\pm, k_\pm $ are integration constants, $E$ corresponds to the energy eigenvalue, $\nu = i\frac{k}{r}$ and $r = \frac{3}{2}(1 - \alpha)$.
\par
This nice solution has one problem: it corresponds to a non-unitary evolution of the system. In fact, constructing the wave packet, its norm is time-dependent.
Hence, no prediction using the many-world interpretation can be obtained.
However, as it happens in the multi-fluid case, predictions can be made by calculating the bohmian trajectories.
This analysis has been performed in Ref. \cite{brasil2}, integrating numerically the differential equations related to the bohmian trajectories. The results indicate
a singularity-free, anisotropic universe that becomes isotropic asymptotically.
However, the violation of the unitarity condition seems to be a general feature of anisotropic quantum cosmological models with a perfect fluid.
\par
In Ref. \cite{majumder} other anisotropic models where the time variable has been established by the Schutz formalism have been studied. In particular, Bianchi V and IX models have been
considered. In general, again the wave function has no unitary evolution. The explicit case of Bianchi V model, filled with a fluid with equation of state such that $\omega_x = - 1/3$ (a fluid formed
by cosmic strings), has been solved completely, revealing the existence of a bounce, approaching the classical behaviour asymptotically. In the asymptotic regime, the unitarity of the wave function is
recovered.

Anisotropic quantum models in cosmology have been studied in many different contexts. Classically, it has been shown in Ref. \cite{lif} that such models generically contains singularities. To give an example of the classical features related to the presence of anisotropies, we may evoke
the results of Ref. \cite{kerner}, where a non-singular isotropic solution has been obtained from the bosonic sector of the 5-dimensional
supergravity theory. Such solution is plagued with instabilities, but the introduction of anisotropy in the model may cure this instability problem leading,
on the other hand, to the emergence of singularities. Even at the semi-classical level, anisotropic models are problematic.
Some discussions on particle production in anisotropic cosmological models can be found, for example, in Ref. \cite{birrell}. The divergences which arise
at the quantum level
are in general more serious than in the isotropic case, and the loss of unitarity seems to be a general feature. This affects, as far as we know, all classes of anisotropic cosmological models,
including the Kantowski-Sachs model, which are, however, very important in quantum gravity models. In Loop Quantum Cosmology, as another example, some isotropic cases
reveal non-singular solutions, while the corresponding Bianchi I models are plagued with instabilities, asking perhaps for a full quantum approach \cite{sake}.

All the problems described above are related to the non-canonical form of the kinetic term
appearing in the equation which governs the dynamics of the wave function. If the kinetic
term is quadratic in the momenta with euclidian metric, these issues do not
appear. We will see more consequences of this kind of situation in the next section.
\newpage

\section{Comparison with other interpretations}

\subsection{The many-worlds theory}

In the many worlds-theory, all possibilities are realized. Hence, one needs a measure that indicates the most frequent
realizations, if any, in order to give some predictive power to the theory. We have discussed this problem in section
3,
where we have shown that this is an open question already when the quantum state evolves according to a Schr\"odinger like
equation, and the goal is to show that this measure should be given by the Born rule.

Due to the issue of time discussed in section 2, it is not easy to put the Wheeler-DeWitt equation
in a Schr\"odinger form, which makes it difficult to envisage a measure for the possible realizations
of the Universe in general.

See eg in CKiefer�s book or other recent in quantum cosmology, on retrieving Hamilton Jacobi (HJ) equations, WKB limit, the induced Schrodinger equation, a relation between identifying a point basis vectors in the quantum setting, interaction terms between modes expressed in a density matrix framework, and as they �decohere� as univ expands (naturally), the (semi)classical universe emerges.

Usually one relies in a WKB approximation in order to define
a measure \cite{hal0}. Probabilities can only be assigned in the semiclassical
limit, when geometry becomes classical and we can recover a Schr\"odinger-like equation for
the matter fields in a classical background gravitational field \cite{kiefer}. Decoherence plays
an important role in this framework, where interference between the many branches are eliminated.
These ideas can also be applied to the description of the evolution of cosmological perturbations
of quantum mechanical origin in classical Friedman models and their classicalization when the
Universe expands \cite{kiefer2}. We will return to this problem in section 6.
However, questions concerning the existence of singularities
cannot be answered in this approximation because they go beyond the semi-classical limit.

For the dBB approach, the existence of a measure is not crucial because the dBB
quantum theory is not essentially probabilistic, as discussed in section 3: the bohmian trajectories
can be calculated without recurring to any probabilistic notion. Then we can see whether these trajectories
are singular or not.

A comparison between the predictions of both quantum theories concerning the existence
of singularities can be made only when a measure can be
naturally postulated in the MW framework in the full quantum regime. We will focus on the situation already discussed in section 4
where a time variable can be defined when matter is described by a perfect hydrodynamical fluid,
and one naturally obtains a Schr\"odinger-like equation with its natural Born measure. A MW interpretation
for this kind of system was first discussed in Ref. \cite{tipler}. Note, however,
that in the case of a perfect fluid with extra degrees of freedom, problems of unitarity can appear,
as discussed in the preceding section, and the MW theory cannot be used in these cases either.

In this subsection we will discuss what kind of information one can extract from
a given quantum state of the Universe in the framework of the dBB and MW theories in the
above-mentioned situations where both theories can be used.
We will show that both approaches
(dBB and MW) lead to the same time evolution of bohmian trajectories and mean values, respectively, when
the kinetic term of the hamiltonian constraint is quadratic in the momenta and the eigenvalues of its metric
have all the same sign. This assure that the evolution of the quantum
state is unitary. However, the meaning of the MW mean values is quite problematic in quantum cosmology.
We will discuss this issue in the following.

As we have shown in section 4, the Schr\"odinger-like equation for quantum cosmology with
a perfect fluid reads,
\begin{equation}
\label{eqs222}
i\frac{\partial\Psi(\chi,T)}{\partial T}= \frac{1}{4}
\frac{\partial^2\Psi(\chi,T)}{\partial \chi^2},
\end{equation}
where
$$\chi\equiv\frac{2}{3} (1-\omega)^{-1} a^{3(1-\omega)/2},$$
and the resulting quantum state calculated in section 3 is given by (\ref{psi1t}).

The Schr\"odinger Eq. (\ref{eqs222}) has a kinetic term with metric with definite signature
(it is one dimensional). Hence the evolution is unitary and we can construct the usual Schr\"odinger
measure in order to evaluate expectation values in the framework of the many-worlds theory. Due to the chosen factor ordering, the mean value of any operator ${\hat{O}}$ can be computed using
the prescription (in the $a$ representation),
\begin{equation}
<{\hat{O}}> = \frac{\int_0^\infty a^{1 - 3\omega}\Psi^*\,{\hat{O}}\,\Psi\,da}{\int_0^\infty a^{1 - 3\omega}\Psi^*\,\Psi\,da}.
\end{equation}
\par
In the case of the scale factor itself one gets
\begin{equation}
\label{atm} <a> =
\left[1+\left(\frac{T}{T_0}\right)^2\right]^\frac{1}{3(1-\omega)}.
\end{equation}
As calculated in section 3, the bohmian trajectory reads
\begin{equation}
\label{atat} a(T) = a_0
\left[1+\left(\frac{T}{T_0}\right)^2\right]^\frac{1}{3(1-\omega)}.
\end{equation}

The main difference of the two results is that the bohmian trajectory
depends on the arbitrary constant $a_0$, which is the value of the scale
factor at the bounce, while the mean value does not. This is because
the mean value can also be calculated in the dBB approach by integrating
the ensemble of bohmian trajectories corresponding to each value of $a_0$
using the quantum state evaluated at the moment of the bounce as its probability distribution,
\begin{equation}
\label{atm0} <a> =
\left[1+\left(\frac{T}{T_0}\right)^2\right]^\frac{1}{3(1-\omega)}
\int_0^\infty a_0^{1 - 3\omega}\Psi^*(a_0,T=0)\,a_0\,\Psi(a_0,T=0)\,da_0,
\end{equation}
yielding the result (\ref{atm}).

If one raises the question concerning the presence of an initial singularity,
one can assert with certainty within the dBB theory that they are eliminated
because there is no quantum bohmian trajectory which is singular. However,
in the MW theory, we cannot arrive at the same conclusion: the mean value
is not singular, but there may exist worlds in which the scale factor vanishes.
Hence, although the dynamics looks like the same, the physical conclusions
one can obtain from the two theories are rather different.
\par

\subsection{The consistent histories approach}

In contrast to what have been shown in the present review, it has been claimed in some recent papers (some few examples are Refs.~\cite{CS,ash1,ash2,ash3,ash41}) that the Wheeler-DeWitt approach to quantum cosmology does not solve the singularity problem of classical cosmology. This claim is usually based on calculations on a very simple model, namely, a free massless scalar field in Friedmann models, one of the models discussed in section 4, and the quantization program which was carried out on those papers is very particular: the Wheeler-DeWitt equation of these models is a Klein-Gordon like equation, and the procedure is to extract a square root of it and work in a single frequency sector. Note that this is not mandatory, and there are other ways to deal with the Klein-Gordon equation working with the two frequency sectors with a well defined inner product, as it can be seen in Refs.~\cite{halli1,halli2}.
Finally, most of these references do not identify
which interpretation of quantum cosmology has been used.

In this subsection, we will focus on Ref.~\cite{CS}, where the interpretation adopted was precisely defined, the consistent histories approach, and the conclusion was that, if one takes family of histories with properties defined in just two moments of time, the infinity past and the infinity future, then the probability of a quantum bounce is null for any state. This was a remarkable result, in contradiction with the results presented in section 4
and in Ref.~\cite{gaussian}, where the dBB quantum theory was used.
Our aim here is to discuss the results of Ref.~\cite{CS} with care, and contextualize it in the framework of the de~Broglie-Bohm theory, and other quantization techniques, as the two frequencies (Klein-Gordon) approach of Ref.~\cite{halli1,halli2}, in order to understand this discrepancy of results.

We will first show that, in the single frequency approach using the consistent histories interpretation, families of histories containing properties defined in one or more moments of time, besides properties defined in the infinity past and in the infinity future, are no longer consistent, unless one takes semi-classical states, which of course corresponds to histories without a bounce. This means that in the framework of these families of histories one cannot answer whether quantum bounces take place because histories involving any genuine quantum states are inconsistent. Hence, the consistent histories approach is silent about quantum bounces happening in family of histories with more than two moments of time. Furthermore, we will show that in the induced Klein-Gordon approach, there are no consistent family of histories involving genuine quantum states. Again, the question about the existence of quantum bounces has no meaning in the induced Klein-Gordon approach.

On the contrary, if one considers the de Broglie-Bohm theory, where trajectories in configuration space are considered to be objectively real, one can show that in the two quantization procedures mentioned above, there exist plenty of non-singular bouncing trajectories which goes to the classical cosmological trajectories when the volume of the universe is big. Hence, the existence of quantum bounces in the Wheeler-DeWitt approach depends strongly on the quantum interpretation one is adopting, and on the quantization procedure one is taking.
In the end of this subsection we will discuss the physical reasons and consequences of this discrepancy.

As we are considering an isotropic and homogeneous model, the corresponding
Wheeler-DeWitt equation can be read from Eq.~(\ref{wdw40}) by freezing the anisotropic
degrees of freedom $y_1$ and $y_2$, and writing $\alpha=y_0$ and $\phi=y_3$:

\begin{equation}
\left(\partial_\phi^2 - \partial_\alpha^2\right) \Psi(\alpha,\phi) = 0,
\end{equation}
defined on the kinematical Hilbert space $L^2(\setR^2,d\alpha d\phi)$.

In order to define a probability measure, the standard procedure is to separate the positive and negative frequency modes and quantize them independently. Taking the square-root of the constraint, we get
\begin{equation}
\pm i\partial_\phi \Psi(\alpha.\phi) = \sqrt{\mathbf{\Theta}} \ \Psi(\alpha,\phi),
\end{equation}
with
 \begin{equation}\mathbf{\Theta} \dee - \partial_\alpha^2.\end{equation}

The action of $\sqrt{\mathbf{\Theta}}$ is best seen on Fourier space. Consider the set of eigenfunctions
\begin{equation}\label{eigenfunc}
e_k(\alpha) = <k|\alpha> = \frac{1}{\sqrt{2\pi}} e^{ik\alpha},
\end{equation}
such that $\mathbf{\Theta} \ e_k = \omega^2 e_k$, with
\begin{equation}
\omega\dee |k|.
\end{equation}
Restricting to the positive frequency sector, evolution is given by the propagator
\begin{equation}
U(\phi-\phi_0)= e^{i\sqrt{\mathbf{\Theta}} (\phi-\phi_0)}.
\end{equation}

The physical scalar product is given by
\begin{equation}
<\Phi|\Psi> \dee \int_{\phi=\phi_0} d\alpha\,  \bar{\Phi}(\alpha,\phi) \Psi(\alpha,\phi),
\end{equation}
and it is independent of the time $\phi_0$ on which it is defined. Positive and negative frequency sectors are orthogonal with respect to this scalar product.

We will now construct the set of histories and the corresponding decoherence functional, as
discussed in section 3.
Following Hartle's approach \cite{har} we are interested in defining a decoherence functional for a set of histories. The decoherence functional is defined as
\begin{equation}
d(h,h')\dee \langle{\Psi_{h'}}|\Psi_h\ra,
\end{equation}
where the branch wave function is given by
\begin{equation}
\Psi_h\dee C_h^\dag\ket{\Psi}.
\end{equation}
In the above formula, $\Psi$ is a given initial state and $C_h$ is the class operator defining the history $h$, given by a product of projectors
\begin{equation}
C_h\dee P^{\mathcal{O}_1}_{\Delta\lambda_{k_1}}(t_1) ... P^{\mathcal{O}_n}_{\Delta\lambda_{k_n}}(t_n),
\end{equation}
where $P^{\mathcal{O}}_{\Delta\lambda_{k}}(t)$ projects onto the subspace for which the $k$th eingenvalue of the observable $\mathcal{O}$ at time $t$ takes values in the interval $\Delta\lambda_{k}$. Here we are using Heisenberg operators for the projectors
\begin{equation}
P^{\mathcal{O}}_{\Delta\lambda_{k}}(t) \dee U^\dag(t) P^{\mathcal{O}}_{\Delta\lambda_{k}} U(t).
\end{equation}

For the present case, we consider the observable given by the scale factor $a$, or $\alpha$, with relational time $\phi$, and we will denote projectors simply by $P_{\Delta\alpha_i}(\phi_i)$. The time independent projector is given explicitly by
\begin{equation}
P_{\Delta\alpha} = \int_{\Delta\alpha} d\alpha \ket{\alpha}\langle{\alpha}|,
\end{equation}
where the ket $\ket{\alpha}$ is defined in \Ref{eigenfunc}, and the normalization is such as to make this basis orthonormal.

The set of histories considered in \cite{CS} are composed by two times, corresponding to the past infinity ($\phi\rightarrow -\infty$) and to the future infinity ($\phi\rightarrow +\infty$). The histories are separated in those where $\alpha$ is bigger or smaller than a given fixed fiducial value $\alpha_*$. Hence, $\Delta\alpha$ is the interval $(-\infty,\alpha_*)$ and $\bar{\Delta}{\alpha}$ is its
complement. The indices $(1,2)$ below designate different fiducial values for $\alpha$, $\alpha_{*1}$ and $\alpha_{*2}$.
There are four possible histories according to these possibilities, described by the following class operators
\begin{eqnarray}
C_{S-S}(-\infty,\infty) &=& P_{\Delta\alpha_1}(-\infty) P_{\Delta\alpha_2}(\infty) \no
C_{S-B}(-\infty,\infty) &=& P_{\Delta\alpha_1}(-\infty) P_{\bar{\Delta}{\alpha_2}}(\infty) \no
C_{B-S}(-\infty,\infty) &=& P_{\bar{\Delta}{\alpha_1}}(-\infty) P_{\Delta\alpha_2}(\infty) \no
C_{B-B}(-\infty,\infty) &=& P_{\bar{\Delta}{\alpha_1}}(-\infty) P_{\bar{\Delta}{\alpha_2}}(\infty) \nonumber
\end{eqnarray}
where $S$ and $B$ subscripts denote, respectively, the domains of the scale factor arbitrarily close to the singularity
or arbitrarily big.

Let us now check whether this set of histories is consistent. Noting that $P_{\Delta\alpha} P_{\bar{\Delta}{\alpha}} = 0$, the only non-trivial terms are $d(h_{S-B},h_{B-B})$ and $d(h_{S-S},h_{B-S})$. Consider for example the first of these terms
\begin{equation}\label{dfunc}
d(h_{S-B},h_{B-B}) = \langle{\Psi} | P_{\bar{\Delta}{\alpha_1}}(\phi_1) P_{\bar{\Delta}{\alpha_2}}(\phi_2) P_{\Delta\alpha_1}(\phi_1) \ket{\Psi}.
\end{equation}
As a first step, let us study the behavior of $P_{\Delta\alpha}(\phi) \ket{\Psi}$ and $P_{\bar{\Delta}{\alpha}}(\phi)$ for $\phi\rightarrow\pm\infty$. We borrow the results without proof from \cite{CS}. We have that
\begin{eqnarray}
&&\lim_{\phi\rightarrow +\infty} P_{\Delta\alpha}(\phi) \ket{\Psi} = \ket{\Psi_L} \no
&&\lim_{\phi\rightarrow -\infty} P_{\Delta\alpha}(\phi) \ket{\Psi} = \ket{\Psi_R} \no
&&\lim_{\phi\rightarrow +\infty} P_{\bar{\Delta}{\alpha}}(\phi) \ket{\Psi} = \ket{\Psi_R} \no
&&\lim_{\phi\rightarrow -\infty} P_{\bar{\Delta}{\alpha}}(\phi) \ket{\Psi} = \ket{\Psi_L}. \no
\end{eqnarray}

In the equation above we have used the left/right-moving decomposition of the wave function
\begin{eqnarray}
\Psi(\alpha,\phi)&&= \frac{1}{\sqrt{2\pi}} \int_\setR dk \, \Psi(k) e^{ik\alpha} e^{i\omega \phi} \no
&\propto& \int_{-\infty}^0 dk \, \Psi(k)e^{ik(\alpha - \phi)}+ \int_0^\infty dk \, \Psi(k)e^{ik(\alpha + \phi)}  = \no
&=& \Psi_R(v_r) + \Psi_L(v_l),
\end{eqnarray}
where $v_r \mathrel{\mathop:}= \alpha - \phi$, $v_l\mathrel{\mathop:}= \alpha+\phi$.

Since right and left moving sectors are orthogonal, the term in \Ref{dfunc} is zero, as is the other term, and the decoherence functional is diagonal for this set of histories.

Craig and Singh (\cite{CS}) go on and recalculate the probabilities of the four histories. They show that $C_{B-B}\ket{\Psi}=0$
(and also that $C_{S-S}\ket{\Psi}=0$) for any state $\ket{\Psi}$, and hence the probability for a non-singular bouncing model is zero.

\subsubsection{Histories with three times}\label{3hist}

Let us now see whether we can obtain a family of consistent histories when we ask about properties concerning the size of the universe in a third moment of time between $\phi_1 \to -\infty$ and $\phi_2\to +\infty$. Thus we want to address the question whether in an arbritary intermediary $\phi$ time
the scale factor of the universe is in the interval $(-\infty, \alpha *)$, or on its complement $[\alpha *,\infty)$.

The new family has now eight histories associated with the following class operators

\begin{eqnarray}
\label{classS2}
C_{S-\Delta\alpha -S}(\phi_1,\phi,\phi_2)&=&P_{\Delta\alpha_1}(\phi_1)P_{\Delta\alpha}(\phi)
P_{\Delta\alpha_2}(\phi_2),\nonumber\\
C_{S-\bar{\Delta}\alpha-S}(\phi_1,\phi,\phi_2)&=&P_{\Delta\alpha_1}(\phi_1)P_{\bar{\Delta}\alpha}(\phi)
P_{\Delta\alpha_2}(\phi_2),\nonumber\\
C_{S-\Delta\alpha -B}(\phi_1,\phi,\phi_2)&=&P_{\Delta\alpha_1}(\phi_1)P_{\Delta\alpha}(\phi)
P_{\bar{\Delta}\alpha_2}(\phi_2),\nonumber\\
C_{S-\bar{\Delta}\alpha -B}(\phi_1,\phi,\phi_2)&=&P_{\Delta\alpha_1}(\phi_1)P_{\bar{\Delta}\alpha}(\phi)
P_{\bar{\Delta}\alpha_2}(\phi_2),\nonumber\\
C_{B-\Delta\alpha -S}(\phi_1,\phi,\phi_2)&=&P_{\bar{\Delta}\alpha_1}(\phi_1)P_{\Delta\alpha}(\phi)
P_{\Delta\alpha_2}(\phi_2),\nonumber\\
C_{B-\bar{\Delta}\alpha -S}(\phi_1,\phi,\phi_2)&=&P_{\bar{\Delta}\alpha_1}(\phi_1)P_{\bar{\Delta}\alpha}(\phi)
P_{\Delta\alpha_2}(\phi_2),\nonumber\\
C_{B-\Delta\alpha -B}(\phi_1,\phi,\phi_2)&=&P_{\bar{\Delta}\alpha_1}(\phi_1)P_{\Delta\alpha}(\phi)
P_{\bar{\Delta}\alpha_2}(\phi_2),\nonumber\\
C_{B-\bar{\Delta}\alpha -B}(\phi_1,\phi,\phi_2)&=&P_{\bar{\Delta}\alpha_1}(\phi_1)P_{\bar{\Delta}\alpha}(\phi)
P_{\bar{\Delta}\alpha_2}(\phi_2)
\end{eqnarray}
with $S$ and $B$ having the same meaning as before.

Each of these class operators is associated with a particular
history. For instance, the class operator $C_{S-\Delta\alpha -B}(\phi_1,\phi,\phi_2)$ is associated
with the history where the universe was singular at $\phi_1\to -\infty$, has a size in a domain $\Delta\alpha$
at the finite time $\phi$, and it will be infinitely large at $\phi_2\to\infty$.

One must now see whether this new family with eight histories is consistent or not. As we have seen above, one must calculate the decoherence functional $d(h,h')$ for the histories associated with the class operators shown in Eq.~(\ref{classS2}).

It is easy to show that, in general, $d(h,h')$ is not approximately zero. This was done in Ref.~\cite{pereira}.
The final result for this off-diagonal term of the decoherence functional is

\begin{equation}
\label{decfinal}
d(h_{S-\bar{\Delta}\alpha -B},h_{S-\Delta\alpha -B}) \propto - i{\rm p.v.}\int_{\alpha * - \phi}^{\infty} dv_r \int_{-\infty}^{\alpha * - \phi} dv_r ''\biggl[\frac{\Psi (v_r)\Psi^{*} (v_r '')}{v_r '' - v_r}\biggr],
\end{equation}
which is not null in general. Due to the disjoint domains of integration, this result can be approximately zero if and only if $\Psi (v_r)$ is concentrated around some fixed value of $v_r$. The classical trajectories are given by $v_r =$ const or $v_l =$ const. Therefore, a wave function sharply concentrated around some fixed value of $v_r$ must describe a semiclassical state. It is straightforward to show that other off-diagonal terms of the decoherence functional,
e.g., $d(h_{B-\Delta\alpha -S},h_{B-\bar{\Delta}\alpha -S})$, are approximately zero only if the wave function $\Psi (v_l)$ is concentrated around the other class of classical trajectories $v_l =$ const.

Concluding, the family of histories described by the class operators (\ref{classS2}) can be made consistent only
for semiclassical states. In that case, of course, the probability of occurrence of a quantum bounce
is null, as before, but the reason for that comes from the fact that we are not allowed to calculate
probabilities in a family of cosmological histories where quantum effects are relevant. Probabilities
are calculable only for semiclassical histories, where bounces cannot occur.
More generally, if quantum effects are important in any family of cosmological histories then,
under the consistent histories approach, one cannot ask questions about properties of the
Universe at an arbitrary finite $\phi$. This is of course a limitation on the applicability of the
consistent histories approach to cosmology, at least for the present simple model. We are simply
prohibited to study the quantum properties of a cosmological model, unless one considers just
two moments of its history, at $\phi \pm \infty$, and nothing more than that.

Let us now see the case where the two frequencies sectors are considered.

\subsubsection{The Klein-Gordon Approach}\label{KGapp}

In the Wheeler-De Witt equation for a free massless scalar field, one can define its square-root and construct a Schr\"odinger-like equation as we have discussed above. However, there are other quantization schemes where the restriction to a single frequency sector is not necessary. A promising alternative approach to quantum cosmology using the consistent histories quantization is to consider the full Klein-Gordon equation. In this approach both energy sectors, positive and negative, are simultaneously taken into account but the Hilbert space is defined with a different inner product (see \cite{halli2} and references therein).

Following closely Ref.~\cite{halli1}, one can define the  eigenstates associated to the position operator as
\begin{eqnarray}
\ket{x}
&=& \frac{1}{\sqrt{2\pi}} \int_{-\infty}^{\infty} \frac{dk}{2|k|} \ e^{i|k |\phi -ik\alpha} \ket{k_{+}}\nonumber\\
&&+ \frac{1}{\sqrt{2\pi}} \int_{-\infty}^{\infty} \frac{dk}{2|k|} \ e^{-i|k |\phi -ik\alpha}\ket{k_{-}}\nonumber\\
&=&  \ket{x_{+}} + \ket{x_{-}}  \label{position}\quad,
\end{eqnarray}
where $\ket{k_{\pm}}$ are eigenstates of the $\hat{k}$ operator such that $\hat{k}\ket{k_{\pm}} = k \ket{k_{\pm}}$ and $\hat{k}_0\ket{k_{\pm}} = \pm |k|\ \ket{k_{\pm}}$. Note that the position eigenstates are not orthogonal,  i.e.
\[
\langle{x}|x'\ra=G^{(+)}(x,x')+G^{(-)}(x,x')
\]
where
\begin{equation}\label{Gpm}
G^{(\pm)}(x,x')\mathrel{\mathop:}= \frac{1}{2\pi} \int_{-\infty}^{\infty} \frac{dk}{2|k|} \ e^{\mp i |k|(\phi-\phi') \pm i k(\alpha-\alpha')}\quad ,
\end{equation}
are respectively the positive and negative Wightman functions. The positive and negative position eigenstates satisfy a completeness relation that reads
\[
\1 = i\int d\alpha \ \Big(
\ket{x_{+}}
\overleftrightarrow{\partial}_\phi
\langle{x_{+}}|-
\ket{x_{-}}\overleftrightarrow{\partial_{\phi}}\langle{x_{-}}|
\Big)\quad.
\]

Given these position eigenstates we can define the induced Klein-Gordon inner product as

\begin{equation}
\label{inner2}
(\Psi,\Phi)\mathrel{\mathop:}= i\int d\alpha \ \Big(\Psi_{+}^{*} \overleftrightarrow{\partial_{\phi}} \Phi_+ -
\Psi_{-}^{*} \overleftrightarrow{\partial_{\phi}} \Phi_-\Big)
\end{equation}
where $\Psi_{\pm}(\alpha,\phi)$ denotes the positive (negative) frequency solutions of the Klein-Gordon equation which are given by the projection of the wave function in the position eigenstates. Recalling that $v_l = \alpha + \phi$ and $v_r = \alpha - \phi$, we have
\begin{align}
\label{wavefunction+}
\Psi_{+} (\phi,\alpha) &= \langle{x_{+}} |\Psi\ra \nonumber\\
=&\frac{1}{\sqrt{2\pi}}
\biggl[\int_{0}^{\infty}{dk} \ e^{i k v_r} \Psi_{+} (k)+
\int_{-\infty}^{0} dk \ e^{i k v_l} \Psi_{+} (k)\biggr] \nonumber\\
=&\Psi_{+}^r (v_r) + \Psi_{+}^l (v_l)
\end{align}
and
\begin{align}
\label{wavefunction-}
\Psi_{-} (\phi,\alpha) &= \langle{x_{-}} |\Psi\ra \nonumber\\
=&\frac{1}{\sqrt{2\pi}}
\biggl[\int_{0}^{\infty} dk \ e^{i k v_l} \Psi_{-} (k)+
\int_{-\infty}^{0} dk \ e^{i k v_r} \Psi_{-} (k)\biggr] \nonumber \\
=&\Psi_{+}^r (v_r) + \Psi_{+}^l (v_l)
\end{align}
with
\begin{equation}
\label{wavefunctionk}
\Psi_{\pm} (k) = \frac{\langle{k_{\pm}} |\Psi\ra }{2|k|}\quad .
\end{equation}

One of the key features of this inner product is that for an arbitrary wave function the quantity
\begin{eqnarray}
\label{posinner}
(\Psi,\Psi)&=&
i\int d\alpha \ \Big(\Psi_{+}^{*} \overleftrightarrow{\partial_{\phi}} \Psi_+ -
\Psi_{-}^{*} \overleftrightarrow{\partial_{\phi}} \Psi_-\Big)\nonumber\\
&=&2\int_{-\infty}^{\infty} dk |k| \Big(\big|\Psi_{+}{(k)}\big|^2 + \big|\Psi_{-}{(k)}\big|^2\Big)\quad ,\quad
\end{eqnarray}
is positive definite.

Once again we shall be interested in calculating the probability of the universe in a given time ($\phi$) to have a size within the range $\Delta$ or in its complement $\bar{\Delta}$. For a given initial state $\Psi (\phi,\alpha)$, we can construct the decoherence functional for a set of histories as proposed in Ref.~\cite{halli1} and then take the limit of infinite past $\phi_1 \to -\infty$ and infinite future $\phi_2\to +\infty$.

The off-diagonal terms of decoherence between histories that cross the surface $\phi=$const. within region $\Delta$ or in $\bar{\Delta}$ is given by
\begin{eqnarray}
\label{decoherence}
&&D(\Delta,\bar{\Delta})=\int_{\Delta}d\alpha\int_{\bar{\Delta}} d\alpha '
\Big[\Psi_{+}^{*}(\alpha ',\phi ') \overleftrightarrow{\partial_{\phi '}} G^{(+)}(\alpha ',\phi ';\alpha ,\phi)
\overleftrightarrow{\partial_{\phi }}\Psi_+(\alpha,\phi) \nonumber\\ &+&\Psi_{-}^{*}(\alpha ',\phi ') \overleftrightarrow{\partial_{\phi '}} G^{(-)}(\alpha ',\phi ';\alpha ,\phi)
\overleftrightarrow{\partial_{\phi }}\Psi_-(\alpha,\phi)\Big]\quad.
\end{eqnarray}

Omitting the negative frequency terms and defining the region $\Delta=(-\infty, \alpha *)$ we have
\begin{eqnarray}
\label{deco2}
D(\Delta,\bar{\Delta})&=& \int_{-\infty}^{\alpha *}d\alpha\int_{\alpha *}^{\infty} d\alpha '
\bigg[\Psi_{+}^{*}(\alpha ',\phi ') \partial_{\phi '} G^{(+)}(\alpha ',\phi ';\alpha ,\phi)
\partial_{\phi }\Psi_+(\alpha,\phi) + \nonumber\\
&&\Psi_+(\alpha,\phi)\partial_{\phi} G^{(+)}(\alpha ',\phi ';\alpha ,\phi)
\partial_{\phi '}\Psi_{+}^{*}(\alpha ',\phi ')-\nonumber\\
&&\Psi_{+}^{*}(\alpha ',\phi ') \Psi_+(\alpha,\phi)
\partial_{\phi} \partial_{\phi '} G^{(+)}(\alpha ',\phi ';\alpha ,\phi) +
\nonumber\\
&& G^{(+)}(\alpha ',\phi ';\alpha ,\phi) \partial_{\phi '}\Psi_{+}^{*}(\alpha ',\phi ')\partial_{\phi}\Psi_+(\alpha,\phi)\bigg]\quad,
\end{eqnarray}
where in our specific case, the Green functions $G^{(\pm)}$ eq.~\eqref{Gpm} read
\begin{equation}
\label{green}
G^{(\pm)} = \frac{1}{2\pi} \biggl\{\int_{0}^{\infty} \frac{dk}{2|k|} e^{\pm i k(v_r-v_r ')}
+ \int_{-\infty}^{0} \frac{dk}{2|k|} \ e^{\pm  i k(v_l-v_l ')}\biggr\}.
\end{equation}

In Ref.~\cite{pereira} Eq.~(\ref{deco2}) has been evaluated term by term.
The integral
\begin{equation}
\int_{0}^{\infty} dk '' \frac{k ''}{(k''-k')(k''-k)}\quad ,
\end{equation}
has been obtained,
which has an ultra-violet logarithmic divergence at infinity, and the integral
\begin{equation}
\int_{0}^{\infty} \frac{dk ''}{k''(k''-k')(k''-k)}\quad,
\end{equation}
was also obtained, which now presents an infra-red logarithmic divergence at the origin. The crucial point is that these are different divergencies which cannot cancel each other out. In this way, the decoherence functional cannot be made diagonal, and hence we cannot construct consistent histories.

In fact, one could have anticipated this result. Note that the Wheeler-DeWitt equation we are considering is completely analogous to the Klein-Gordon equation for a massless relativistic particle. However,
as pointed out in Ref.~\cite{halli1}, where the decoherence functional was constructed for a massive relativistic particle, it was observed that the off-diagonal terms $D(\Delta,\bar{\Delta})$ may become negligible only if the region $\Delta$ and its complement are much larger than the Compton wavelength $m^{-1}$ of the particle. If we naively take the limit $m\rightarrow0$, there is no region $\Delta$ in which the off-diagonal terms can become negligible. Note, however, that the $m\rightarrow0$ limit of a Klein-Gordon particle is tricky and subtle. That is why we have constructed the decoherence functional for the equivalent of a massless scalar field from the beginning.

The conclusion is that in this approach we are stuck. Are there any other
approaches to quantum cosmology where one can go further?
Note that if one applies the de Broglie-Bohm quantum theory to the same problem, one can obtain information about the behavior of the early universe, and bohmian bouncing trajectories appear in many circumstances. We will
show now that these bohmian trajectories are non-singular.

\subsubsection{The de Broglie-Bohm approach}

From the Wheeler-DeWitt equation
\begin{equation}
\label{wdw}
-\frac{\partial ^2\Psi}{\partial \alpha ^2} +  \frac{\partial
^2\Psi}{\partial \phi ^2} = 0 \quad.
\end{equation}
we get the two real equations
\begin{equation}
\label{hoqgp}
- \biggl(\frac{\partial S}{\partial \alpha}\biggr)^2 + \biggl(\frac{\partial S}{\partial \phi}\biggr)^2
+ Q(q_{\mu}) = 0 \quad,
\end{equation}
\begin{equation}
\label{hoqg2p}
\frac{\partial}{\partial \phi} \biggl(R^2\frac{\partial S}{\partial \phi}\biggr)
- \frac{\partial}{\partial \alpha} \biggl(R^2\frac{\partial S}{\partial \alpha}\biggr) = 0 \quad,
\end{equation}
where the quantum potential reads
\begin{equation}
\label{qp1}
Q(\alpha ,\phi )\dee \frac{1}{R}\biggr[\frac{\partial^{2}R}
{\partial \alpha^{2}}-\frac{\partial^{2}R}{\partial \phi^{2}}\biggl]\quad .
\end{equation}

The guidance relations are

\begin{equation}
\label{guialpha}
\frac{\partial S}{\partial \alpha}=-\frac{e^{3\alpha}\dot{\alpha}}{N}\quad ,
\end{equation}
\begin{equation}
\label{guiphi}
\frac{\partial S}{\partial \phi}=\frac{e^{3\alpha}\dot{\phi}}{N}\quad .
\end{equation}

We can write equation Eqs.~(\ref{hoqgp}) in null coordinates,
\begin{eqnarray}
\label{nulas}
v_l\mathrel{\mathop:}=\frac{1}{\sqrt{2}}(\alpha+\phi) \quad & &\quad  \alpha\mathrel{\mathop:}=\frac{1}{\sqrt{2}} \left(v_l+v_r\right)\nonumber\\
v_r\mathrel{\mathop:}=\frac{1}{\sqrt{2}}(\alpha-\phi) \quad  & &\quad  \phi\mathrel{\mathop:}=\frac{1}{\sqrt{2}} \left(v_l-v_r\right)
\end{eqnarray}
yielding,
\begin{equation}
\left(-\frac{\partial^{2} }{\partial v_l \partial v_r }\right) \Psi \left(v_l,v_r \right) =0 \quad.
\end{equation}
The general solution is
\begin{equation}
\label{sol0}
\Psi(u,v) = F(v_l) + G(v_r) \quad,
\end{equation}
where $F$ and $G$ are arbitrary functions.
Using a separation of variable method, one can write these solutions
as Fourier transforms given by
\begin{equation}
\label{sol0k}
\Psi(v_l,v_r) = \int_{-\infty}^{\infty} d k U(k)\ e^{ikv_l} + \int_{-\infty}^{\infty} d k V(k)\ e^{ikv_r} \quad,
\end{equation}
with $U$ and  $V$ also being two arbitrary functions. If one restricts the wave function
Eq.~(\ref{sol0k}) to left or right moving components only, the $R$ function
will necessarily be a function of either $v_l$ or $v_r$, and hence the quantum potential will be null
(see Eq.~(\ref{qp1})). In this case, only classical trajectories, which are of course
singular, are allowed. This is a trivial proof that avoidance of singularities is possible if and only
if the wave function Eq.~(\ref{sol0k}) depends on both left and right moving components.

Under restriction to positive frequency solutions, one gets a subclass of the general solution Eq.~(\ref{sol0k})

\begin{equation}
\label{solf+}
\Psi(v_l,v_r) = \int_{0}^{\infty} d k \Psi(k)\ e^{ikv_l} + \int_{-\infty}^{0} d k \Psi(k)\ e^{ikv_r} \quad .
\end{equation}

Given an arbitrary  wave function $\Psi(v_l,v_r)$, the phase $S$ can be written in terms of the wave function and its complex conjugate as
\begin{equation}
S=\frac{i}{2}\ln (\Psi^{*} \Psi)-i\ln(\Psi)\quad.
\end{equation}
Using this result in the guidance relations Eq.'s~(\ref{guialpha}) and (\ref{guiphi}), one obtains that
\begin{equation}
\frac{d\alpha}{d\phi} = -\frac{\partial S/\partial \alpha}{\partial S /\partial \phi}=
-\left(\Psi\frac{\partial \Psi^{*}}{\partial \alpha}-\Psi^{*}\frac{\partial \Psi}{\partial \alpha}\right)
\left(\Psi\frac{\partial \Psi^{*}}{\partial \phi}-\Psi^{*}\frac{\partial \Psi}{\partial \phi}\right)^{-1}
\end{equation}
For the particular case of the positive frequency restriction given in Eq. (\ref{solf+}), we have

\begin{eqnarray}
\label{guitotal}
\frac{d\alpha}{d\phi} &=& -\frac{\partial S/\partial \alpha}{\partial S /\partial \phi}
\nonumber \\
&=& - \Biggl\{\int_{0}^{\infty} dk  \int_{0}^{\infty} dk ' \biggl[\Psi (k) \Psi^{*} (k') \ e^{i v_l(k-k ')}
- \Psi (-k) \Psi^{*} (-k') \ e^{-i v_r(k-k ')}\biggr] (k+k ')\nonumber \\
&&\qquad - \biggl[\Psi (-k) \Psi^{*} (k') \ e^{-i v_r k}\ e^{-i v_l k '}
- \Psi (k) \Psi^{*} (-k') \ e^{i v_l k}]\ e^{i v_r k '}\biggr] (k-k ')\Biggr\}\Bigg/\nonumber \\
&& \Biggl\{\int_{0}^{\infty} dk  \int_{0}^{\infty} dk ' \biggl[\Psi (k) \Psi^{*} (k') \ e^{i v_l(k-k ')}
+ \Psi (-k) \Psi^{*} (-k') \ e^{-i v_r(k-k ')} \nonumber \\
&&\qquad  + \Psi (-k) \Psi^{*} (k') \ e^{-i v_r k}\ e^{-i v_l k '}
+ \Psi (k) \Psi^{*} (-k') \ e^{i v_l k}]\ e^{i v_r k '}\biggr] (k+k ')\Biggr\} .
\end{eqnarray}

Let us analize Eq.~(\ref{guitotal}) in the limits $v_r\to\pm\infty$ or $v_l\to\pm\infty$.
When $v_r\to\pm\infty$, the integrals involving $\int_{0}^{\infty} dk \Psi (k) \ e^{i v_r k}$,
$\int_{0}^{\infty} dk k \Psi (k) \ e^{i v_r k}$ correspond to a Fourier transform
of square integrable functions which are null when evaluated at $v_r\to\pm\infty$. Hence we obtain
from Eq.~(\ref{guitotal}), in this limit, that

\begin{equation}
\label{inftyr}
\frac{d\alpha}{d\phi} = -1 \ \Rightarrow \ \alpha + \phi = v_l = {\rm {const}}\quad.
\end{equation}

For $v_l\to\pm\infty$, an analogous reasoning yields

\begin{equation}
\label{inftyl}
\frac{d\alpha}{d\phi} = 1 \ \Rightarrow  \ \alpha - \phi = v_r = {\rm {const}}\quad.
\end{equation}

Hence, in the regions $v_r\to\pm\infty$ and $v_l\to\pm\infty$, the bohmian trajectories emerging
from Eq.~(\ref{guitotal}) are the classical trajectories, irrespectively of the wave function.

We will now see, however, that there are a huge
class of states where the bohmian trajectories are not classical in other regions of the
$(\alpha, \phi)$ plane. For instance,
when $\Psi(k)$ is even on $k$, Eq.~(\ref{guitotal}) reads

\begin{equation}
\label{guitotalp}
\frac{d\alpha}{d\phi} = - i\frac{\int_{0}^{\infty} dk  \int_{0}^{\infty} dk ' \Psi (k) \Psi^{*} (k') \ e^{i \phi (k-k ')}
\{\sin [\alpha (k-k ')](k+k')+\sin [\alpha (k+k ')](k-k')\}}
{\int_{0}^{\infty} dk  \int_{0}^{\infty} dk ' \Psi (k) \Psi^{*} (k') \ e^{i \phi (k-k ')}
\{\cos [\alpha (k-k ')]+\cos [\alpha (k+k ')]\}(k+k')} .
\end{equation}

Note that Eq.~(\ref{guitotalp}) is anti-symmetric under the change $\alpha \rightarrow -\alpha$ and also $d\alpha / d\phi =0$ at $\alpha =0$. Consequently, the bohmian trajectories that start at $v_r\to\infty$ (infinitely big universe) cannot cross the line $\alpha =0$ and goes to the singularity at $v_r\to -\infty$ in the same way as the classical trajectories do. These bohmian trajectories are non-singular. On the other hand, if they start at the singularity in $v_l\to -\infty$, they cannot become infinitely big at $v_l\to \infty$.

Note that this result is in opposition to the consistent histories conclusion. As just argued, there is no single trajectory that can start infinitely big in the far past and goes to a singularity in the far future or the reverse because the line $\alpha=0$ cannot be crossed,
and these are exactly the only histories that the CH approach claims to have non-null probability.
Also, it is certain that there exist non-singular
bohmian trajectories, which describe exactly what the consistent histories approach has claimed to be impossible, namely, universe histories that start infinitely big in the far past and go infinitely big also in the far future. In fact, the even states shown above within the de Broglie-Bohm scenario violate the consistent histories description for all trajectories: the bohmian trajectories describe exactly what the consistent histories approach has claimed to be impossible.

One can also obtain bounces in the situation where $\Psi(k)$ is not only even on $k$ but it is also real. In that case
Eq.~(\ref{guitotalp}) simplifies to
\begin{equation}
\label{guitotalpr}
\frac{d\alpha}{d\phi} = \frac{\int_{0}^{\infty} dk  \int_{0}^{\infty} dk ' \Psi (k) \Psi^{*} (k') \sin [\phi (k-k ')]
\{\sin [\alpha (k-k ')](k+k')+\sin [\alpha (k+k ')](k-k')\}}
{\int_{0}^{\infty} dk  \int_{0}^{\infty} dk ' \Psi (k) \Psi^{*} (k') \cos [\phi (k-k ')]
\{\cos [\alpha (k-k ')]+\cos [\alpha (k+k ')]\}(k+k')} ,
\end{equation}
where we have used the fact that only even integrands can survive. This can be seen by performing a
coordinate transformation in $k$ space,
\begin{eqnarray}
\label{nulask}
u\mathrel{\mathop:}=\frac{1}{\sqrt{2}}(k+k') &\qquad  & k\mathrel{\mathop:}=\frac{1}{\sqrt{2}} \left(u+w\right)\nonumber\\
w\mathrel{\mathop:}=\frac{1}{\sqrt{2}}(k-k') & \qquad & k'\mathrel{\mathop:}=\frac{1}{\sqrt{2}} \left(u-w\right) ,
\end{eqnarray}
changing the integral domains accordingly, $\int_{0}^{\infty} du  \int_{-u}^{u} dw$, and noting
that $\Psi (u+w) \Psi (u-w)$ is even under the change $w\to -w$. Note that now Eq.~(\ref{guitotalpr}) is anti-symmetric under the change $\phi \to -\phi$ and again we have that $d\alpha / d\phi =0$, but now also at $\phi =0$. Hence, the bohmian trajectories must certainly present a bounce when they cross the line $\phi =0$, and
if they start at $v_r\to\infty$ in a classical contraction from infinity, they must necessarily end at $v_l\to \infty$ in
classical expansion to infinity, realizing a bounce at $\phi=0$ and never reaching the singularity, in a symmetric trajectory in $\phi$.
On the other hand,
if they start at the singularity in $v_l\to -\infty$, they come back to the singularity at $v_r\to -\infty$, with the turning
point taking place at $\phi=0$. Again, contrary to the consistent history conclusion, any universe history as described by
these bohmian trajectories coming from infinity must go back to infinity, and any bohmian trajectory coming from the singularity
must go back to the singularity.

Note that the line $\alpha =0$, where these non-classical behaviors are strong, corresponds, in our units, to $a_{\rm phys} = l_{\rm pl}$, where $l_{\rm pl}$ is the Planck length. All these features can be seen numerically with a particular example. Let us take, for instance,
\begin{equation}
\label{psipr}
\Psi (k) = \ e^{-(|k| - d)^2/\sigma ^2} ,
\end{equation}
with $\sigma << 1$ and $d\geq 1$. This is a real and even $\Psi (k)$, consisting of two sharply
peaked gaussians centered at $\pm d$. The wave function reads

\begin{eqnarray}
\label{solf+2}
\Psi(v_l,v_r) &=& \int_{0}^{\infty} d k \Psi(k)\ e^{ikv_l} + \int_{-\infty}^{0} d k \Psi(k)\ e^{ikv_r} \nonumber \\
&\approx& \int_{-\infty}^{\infty} d k \ e^{-(k - d)^2/\sigma ^2}\ e^{ikv_l} +
\int_{-\infty}^{\infty} d k \ e^{-(k + d)^2/\sigma ^2}\ e^{ikv_r}\quad.
\end{eqnarray}

\begin{figure}
\begin{center}
\begin{minipage}[t]{0.55\linewidth}
\includegraphics[width=\linewidth]{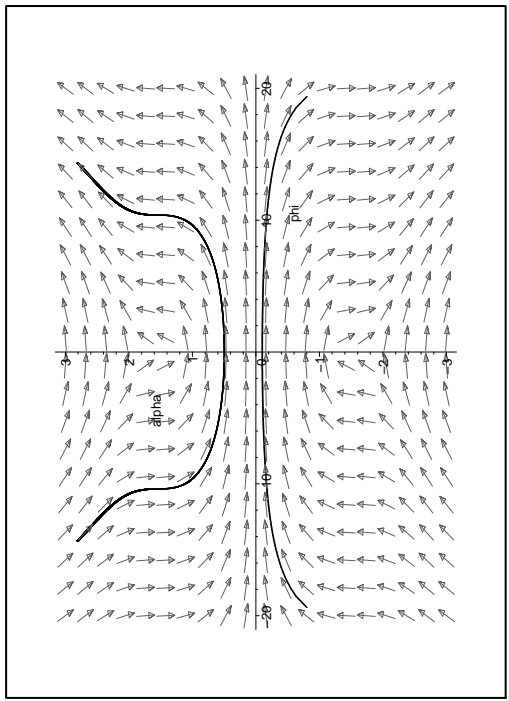}
\end{minipage} \hfill
\end{center}
\caption{The field plot shows the family of trajectories for the planar system given by (\ref{guialpha}) (\ref{guiphi}) for the wave function (\ref{solf+2}). Two of them that describe their general behavior are depicted in solid line: the first one represents a bouncing universe, while the second one corresponds to a universe which begins and ends in singular states (``big bang - big crunch'' universe). }
\label{traj}
\end{figure}
The bohmian trajectories associated with this wave function can be seen from figure 1. We can distinguish two kind of trajectories. The left half of the figure contains trajectories describing bouncing universes while the right half corresponds to universes that begins and ends in singular states (``big bang - big crunch'' universe).

In general, there is also the possibility of trajectories describing cyclic universes. In Ref.~\cite{gaussian}, it was considered bohmian trajectories associated with wave functions similar to the above one, but without the restriction to positive frequencies only. Considering both positive and negative frequencies, there are oscillatory trajectories in $\phi$. In this case, if one wishes to interpret
$\phi$ as time, this corresponds to creation and annihilation of expanding and contracting universes that exist for a very short duration. This fact suggests that one cannot understand Eq.~(\ref{hoqg2p}) as a continuity equation for an ensemble of trajectories with a distribution of initial conditions given by $R^2$ in this bohmian approach with guidance relations defined as in Eqs. (\ref{guialpha}) and (\ref{guiphi}).
In fact, this interpretation of a continuity equation would be possible only if Eq.~(\ref{hoqg2p}) could be reduced to the form
\begin{equation}
\label{cont}
\frac{\partial R^2}{\partial\phi}+\frac{\partial}{\partial\alpha}\biggl(R^2\frac{d\alpha}{d\phi}\biggr) = 0 ,
\end{equation}
with $d\alpha/d\phi$ given by Eq.~(\ref{guitotal}),

\begin{equation}
\label{fluidvelocity}
\frac{d\alpha}{d\phi} = -\frac{\partial S/\partial \alpha}{\partial S /\partial \phi} .
\end{equation}
It can be shown, using Eqs.~(\ref{hoqgp},\ref{hoqg2p},\ref{guialpha},\ref{guiphi}), that this is
possible if and only if

\begin{equation}
\frac{\partial S}{\partial \alpha}\frac{\partial^2 S}{\partial\alpha\partial\phi}=
\frac{\partial S}{\partial \phi}\frac{\partial^2 S}{\partial\phi\partial\phi} ,
\end{equation}
which implies that $ {\dot\phi}=0$, stating that $\phi$ is a monotonic function
of coordinate time. This is a strong restrictive condition, which cannot be satisfied
by general quantum states as the ones discussed above. In general, $\dot{\phi}\neq 0$.
Hence, Eq.~(\ref{hoqg2p}) cannot be interpreted as a continuity equation in $\phi$ time for the ensemble of trajectories given by
Eq.~(\ref{fluidvelocity}) with distribution $R^2$, even in the single frequency approach
where one has a Schr\"odinger-like equation.

If, however, one insists in interpreting $\phi$ as the time variable, then
one would have to face the situation of creation and annihilation of universes which is a typical
feature of relativistic quantum theory. Accordingly, the lost of a continuity equation for $R^2$ can be associated with the non-conservation of the number of trajectories of this ensemble in the $\phi$ time.

Normally the de~Broglie-Bohm theory of a Schr\"odinger-like equation
furnishes, besides the quantum trajectories, a probabilistic Born measure for these trajectories.
This is not the case here since the kinetic term present in the Schr\"odinger-like equation in the
single frequency approach
is not canonical, it is not of the form $g_{ij}(x) p^i p^j$, where $g_{ij}(x)$ has an euclidean
signature. This is crucial to obtain a continuity equation in the form of Eq. (\ref{cont}).

Concluding this subsection, in the consistent histories approach we may have the notion
of probabilities but we are not allowed to investigate non-classical properties of the
universe in any finite $\phi$ time, or to have more than two snapshots of any non-classical
universe. We have shown here that the answers given by the consistent histories approach are quite fragile. In fact, the existence of a quantum bounce strongly depends on the family of histories one is taking. One can argue that the family with only two moments of time, where quantum bounces do not exist, encompass the families of histories with three moments of time. Take, however, a genuine quantum state. In the two-time family we are sure that there is no quantum bounce, but in the three-time family this question cannot even be posed. This is characteristic of the consistent histories approach: the notion of truth depends on the family of histories one is taking. This ambiguity on the notion of a true statement in the consistency histories approach can be made quite dramatic in other circumstances \cite{ghirardi,ghirardi2}.
It seems to us that family of histories with only two times at $\pm\infty$ is so much coarse grained that no quantum effects can be
seen in between.

Finally, we would like to stress that the results of this section go much beyond the question about the existence of a quantum bounce. It shows that different quantum theories may present quite discrepant results when this system is the Universe, and the mathematical
reasons for that. This finding points out to a hope that maybe in cosmology, or in some analog model to it, one can find a way do discriminate between the many proposed quantum theories which, asides subjective and philosophical preferences, have all the same scientific status in the laboratory.
\newpage

\section{Cosmological perturbations in the de Broglie-Bohm quantum cosmology}

The conventional approach to deal with quantum cosmological perturbations is to consider a semi-classical treatment that quantize only the first order perturbations while the background
is treated classically. This was largely explored in inflationary models in order to calculate the primordial power spectrum
of scalar and tensor cosmological perturbations coming from these models, and evaluate their observational consequences.
Note, however, that such classical cosmological models generally contain initial singularities, a point where no physics is possible,
rendering them incomplete.

In the previous sections, we have obtained quantum cosmological background models which are free of singularities and particle horizons,
and can reproduce the physical properties of the standard classical cosmological model. Hence, we would like to extend the usual approach to cosmological perturbations in order to consider quantum corrections to the background evolution itself.

One very important application of the de Broglie-Bohm theory to quantum cosmology is
its use to the investigation of the dynamics of cosmological perturbations
of quantum mechanical origin in quantum cosmological backgrounds, in order to infer their consequences to the formation
of structures in the Universe, and on the anisotropies of the cosmic background radiation.
Early attempts on this approach resulted in very complicate and intractable
equations \cite{halli}. We will show in this section how the de Broglie-Bohm theory can be used
in this research in order to tremendously simplify the evolution equations of quantum
cosmological perturbations in quantum backgrounds, rendering them into a simple and solvable form,
suitable for the calculation of their observational consequences
(see Refs.~\cite{PPN1,PPN2,PPN3} for details).

\subsection{Theory of cosmological perturbations in a quantum cosmological background}

The minisuperspace bouncing non singular models we obtained in the previous sections considered
hydrodynamical fluids and scalar fields as their matter contents. In the following,
we will present the main features for the quantization of perturbations and background
in the case of perfect hydrodynamical fluids ($p=w\rho$). The case of the scalar field
is similar in many aspects, except for some conceptual issues which will be discussed afterwards.

The action we shall begin with is that of general relativity with
a perfect fluid, the latter being described in previous sections using the
formalism of Refs.~\cite{schutz1,schutz2},  i.e.
\begin{equation}
\mathcal{S}= \mathcal{S}_{_\mathrm{GR}} + \mathcal{S}_\mathrm{fluid}
= -\frac{1}{6 l_{P}^2} \int \sqrt{-g} R  {\rm {d}}^4 x - \int \sqrt{-g} p  {\rm {d}}^4 x, \label{action}
\end{equation}
where $ l_{P}=(8\pi G_N/3)^{1/2}$ is the Planck length in natural units
($\hbar=c=1$), $\rho$ is the perfect fluid energy density whose
pressure $p$ is provided by the relation $p=\omega\rho$,
$\omega$ being a nonvanishing constant.

Let the geometry of spacetime be given by
\begin{equation}
\label{perturb}
g_{\mu\nu}=g^{(0)}_{\mu\nu}+h_{\mu\nu},
\end{equation}
where $g^{(0)}_{\mu\nu}$ represents a homogeneous and isotropic
cosmological background,
\begin{equation}
\label{linha-fried}
 {\rm {d}} s^{2}=g^{(0)}_{\mu\nu} {\rm {d}} x^{\mu} {\rm {d}} x^{\nu}=N^{2}(t) {\rm {d}} t^2 -
a^{2}(t)\delta_{ij} {\rm {d}} x^{i} {\rm {d}} x^{j},
\end{equation}
where we are restricted to a flat spatial metric, and the $h_{\mu\nu}$
represents linear scalar perturbations around it, which we decompose
into
\begin{eqnarray}
\label{perturb-componentes}
h_{00}&=&2N^{2}\phi \nonumber \\
h_{0i}&=&-NaB_{,i} \\
h_{ij}&=&2a^{2}(\psi\gamma_{ij}-E_{,ij}). \nonumber
\end{eqnarray}

The case of primordial gravitational waves is very similar and easier.
For details, see Refs~\cite{PPN1,PPN2}.

Substituting Eqs.~(\ref{perturb-componentes}) and
(\ref{linha-fried}) into the Einstein-Hilbert action (\ref{action}),
performing Legendre and canonical transformations, redefining $N$
with terms which do not alter the equations of motion up to first
order, all this without ever using the background equations of
motion, the Hamiltonian up to second order is simplified to (see
Ref.~\cite{PPN3} for details)
\begin{equation}
\label{hfinal-vinculos-escalares-hidro} H=N\left[ H_0^{(0)} +
H_0^{(2)}\right] +\Lambda_N P_N + \int  {\rm {d}}^{3}x\phi\pi_{\psi}+ \int
 {\rm {d}}^{3}x\Lambda_{\phi}\pi_{\phi},
\end{equation}
where
\begin{equation}
\label{h00} H_0^{(0)}\equiv
-\frac{l^{2}P_{a}^{2}}{4aV}+\frac{P_{T}}{a^{3\omega}},
\end{equation}
and
\begin{equation}
\label{h02} H_0^{(2)}\equiv \frac{1}{2a^{3}}\int
 {\rm {d}}^{3}x\pi^{2}+\frac{a\omega}{2} \int  {\rm {d}}^{3}x v^{,i}v_{,i}.
\end{equation}

One can recognize $H_0^{(0)}$ as the hamiltonian describing the background FLRW model
with perfect fluids presented in the section 4. The second order Hamiltonian $H_0^{(2)}$ describes the dynamics of
the perturbations on this background. The conjugate variables $P_a$, $P_N$, $P_T$, $a$, $N$ and $T$ are
the degrees of freedom of the backgound, while the set of conjugate fields $\pi_{\phi}({\bf x})$,
$\pi_{\psi}({\bf x})$, $\pi({\bf x})$, $\phi({\bf x})$, $\psi({\bf x})$ and $v({\bf x})$
are the degrees of freedom describing the perturbations.

The quantities $N$, $\phi$, $\Lambda_N$ and $\Lambda_{\phi}$ play the
role of Lagrange multipliers of the constraints $H_0^{(0)} +
H_0^{(2)}\approx 0$, $\pi_{\psi}\approx 0$, $P_{N}\approx0$, and
$\pi_{\phi}\approx 0$, respectively, with $T$ playing the role of time.
The constraint $H_0^{(0)} + H_0^{(2)}$ is the one which generates the
dynamics, yielding the correct Einstein equations both at zeroth and
first order in the perturbations, as can be checked explicitly.  The
others imply that $N$, $\phi$, and $\psi$ are not relevant.
Hence, $a$ is the unique genuine degree of freedom
describing the background, while $v({\bf x})$ is the unique
field left to describe the evolution of scalar perturbations evolving on this
background, as expected. In fact $v({\bf x})$ is the usual gauge invariant Mukhanov-Sasaki variable \cite{luka1,luka2,mukha,sasa}
\begin{equation}
\label{vdefinition}
v=\frac{{a}^{\frac{1}{2}(3w-1)}}{\sqrt{6}l_P}\biggl( \varphi_{(c)}+\frac{2\sqrt{6}\sqrt{(w+1)P_{T}}\sqrt{V}}{l_P {P}_{a}\sqrt{w}} {a}^{2-3w} {\psi}\biggr)
\end{equation}
written in terms of of our background variables, where $V$ is
the comoving volume of the background spacelike hypersurfaces,
which we suppose to be compact, and $\varphi_{(c)}$ is related to
the perturbation of the velocity field describing the fluid.
The $v$ field is connected to the gauge invariant Bardeen
potential $\Phi$ (see Ref.~\cite{PPN2}) through
\begin{equation}
\label{vinculo-simples}
\Phi^{,i}\,_{,i} =
-\frac{3 l_{P}^2\sqrt{(\omega+1)\epsilon_0}}{2\sqrt{\omega}}a
\biggl(\frac{v}{a}\biggr)' .
\end{equation}

We would like to emphasize again that in order to
obtain the above results, no assumption has been made about the
background dynamics:
Hamiltonian~(\ref{hfinal-vinculos-escalares-hidro}) is ready to be
applied in the quantization procedure.

In the Dirac quantization procedure, the first class constraints
must annihilate the wave functional
$\Psi[N,a,\phi(x^i),\psi(x^�),v(x^i),T]$, yielding
\begin{eqnarray}
\label{vinculos-quanticos}
\frac{\partial}{\partial N}\Psi&=&0, \nonumber \\
\frac{\delta}{\delta\phi}\Psi&=&0, \nonumber \\
\frac{\delta}{\delta\psi}\Psi&=&0, \nonumber \\
H\Psi&=&0.
\end{eqnarray}

The first three equations impose that the wave functional does not
depend on $N$, $\phi$ and $\psi$: as mentioned above, $N$ and $\phi$
are, respectively, the homogeneous and inhomogeneous parts of the
total lapse function, which are just Lagrange multipliers of
constraints, and $\psi$ has been substituted by $v(x^i)$, the unique
degree of freedom of scalar perturbations, as expected.

As $P_T$ appears linearly in $H$, and making the gauge choice
$N=a^{3w}$, one can interpret the $T$ variable as a time
parameter. Hence, the equation
\begin{equation}
\label{schroedinger} H\Psi=0
\end{equation}
assumes the Schr\"odinger form
\begin{equation}
\imath\frac{\partial}{\partial T}\Psi =\frac{1}{4} \left\{
a^{(3w-1)/2}\frac{\partial}{\partial a} \left[
a^{(3w-1)/2}\frac{\partial}{\partial a}\right]
\right\}\Psi-\biggl[\frac{a^{3w-1}}{2}\int
d^3x\frac{\delta^2}{\delta v^2}-\frac{a^{3w+1}w}{2}\int
d^3x v^{,i}v_{,i}\biggr]\Psi ,
\end{equation}
where we have chosen the same factor ordering in $a$ as in the previous
section, and
$V$ and $l_P$ have been absorbed in redefinitions of the fields
in order to make them
dimensionless. For instance, the physical scale factor $a_{\rm
phys}$ can be obtained from the dimensionless $a$ present
in~(\ref{h00}) through $a_{\rm phys}= l_{P} a/\sqrt{V}$.

If one makes the ansatz
\begin{equation}
\label{ansatz} \Psi[a,v,T]=\Psi_{(0)}(a,T)\Psi_{(2)}[v,T]
\end{equation}
where $\Psi_{(0)}(a,T)$ satisfies the equation,
\begin{eqnarray}
\label{scrhoedinger-separado-fundo} &&\imath\frac{\partial}{\partial
T} \Psi_{(0)}(a,T)=\nonumber\\&&\frac{1}{4} \left\{
a^{(3w-1)/2}\frac{\partial}{\partial a} \left[
a^{(3w-1)/2}\frac{\partial}{\partial a}\right] \right\}
\Psi_{(0)}(a,T),
\end{eqnarray}
then we obtain for $\Psi_{(2)}(a,v,T)$ the equation
\begin{equation}
\label{scrhoedinger-separado-perturb} \imath\frac{\partial}{\partial
T} \Psi_{(2)}(a,v,T)=-\frac{a^{(3w-1)}}{2}\int
d^3x\frac{\delta^2}{\delta v^2}\Psi_{(2)}(a,v,T)+\frac{w
a^{(3w+1)}}{2}\int d^3x v^{,i}v_{,i}\Psi_{(2)}(a,v,T)
\end{equation}

Solutions of the zeroth order equation
(\ref{scrhoedinger-separado-fundo}) are known \cite{pinto,lemos-ini,fabris}
and presented in the previous sections. In the next sub-section
we will show how the second order part of this equation can
be very simplified using the de Broglie-Bohm quantum theory.

\subsection{Application of the de Broglie-Bohm theory and its consequences}

\subsubsection{The perfect fluid case}

If one uses the ontological de Broglie-Bohm quantum theory in order to obtain the bohmian trajectories $a(T)$
from Eq.~(\ref{scrhoedinger-separado-fundo}), this $a(T)$ can be
viewed as a given function of time in the second equation
(\ref{scrhoedinger-separado-perturb}). Going to conformal time
$d\eta=a^{3w-1}dT$, and performing the unitary transformation
\begin{equation}
\label{unitarias} U=e^{\{\imath[\int
d^3x\gamma^{\frac{1}{2}}\frac{\dot{a}v}{2a}]\}}e^{\{\imath[\int
d^3x(\frac{v\pi+\pi v}{2})\ln (\frac{1}{a})]\}},
\end{equation}
the Schr\"odinger functional equation for the perturbations is
transformed to
\begin{equation}
i\frac{\partial\Psi_{(2)}[v,\eta]}{\partial \eta}= \int \dd^3 x
\left(-\frac{1}{2} \frac{\delta^2}{\delta v^2} +
\frac{w}{2}v_{,i} v^{,i} - \frac{{a''}}{2a}v^2 \right)
\Psi_{(2)}[v,\eta], \label{schroedinger-conforme}
\end{equation}
where we have gone to the new quantum variable $\bar{v}=av$, the
Mukhanov-Sasaki variable defined in Ref.~\cite{MFB}, after
performing transformation (\ref{unitarias}) (we have omitted the
bars).

This is the most simple form of the Schr\"odinger equation which
governs scalar perturbations in a quantum minisuperspace model with
fluid matter source. Note that it was crucial to assume that $a(T)$
is a function of time coming from the bohmian quantization of the
background, not an operator. That is why we treated the unitary
transformation (\ref{unitarias}) as a time-dependent unitary transformation.
We do not know how to implement this
simplification without this assumption.

The corresponding time evolution equation for
the operator $v$ in the Heisenberg picture is given by
\begin{equation}
\label{ff}
v''-\omega v^{,i}_{\ ,i}-\frac{a''}{a}v=0,
\end{equation}
where a prime means derivative with respect to conformal time. In
terms of the normal modes $v_k$, the above equation reads
\begin{equation}
\label{equacoes-mukhanov} v''_k+\biggl(\omega
k^2-\frac{{a''}}{a}\biggr)v_k=0.
\end{equation}
These equations have the same form as the equations for scalar
perturbations obtained in Ref.~\cite{MFB}. This is quite natural since
for a single fluid with constant equation of state $\omega$, the pump
function $z''/z$ obtained in Ref.~\cite{MFB} is exactly equal to the term
$a''/a$ obtained here. The difference is that the function $a(\eta)$
is no longer a classical solution of the background equations but a
quantum bohmian trajectory of the quantized background, which may lead
to different power spectra.

\subsubsection{The scalar field case and the time issue again}

The technical treatment of the scalar field is rather similar.
It is also possible to simplify the full hamiltonian system by a series of canonical transformations, as shown in
Refs.~\cite{PPN2,falciano}, independently of the background equations of motion.
There is, however, a new conceptual problem because in the hamiltonian of the scalar field case
there is no obvious variable which can play the role of time in order to describe the
evolution of cosmological perturbations. Again, in the framework of the de Broglie-Bohm theory,
one can obtain a Schr\"odinger like equation for the perturbations.
We will focus on the case of a vanishing potential $U(\varphi)$,
and show how it is possible to consistently quantize simultaneously both the background and the perturbations.

The system is composed of a spatially flat Friedmann-Lema\^itre-Robertson-Walker metric (FLRW) together with
its scalar perturbations, and a free massless scalar field $\varphi\left(t,x\right)=\varphi_0 \left(t\right) +\delta \varphi\left(t,x\right)$, where $\varphi_0$ is the background homogeneous scalar field and $\delta \varphi\left(t,x\right)$ is its linear perturbation.

Using these definitions in the lagrangian density for the scalar field, namely ${\cal{L}}_{m} =\frac{1}{2}\varphi_{;\mu}\varphi^{;\mu}$, we find
for the hamiltonian constraint, after some canonical transformations (see \cite{falciano}),

\begin{equation}
\label{h0}
H_0=\frac{1}{2 e^{3\alpha}}\left[-P_{\alpha}^{2} +P_{\varphi}^2+\int d ^3x \left( \frac{\pi^2}{\sqrt{\gamma}}+\sqrt{\gamma}e^{4\alpha}v^{,i}v_{,i}\right)\right] \; ,
\end{equation}
where $v$ is the usual Mukhanov-Sasaki variable \cite{MFB} divided by $a$.

The system described by the hamiltonian $H=NH_0$ can be immediately quantized.
The Dirac's quantization procedure for constrained hamiltonian systems requires that the first class constraints must annihilate the wave-function

\begin{equation}\label{vinculoh0}
\hat{H}_0\Psi\left(\alpha,\varphi,v\right)=0 \quad ,
\end{equation}
which has only quadratic terms in the momenta.

Contrary to the hydrodynamical fluid case, the hamiltonian (\ref{h0}) does not possess any linear term in the momenta, rendering ambiguous
the choice of an intrinsic time variable. Notwithstanding, we still can define an evolutionary time for the perturbations if we use the dBB theory. The procedure is similar to what is usually done in a semiclassical approach, where a time evolution for the quantum perturbations
is induced from the classical background trajectory.
Let us summarize it in the following paragraphs.

Take the hamiltonian
$NH_0$,
with $H_0$ given in Eq. (\ref{h0}) satisfying the hamiltonian constraint
$H_0\approx 0$, and let
us solve it classically using the Hamilton-Jacobi theory. The respective Hamilton-Jacobi equation
reads
\begin{eqnarray}
&&-\frac{1}{2}\left(\frac{\partial S_T}{\partial \alpha}\right)^2+\frac{1}{2}\left(\frac{\partial S_T}{\partial \varphi}\right)^2 \nonumber \\
&&+ \frac{1}{2}\int d ^3x \left[ \frac{1}{\sqrt{\gamma}}
\left(\frac{\delta S_T}{\delta v}\right)^2+\sqrt{\gamma}e^{4\alpha}v^{,i}v_{,i}\right]
\; ,\quad \label{hjclass}
\end{eqnarray}
where the classical trajectories can be obtained from a solution $S_T$ of Eq.~(\ref{hjclass}) through
\begin{eqnarray}
\dot{\alpha}&=&- P_{\alpha}=-\frac{\partial S_T}{\partial \alpha}\quad , \nonumber \\
\dot{\varphi}&=& P_{\varphi}=\frac{\partial S_T}{\partial \varphi}\quad , \nonumber \\
\dot{v}&=&\frac{1}{\sqrt{\gamma}}\pi =\frac{1}{\sqrt{\gamma}}\frac{\delta S_T}{\delta v}
\quad , \label{guiclass}
\end{eqnarray}
where we have chosen $N=e^{3\alpha}$, and hence a time parameter $t$ (a dot means derivative
with respect to this parameter), related to conformal time
through $dt \propto a^2 d\eta$.

We will now use the fact that the $v$ variable is a small perturbation over the background
variables $\alpha$ and $\varphi$, and that its back-reaction in the dynamics of the background
is negligible. In this case, one can write $S_T(\alpha,\varphi,v)$ as
\begin{equation}
S_T(\alpha,\varphi,v)=S_0(\alpha,\varphi)+S_2(\alpha,\varphi,v) ,
\label{split}
\end{equation}
where it is assumed
that $S_2(\alpha,\varphi,v)$
cannot be splitted again into a sum involving a function of the background variables alone
(which would just impose a redefinition of $S_0$). Noting that, in order to be a solution
of the Hamilton-Jacobi Eq. (\ref{hjclass}), $S_2$ must be at least a second order functional
of $v$, then $S_2<<S_0$ as well as their partial derivatives with respect
to the background variables. Hence one obtains for the background that

\begin{eqnarray}
\dot{\alpha}&\approx &-\frac{\partial S_0}{\partial \alpha}\quad , \nonumber \\
\dot{\varphi}&\approx &\frac{\partial S_0}{\partial \varphi} \quad .
\label{guizeroth}
\end{eqnarray}

Inserting the splitting given in Eq. (\ref{split}) into Eq. (\ref{hjclass}), one
obtains, order by order:

\begin{equation}
-\frac{1}{2}\left(\frac{\partial S_0}{\partial \alpha}\right)^2+\frac{1}{2}\left(\frac{\partial S_0}{\partial \varphi}\right)^2 = 0 ,
\label{split0}
\end{equation}

\begin{equation}
-\left(\frac{\partial S_0}{\partial \alpha}\right)\left(\frac{\partial S_2}{\partial \alpha}\right)+\left(\frac{\partial S_0}{\partial \varphi}\right)\left(\frac{\partial S_2}{\partial \varphi}\right)+ \frac{1}{2}\int d ^3x \left[ \frac{1}{\sqrt{\gamma}}
\left(\frac{\delta S_2}{\delta v}\right)^2+\sqrt{\gamma}e^{4\alpha}v^{,i}v_{,i}\right]=0,
\label{split2}
\end{equation}

\begin{equation}
-\frac{1}{2}\left(\frac{\partial S_2}{\partial \alpha}\right)^2+\frac{1}{2}\left(\frac{\partial S_2}{\partial \varphi}\right)^2 + O(4) = 0 .
\label{split4}
\end{equation}

In Eq. (\ref{split4}), the symbol $O(4)$ represents terms coming from high order corrections
to the hamiltonian (\ref{h0}). As we are interested only on linear perturbations,
this term will not be relevant. The first Eq. (\ref{split0}) is the Hamilton-Jacobi
equation of the background, which solution yields, together with Eqs. (\ref{guizeroth}), the background
classical trajectories. Once one obtains the classical trajectories $\alpha(t), \varphi(t)$,
the functional $S_2(\alpha,\varphi,v)$ becomes a functional of $v$ and a function of $t$,
$S_2(\alpha,\varphi,v)\rightarrow S_2(\alpha(t),\varphi(t),v)={\bar{S}}_2(t,v)$.
Hence Eq. (\ref{split2}), using Eqs. (\ref{guizeroth}), can be written as

\begin{equation}
\frac{\partial S_2}{\partial t} + \frac{1}{2}\int d ^3x \left( \frac{1}{\sqrt{\gamma}}
\left(\frac{\delta S_2}{\delta v}\right)^2+\sqrt{\gamma}e^{4\alpha(t)}v^{,i}v_{,i}\right)=0.
\label{split2t}
\end{equation}

Equation (\ref{split2t}) can now be understood as the Hamilton-Jacobi equation coming from the
hamiltonian
\begin{equation}
\label{htau}
H_2=\frac{1}{2}\int d ^3x \left( \frac{\pi^2}{\sqrt{\gamma}}+\sqrt{\gamma}e^{4\alpha(t)}v^{,i}v_{,i}\right) ,
\end{equation}
which is the generator of time $t$ translations (and not anymore constrained to be null).

If one wants to quantize the perturbations, the correspoding Schr\"odinger equation should be
\begin{equation}
i\frac{\partial \chi}{\partial t}=\hat{H}_2\chi\, ,
\end{equation}
where $\chi$ is a wave functional depending on $v$ and $t$, and the dependence
of $\hat{H}_2$ on the background variables are understood as a dependence on $t$.

Once one has obtained the trajectories for the background variables, they can be used to define a time dependent unitary transformation for the perturbative sector in order to put $\hat{H}_2$ in a familiar form. This unitary transformation takes the vector $|\chi\rangle $ into $|\xi\rangle=U|\chi\rangle$, i.e. $|\chi\rangle=U^{-1}|\xi\rangle$. With respect to this transformation the hamiltonian is taken into $\hat{H}_2\longrightarrow \hat{H}_{2U}$ with
\begin{equation}
i\frac{d}{d t}|\xi\rangle=\hat{H}_{2U}|\xi\rangle =\left(U\hat{H}_2U^{-1}-iU\frac{d }{d t}U^{-1} \right)|\xi\rangle\quad.
\end{equation}

Let us define this unitary transformation by
\begin{equation}\label{Utransf}
U=e^{iA}e^{-iB}
\end{equation}
with,
\begin{eqnarray}
&&A=\frac{1}{2}\int d^3x\sqrt{\gamma}\frac{\dot{a}}{a^3}\hat{v}^2 \qquad ,\\
&&B=\frac{1}{2}\int d^3x\left(\hat{\pi} \hat{v}+\hat{v}\hat{\pi}\right)\log(a) \qquad .
\end{eqnarray}

Remember that the time derivative, $\dot{a}=\frac{d a}{d t}$, is taken with respect to the parametric time $t$ related to the cosmic time $\tau $ by $d \tau=N d t\propto a^3d t$. In these expressions, the scale factor $a=a(t)$ should be understood as a function of time since we suppose that the background equations have already been solved.

Naturally, the $\hat{\pi}$ and $\hat{v}$ operators do not commute with the unitary transformation. Using the following relations
\begin{eqnarray*}
e^{iA}\, \hat{v}\, e^{-iA}=\hat{v} &\quad , \quad &  e^{iA}\, \hat{\pi}\, e^{-iA}=\hat{\pi}-\frac{\dot{a}}{a^3}\sqrt{\gamma}\, \hat{v}\\
e^{-iB}\,\hat{v}\, e^{iB}=a^{-1} \, \hat{v}& \quad , \quad &e^{-iB}\, \hat{\pi}\, e^{iB}=a\hat{\pi}  \quad .
\end{eqnarray*}
we can calculate the transformed hamiltonian as
\begin{equation}\label{h2u}
\hat{H}_{2U}=\frac{a^2}{2}\int d^3x \left[ \frac{\hat{\pi}^2}{\sqrt{\gamma}}+\sqrt{\gamma}\,\hat{v}^{,i}\hat{v}_{,i}-\left(\frac{\dot{a}}{a^5}-2\frac{\dot{a}^2}{a^6} \right)\sqrt{\gamma}\,\hat{v}^2\right] \qquad
\end{equation}

Note that the unitary transformation $U$ takes us back to the Mukhanov-Sasaki variable.

Recalling that $d t=a^{-2}d \eta$, where $\eta$ is the conformal time, we have $\dot{a}=a^2a'$ and $\dot{a}=a^4a''+2a^3a'^2$, and the hamiltonian can be recast as
\begin{equation}\label{hquantica}
\hat{H}_{2U}= \frac{a^2}{2}\int d^3x \left[ \frac{\hat{\pi}^2}{\sqrt{\gamma}}+\sqrt{\gamma}\,\hat{v}^{,i}\hat{v}_{,i}-\frac{a''}{a}\sqrt{\gamma}\,\hat{v}^2\right]\; .
\end{equation}

So far our analysis has been made in the Schr\"odinger picture but now it is convenient to describe the dynamics using the Heisenberg representation. The equations of motion for the Heisenberg operators
are written as
\begin{eqnarray*}
&&\dot{\hat{v}}=-i\left[\hat{v},\hat{H}_{2U}\right]=a^2\frac{\hat{\pi}}{\sqrt{\gamma}} \qquad ,\\
&&\dot{\hat{\pi}}=-i\left[\hat{\pi},\hat{H}_{2U}\right]=a^2 \sqrt{\gamma}\left(\hat{v}^{,i}_{\phantom a ,i}+\frac{a''}{a}\hat{v}\right) \qquad.
\end{eqnarray*}

Combining these two equations and changing to conformal time, we find
the following equations for the operator modes of wave number $k$, $v_k$:
\begin{equation}\label{eqv}
v_k''+ \left(k^2-\frac{a''}{a}\right)v_k \quad =0 \qquad .
\end{equation}

This is the same equation of motion for the perturbations known in the literature, in the absence of a scalar field potential,
which appears in Ref.~\cite{MFB}.

Let us now go one step further and quantize both the background and perturbations.
When the background is also quantized, this procedure can also be implemented
in the framework of the dBB interpretation of quantum theory, where there is a definite notion of trajectories as well, the bohmian trajectories. In order to do that, we
first note that Eqs. (\ref{vinculoh0}) and (\ref{h0}) imply that
\begin{equation}
\label{split-h0}
(\hat{H}_0^{(0)}+\hat{H}_2) \Psi = 0,
\end{equation}
where
\begin{eqnarray}
&&\hat{H}_0^{(0)}=-\frac{\hat{P}_\alpha^2}{2}+\frac{\hat{P}_\varphi^2}{2} \quad , \\
&&\hat{H}_2=\frac{1}{2}\int d ^3x \left( \frac{\hat{\pi}^2}{\sqrt{\gamma}}+\sqrt{\gamma}e^{4\hat{\alpha}}\hat{v}^{,i}\hat{v}_{,i}\right) \quad .
\end{eqnarray}

We write the wave functional $\Psi$ as $\Psi=\exp(A_T+iS_T)\equiv R_T\exp(iS_T)$,
where both $A_T$ and $S_T$ are real functionals.
Inserting it in the Wheeler-DeWitt Eq. (\ref{split-h0}), the two real
equations we obtain are

\begin{equation}
\label{Thoqg}
-\frac{\partial}{\partial \alpha}
\biggl(R_T^2\frac{\partial S_T}{\partial \alpha}\biggr)
+\frac{\partial}{\partial \varphi}
\biggl(R_T^2\frac{\partial S_T}{\partial \varphi}\biggr)
+\int\frac{d^3 x}{\sqrt{\gamma}}\frac{\delta}{\delta v}
\biggl(R_T^2\frac{\delta S_T}{\delta v}\biggr)= 0 \quad,
\end{equation}

\begin{eqnarray}
&-&\frac{1}{2}\left(\frac{\partial S_T}{\partial \alpha}\right)^2+\frac{1}{2}\left(\frac{\partial S_T}{\partial \varphi}\right)^2
+ \frac{1}{2}\int d ^3x \left( \frac{1}{\sqrt{\gamma}}
\left(\frac{\delta S_T}{\delta v}\right)^2+\sqrt{\gamma}e^{4\alpha}v^{,i}v_{,i}\right)\nonumber \\
&+&\frac{1}{2R_T}\left(\frac{\partial ^2 R_T}{\partial \alpha ^2}-
\frac{\partial ^2 R_T}{\partial \varphi ^2}\right) -
\frac{1}{2}\int\frac{d^3 x}{\sqrt{\gamma}}\frac{1}{R_T}\frac{\delta^2 R_T}{\delta v^2}= 0
\; .\quad \label{hjquant}
\end{eqnarray}

The bohmian guidance relations are the same as in the classical case,
\begin{eqnarray}
\dot{\alpha}&=&- P_{\alpha}=-\frac{\partial S_T}{\partial \alpha}\quad , \nonumber \\
\dot{\varphi}&=& P_{\varphi}=\frac{\partial S_T}{\partial \varphi}\quad , \nonumber \\
\dot{v}&=&\frac{1}{\sqrt{\gamma}}\pi =\frac{1}{\sqrt{\gamma}}\frac{\delta S_T}{\delta v}
\quad , \label{guiq}
\end{eqnarray}
with the difference that the new $S_T$ satisfies a Hamilton-Jacobi equation different
from the classical one due to the presence of the quantum potential terms (the
last two terms in Eq. (\ref{hjquant})), which are responsible for the quantum effects.
We have again made the choice $N\propto e^{3\alpha}$.

Let us assume, as in the classical case, that we can split
$A_T(\alpha,\varphi,v)=A_0(\alpha,\varphi)+A_2(\alpha,\varphi,v)$
implying that $R_T(\alpha,\varphi,v)=R_0(\alpha,\varphi)R_2(\alpha,\varphi,v)$, and
$S_T(\alpha,\varphi,v)=S_0(\alpha,\varphi)+S_2(\alpha,\varphi,v)$, and that $A_2 << A_0$,
$S_2 << S_0$, together with their derivatives with respect to the background variables.
The approximate guidance relations are
\begin{eqnarray}
\dot{\alpha}&\approx &-\frac{\partial S_0}{\partial \alpha}\quad , \nonumber \\
\dot{\varphi}&\approx &\frac{\partial S_0}{\partial \varphi}\quad \; \; ,
\label{guizerothq}
\end{eqnarray}
and the zeroth order terms of Eqs. (\ref{Thoqg}) and (\ref{hjquant}) read

\begin{equation}
\label{0Thoqg2}
-\frac{\partial}{\partial \alpha}
\biggl(R_0^2\frac{\partial S_0}{\partial \alpha}\biggr)
+\frac{\partial}{\partial \varphi}
\biggl(R_0^2\frac{\partial S_0}{\partial \varphi}\biggr)
\approx 0 \quad,
\end{equation}
\begin{equation}
-\frac{1}{2}\left(\frac{\partial S_0}{\partial \alpha}\right)^2+\frac{1}{2}\left(\frac{\partial S_0}{\partial \varphi}\right)^2
+\frac{1}{2R_0}\left(\frac{\partial ^2 R_0}{\partial \alpha ^2}-
\frac{\partial ^2 R_0}{\partial \varphi ^2}\right) \approx 0
\; .\quad \label{0hjquant}
\end{equation}

A solution $(S_0,R_0)$ of Eqs. (\ref{0Thoqg2}) and (\ref{0hjquant}) yield a bohmian quantum
trajectory for the background through Eq. (\ref{guizerothq}) as those obtained in Ref. \cite{gaussian}.

As in the classical case, once one obtains the bohmian quantum trajectories $\alpha(t), \varphi(t)$,
the functionals $S_2(\alpha,\varphi,v)$, $A_2(\alpha,\varphi,v)$ become  functionals of $v$ and functions of $t$,
$S_2(\alpha,\varphi,v)\rightarrow S_2(\alpha(t),\varphi(t),v)={\bar{S}}_2(t,v)$,
$A_2(\alpha,\varphi,v)\rightarrow A_2(\alpha(t),\varphi(t),v)={\bar{A}}_2(t,v)$.

Defining $\chi(\alpha,\varphi,v)\equiv R_2(\alpha,\varphi,v)\exp(iS_2(\alpha,\varphi,v))$,
writing it as
\begin{equation}
\label{defF}
\chi(\alpha,\varphi,v)=\int{d \lambda}\,G(\lambda,v)F(\lambda,\alpha,\phi)\quad ,
\end{equation}
where $F$ satisfies
\begin{equation}
\label{F}
\frac{1}{2}\left(\frac{\partial^2 F}{\partial \alpha^2}-\frac{\partial^2 F}{\partial \varphi^2}\right)+\frac{1}{R_0}\left(\frac{\partial R_0}{\partial \alpha}\frac{\partial F}{\partial \alpha}-\frac{\partial R_0}{\partial \varphi}\frac{\partial F}{\partial \varphi}\right)=0\, ,
\end{equation}
and $G$ is an arbitrary functional of $v$, which also depends on an integration constant $\lambda$,
then the next-to-leading-order terms of Eqs. (\ref{Thoqg}) and (\ref{hjquant}) read

\begin{equation}
\label{Thoqg2}
\frac{\partial {\bar{R}}_2^2}{\partial t}
+\int\frac{d^3x}{\sqrt{\gamma}}\frac{\delta}{\delta v}
\biggl({\bar{R}}_2^2\frac{\delta {\bar{S}}_2}{\delta v}d^3 x\biggr)= 0 \quad,
\end{equation}

\begin{equation}
\frac{\partial {\bar{S}}_2}{\partial t}
+ \frac{1}{2}\int d ^3x \left[ \frac{1}{\sqrt{\gamma}}
\left(\frac{\delta {\bar{S}}_2}{\delta v}\right)^2+\sqrt{\gamma}e^{4\alpha(t)}v^{,i}v_{,i}\right]
-\frac{1}{2}\int\frac{d^3 x}{{\bar{R}}_2\sqrt{\gamma}}\frac{\delta^2 {\bar{R}}_2}{\delta v^2}
= 0
\; ,\quad \label{hjquant2}
\end{equation}
where ${\bar{R}}_2(t,v)\equiv\exp({\bar{A}}_2(t,v))$, and we have used that
\begin{equation}
\label{essential}
-\left(\frac{\partial S_0}{\partial \alpha}\right)\left(\frac{\partial S_2}{\partial \alpha}\right) +
\left(\frac{\partial S_0}{\partial \varphi}\right)\left(\frac{\partial S_2}{\partial \varphi}\right)
= \frac{\partial {\bar{S}}_2}{\partial t},
\end{equation}
\begin{equation}
\label{essential2}
-\left(\frac{\partial R_0}{\partial \alpha}\right)\left(\frac{\partial R_2}{\partial \alpha}\right) +
\left(\frac{\partial R_0}{\partial \varphi}\right)\left(\frac{\partial R_2}{\partial \varphi}\right)
= \frac{\partial {\bar{R}}_2}{\partial t},
\end{equation}
which are possible only because of the guidance
relations, a feature of the dBB theory.

These two equations can be grouped into a single Schr\"odinger equation
\begin{equation}
\label{xo}
i\frac{\partial \bar{\chi}}{\partial t}=\hat{H}_2\bar{\chi}\, ,
\end{equation}
where $\bar{\chi}(t,v)=\chi(\alpha(t),\varphi(t),v)$ is a wave functional depending on $v$ and $t$,
and, as before, the dependence
of $\hat{H}_2$ on the background variables are understood as a dependence on $t$.

From here on we implement the same unitary transformation as the one defined in Eqs. (\ref{Utransf}),
but now the time function $a(t)$ is the calculated background bohmian trajectory of the scale factor
associated with the zeroth order equation $\hat{H}_0^{(0)}|\Psi\rangle=0$, which does not follow the classical
scale factor evolution. As before, we obtain the equation
\begin{equation}\label{eqv2}
v_k''+ \left(k^2-\frac{a''}{a}\right)v_k \quad =0 \qquad .
\end{equation}

The crucial point is that we have not used the background equations of motion. Thus we have shown that Eq. (\ref{eqv2}) is well defined, independently of the background dynamics, and it is correct even if we consider quantum background trajectories.
Hence, in spite of the fact that Eq. (\ref{eqv}) of the semi-classical limit and Eq. (\ref{eqv2}) have the same form, the time dependent
potential $a''/a$ in Eq. (\ref{eqv2}) can be rather different because it is calculated from bohmian trajectories,
not from the classical ones. This can give rise to different effects in the region where the quantum effects
on the background are important, which can propagate to the classical region.

During our procedure, we have supposed that the evolution of the background is independent of the perturbations.
This no back-reaction assumption is based on the fact that
terms induced by the linear perturbations in the zeroth order hamiltonian
are negligible, which should be the case when one assumes that quantum perturbations are initially in
a vacuum quantum state.

Note, however, that this result was obtained using a specific subclass of wave functionals which satisfies
the extra condition Eq. (\ref{F}). What are the physical assumptions behind this choice?

When one approaches the classical limit, where $R_0$ is a slowly varying function of
$\alpha$ and $\varphi$, condition (\ref{F}) reduces to

\begin{equation}
\label{Fc}
\frac{\partial^2 F}{\partial \alpha^2}-\frac{\partial^2 F}{\partial \varphi^2}\approx 0\, .
\end{equation}
If Eq. (\ref{Fc}) were not satisfied, one would not obtain anymore the usual Schr\"odinger equation
for quantum perturbations in a classical background (which arises when $R_0$ is a slowly varying
function of $\alpha$ and $\varphi$), due to extra terms in Eqs.~(\ref{Thoqg2}) and
(\ref{hjquant2}): there would be corrections originated
from some quantum entanglement between the background and the perturbations, even when
the background is already classical, which would spoil
the usual semiclassical approximation. This could be a viable possibility driven by a different
type of wave functional than the one considered here, but it seems that our Universe is not so complicate.
In fact, the observation that the simple semiclassical model without this sort of
entanglement works well in the real Universe indicates something about the wave functional of
the Universe. In other words, the validity
of the usual semiclassical approximation imposes Eq. (\ref{Fc}).

When $R_0$ is not slowly varying and quantum effects on the background become important
causing the bounce, the two last terms of condition (\ref{F}) cannot be neglected. They would also
induce extra terms in Eqs. (\ref{Thoqg2}) and
(\ref{hjquant2}), again originated
from some quantum entanglement between the background and the perturbations, but now in
the background quantum domain, and the
final quantum Eq. (\ref{eqv}) for the perturbations we obtained would not be valid around the
bounce. In this case, there is no observation indicating us which class of wave functionals
one should take, and our choice resides only on assumptions of simplicity:
there is no quantum entanglement between the background and the perturbations in the entire
history of the Universe. This is the physical hypothesis behind the choice of the specific
class of wave functionals satisfying condition (\ref{F}). Relaxing this hypothesis may lead
to new effects which should be investigated.

Note that, although the original Wheeler-DeWitt equation has no time in it, we were able
to construct a Schr\"dinger equation for the perturbations. In this derivation, the assumption
of the existence of a bohmian background quantum trajectory was essential, see Eq.~(\ref{essential}).
Hence the picture is the following: in the dBB approach no probability measure is required a
priory. In fact, there is no notion of probability for the possible backgrounds. However,
this does not forbid us to obtain the bohmian trajectories for the background through the
guidance relations. The background will follow one of the possible solutions yielding a
definite background trajectory, which can then be used in the equations for the perturbations
in order to yield a Schr\"odinger equation for them. Then, using the result of Ref.~\cite{valentini}
discussed in section 3,
the distribution of initial conditions for the perturbations will soon converge to satisfy the
Born rule, and we are back to standard quantum theory of cosmological perturbations, now evolving
in a background which does not always present the classical behaviour and, thanks to that,
is free of singularities.

In the next section we will obtain the power spectrum for the perfect fluid case using Eq. (\ref{equacoes-mukhanov})
and the background quantum solution (\ref{at}). We will also discuss the results for the case of
tensor perturbations and in the situation where a cosmological constant is present. As we
shall see, we obtain some slight different results from the ones obtained in inflationary models.

\subsection{Observational aspects}

Having obtained in the previous section the propagation equation for
the full quantum scalar modes
in the bohmian picture with the scale factor assuming the form
(\ref{at}), it is our goal now to solve this equation in
order to obtain the scalar perturbation power spectrum as predicted
by such models.

We shall begin with the asymptotic behaviors. When $|T|\gg |T_0|$,
far from the bounce, one can write Eq.~(\ref{equacoes-mukhanov}) as

\begin{equation}
\label{Modes} v'' +\left[ \omega k^2
+\frac{2(3\omega-1)}{(1+3\omega)^2\eta^2}\right]\mu = 0,
\end{equation}
whose solution is
\begin{equation}
\label{Bessel} v = \sqrt{\eta} \left[ c_1(k) H^{(1)}_\nu
(\bar{k}\eta)+ c_2(k)
  H^{(2)}_\nu(\bar{k}\eta)\right],
\end{equation}
with
$$ \nu = \frac{3(1-\omega)}{2(3\omega+1)}, $$ $c_1$ and $c_2$ being
two constants depending on the wavelength, $H^{(1,2)}$ being Hankel
functions, and $\bar{k}\equiv \sqrt{\omega}k$.

This solution applies asymptotically, where one can impose vacuum
initial conditions as in \cite{MFB}

\begin{equation}
v_\mathrm{ini} =
\frac{\exp {i \bar{k} \eta}}{\sqrt{\bar{k}}},
\label{v}
\end{equation}
which
implies that
$$ c_1=0 \quad \hbox{and} \quad c_2= l_{P} \sqrt{\frac{3\pi}{2}}
 \exp^{-i\frac{\pi}{2} \left(\nu+\frac{1}{2}\right)}.
$$

The solution can also be expanded in powers of $k^2$ according to
the formal solution~\cite{MFB}
\begin{eqnarray}
\frac{v}{a} & \simeq & A_1(k)\biggl[1 - \omega k^2 \int^t \frac{d\bar
  \eta}{a^2\left(\bar \eta\right)} \int^{\bar{\eta}}
  a^2\left(\bar{\bar{\eta}}\right)d\bar{\bar{\eta}}\biggr]\nonumber
  \\ &+& A_2(k) \biggl[\int^\eta\frac{d\bar{\eta}}{a^2} - \omega k^2
  \int^\eta \frac{d\bar{\eta}}{a^2} \int^{\bar{\eta}} a^2
  d\bar{\bar{\eta}} \int^{\bar{\bar{\eta}}}
  \frac{d\bar{\bar{\bar{\eta}}}}{a^2} \biggr],\cr & & \label{solform}
\end{eqnarray}
up to order $ \mathcal{O}(k^{j\geq 4})$ terms. In Eq.~(\ref{solform}),
the coefficients $A_1$ and $A_2$ are two constants depending only on
the wavenumber $k$ through the initial conditions. Although this form
is particularly valid as long as $\omega k^2\ll a''/a$,  i.e. when the
mode is below its potential, Eq.~(\ref{solform}) should formally apply
for all times. In the matching region $\omega k^2\approx a''/a$, the
$\mathcal{O}(k^2)$ terms may give contributions to the amplitude, but
they do not alter the $k$-dependence of the power spectrum. Using this
procedure, we can evaluate $A_1$ and $A_2$, yielding

\begin{eqnarray}
A_1 &\propto&
\biggl(\frac{\bar{k}}{k_0}\biggr)^\frac{3\left(1-\omega\right)}
{2\left(3\omega+1\right)},\label{A1}\\ A_2 &\propto& \biggl(\frac{\bar{k}}{k_0}
\biggr)^\frac{3\left(\omega-1\right)}{2\left(3\omega+1\right)},\label{A2}
\end{eqnarray}
where $k^{-1}_0=T_0a_0^{3\omega-1}=l_c^B$ and $l_c^B$ is the curvature scale at the bounce.

The relation between the Bardeen potential $\Phi$ and $v$ is given by

\begin{equation}
\label{vinculo-simples2}
\Phi^{,i}\,_{,i} =
-\frac{3 l_{P}^2\sqrt{(\omega+1)\epsilon_0}}{2\sqrt{\omega}}a
\biggl(\frac{v}{a}\biggr)' .
\end{equation}
At large positive values of $T$, when $v_k$ is leaving
the potential and $T\propto\eta^{3(1-\omega)/(1+3\omega)}$, the
constant mode of $\Phi$, like $v$, mixes $A_1$ with $A_2$. In this
region, taking into account that $A_2$ dominates over $A_1$, we
obtain:
\begin{equation}
\Phi \propto
k^\frac{3\left(\omega-1\right)}{2\left(3\omega+1\right)}
\biggl[\mathrm{const.}+\frac{1}{\eta^{(5+3\omega)/(1+3\omega)}}\biggr].
\end{equation}
The power spectrum
\begin{equation}
\label{PS} \mathcal{P}_\Phi \equiv \frac{2 k^3}{\pi^2}
\left| \Phi \right|^2,
\end{equation}
then reads
\begin{equation}
\mathcal{P}_\Phi \propto k^{n_{_\mathrm{S}}-1},
\label{powspec}
\end{equation}
and we get
\begin{equation}
\label{indexS} n_{_\mathrm{S}} = 1+\frac{12\omega}{1+3\omega}.
\end{equation}

For gravitational waves (see Ref.~\cite{PPN2} for details), the
equation for the modes $\mu = h/a$ reads
\begin{equation}
\label{mu} \mu''+\left( k^2 +2Ka -\frac{a''}{a} \right)\mu =0,
\end{equation}
yielding the power spectrum for long wavelengths:
\begin{equation}
\label{PT} \mathcal{P}_h \equiv \frac{2 k^3}{\pi^2}
\left| \frac{\mu}{a} \right|^2.
\end{equation}
In Ref.~\cite{PPN2}, we have obtained
\begin{equation}
\mathcal{P}_h \propto k^{n_{_\mathrm{T}}},
\label{powspect}
\end{equation}
with, as for the scalar modes,
\begin{equation}
\label{indexT} n_{_\mathrm{T}} = \frac{12\omega}{1+3\omega}.
\end{equation}

Note that in the limit $\omega\rightarrow 0$ (dust) we obtain a scale
invariant spectrum for both tensor and scalar perturbations.  This
result was confirmed through numerical
calculations, which also give the amplitudes.  However, it is not
necessary that the fluid that dominates during the bounce be dust. The
dependencies on $k$ of $A_1$ and $A_2$ are obtained far from the
bounce, and they should not change in a transition, say, from matter
to radiation domination in the contracting phase, or during the
bounce. The effect of the bounce is essentially to mix these two
coefficients in order for the constant mode to acquire the scale
invariant piece. Hence, the bounce itself may be dominated by another fluid,
like radiation. The important point is that if while entering the potential ($k^2\approx a''/a)$
the fluid that
dominates is dust-like, then the spectrum of perturbations for such wavelengths should be almost scale
invariant.

We have also solved the equations numerically and obtained the values
of the free parameters which best fit the data. First, we assumed a
spectral index limited by $n_{_\mathrm{S}} \leq 1.01$, an
admittedly conservative constraint, which, given Eq.~(\ref{indexS}),
provide the already severe bound on the equation of state
$\omega \leq 8\times 10^{-4}$. Constraining the amplitude with
WMAP data (see Ref. \cite{PPN4} for details) then implies the characteristic bounce
length-scale $L_0$ (the curvature scale at the bounce) to be
$$
L_0 \geq 1500  l_{P},
$$
a value consistent with our use of quantum cosmology: it is indeed
in this kind of distance scale ranges that one expect quantum effects to be
of some relevance, while at the same time the Wheeler-de Witt equation
to be valid without being possibly spoiled by some discrete nature of
geometry coming from loop quantum gravity, and/or string effects.

An extremely important point in all models of primordial
perturbations is that of comparison with other models. In particular,
one wants to devise tests allowing to discriminate among them.
With this in mind, we have also calculated the amplitude of primordial gravitational waves (tensor perturbations)
in the quantum bouncing models described above \cite{denis}. We have shown that the stochastic background
of relic gravitons
cannot be strong enough to be directly detected by present and future planned experiments.
The strain spectrum is quite different from the cyclic and inflationary
scenarios. While these two models have spectra $\approx k^{-2}, \approx k^{-1}$ at dust domination, and $\approx k^{-1}, \approx k^{0}$ at radiation domination, respectively, our model have spectra $\approx k^{-2}$ and $\approx k^{0}$ at the same eras. Our calculations have shown that the resulting amplitude is too small to be detected by any gravitational wave detector. In particular, the sensitivity of the future third generation of gravitational wave detectors, as for example the Einstein Telescope, could reach $\Omega_{\rm GW} \sim 10^{-12}$ at the frequency range $10-100\,{\rm Hz}$ to an observation time of $\sim 5$ years and with a signal-to-noise ratio $({\rm S/N})\sim 3$. Therefore, any detection of relic gravitons, in this frequency range, will rule out this type of quantum bouncing model as a viable cosmological model of the primordial universe, as it can be seen from Fig. 2.

\begin{figure}
\begin{center}
\begin{minipage}[t]{0.55\linewidth}
\includegraphics[width=\linewidth]{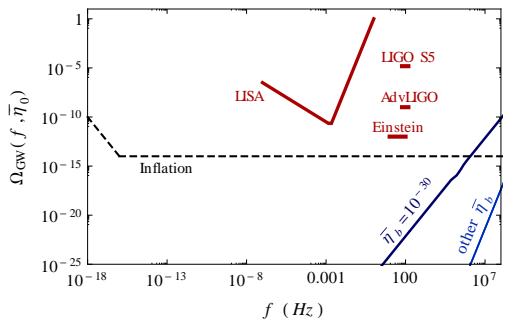}
\end{minipage} \hfill
\end{center}
\caption{The field plot shows a comparison of our results, labeled by ${\bar{\eta}}_b$ (the smaller this
parameter the bigger is the energy scale of the bounce, and the value $10^{-30}$ is just two orders of magnitude
away from the Planck scale) with experimental sensitivities and a prediction of the upper limits on the spectrum of primordial gravitational waves generated in inflationary models. The other show the sensitivities achieved by LIGO's 5th run and the ones predicted for Advanced LIGO and LISA.}
\label{traj2}
\end{figure}

Another investigation concerns the presence of a cosmological constant on such models.
As we have seen, in bouncing models without a cosmological constant,
vacuum initial conditions for quantum cosmological
perturbations are set in the far past of the contracting phase, when
the Universe was very big and almost flat, justifying the choice of an
adiabatic Minkowski vacuum in that phase. However, if a cosmological
constant is present, the asymptotic past of bouncing models will
approach de Sitter rather than Minkowski spacetime. Furthermore, the
large wavelengths today become comparable with the Hubble radius in
the contracting phase when the Universe was still slightly influenced by the
cosmological constant.
Hence, the existence of a cosmological constant
can modify the spectrum and amplitude of cosmological perturbations.
Note that this is not a question for inflation because initial conditions
for quantum perturbations and the moment of Hubble radius
crossing in such models take place when the cosmological constant is completely
irrelevant.

The main difference is originated from processes much before the bounce, when the initial conditions are
set and the cosmological constant is relevant. In that case, a Minkowski adiabatic vacuum can
only be defined in a precise time domain, i.e., at the end of cosmological constant domination,
but when the Universe was still very big and rarefied. However, even in this time domain, as the length scale
associated with the cosmological constant, given by the present acceleration of the Universe,
is not much bigger than the long wavelengths of physical interest today,
the spectrum of these scales can still be slightly affected by the cosmological constant.
And indeed we have shown \cite{bia}, analytically and numerically, that the usual result for
bouncing models, namely, that the fluid should satisfy $w\approx 0$ in order to have an
almost scale invariant spectrum of long wavelength perturbations, still holds,
but the presence of the cosmological constant
induces small oscillations and a small running towards a red-tilted spectrum for these scales.
This may lead to small oscillations in the spectrum of temperature fluctuations in
the cosmic background radiation at large scales superimposed to the usual acoustic oscillations.

Note that all these results were only possible to obtain because we have a scale factor
function defined even at the quantum bounce due to the de Broglie-Bohm prescription
given by guidance relations. Because of that, the evolution of scalar and tensor perturbations through
the quantum bounce is then smooth and suitable for analytical and numerical calculations.

\begin{figure}
\begin{center}
\begin{minipage}[t]{0.55\linewidth}
\includegraphics[width=\linewidth]{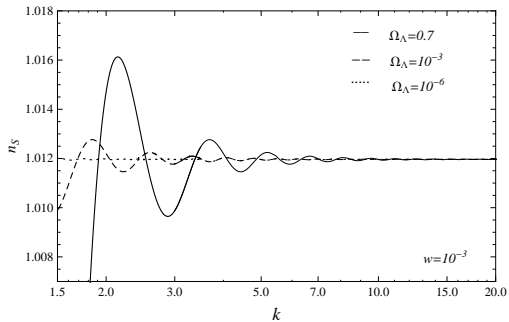}
\end{minipage} \hfill
\end{center}
\caption{Numerical results for $n_S(k)$ in the presence of a cosmological constant. The solid line indicates the result obtained using $\Omega_{\Lambda}=0.7$, the dashed line for $\Omega_{\Lambda}=10^{-3}$ and the dotted line for $\Omega_{\Lambda}=10^{-6}$.
Note that the oscillations become smaller for
smaller $\Omega_\Lambda$, showing that they are due to the presence of the cosmological constant.}
\label{traj3}
\end{figure}

In order to complete this section related to cosmological perturbations, the dBB quantum theory
can also be very useful already at the semi-classical level in order to explain
the quantum to classical
transition of quantum cosmological perturbations propagating in a standard classical cosmological
model with inflation. This transition from quantum to classical fluctuations is plagued with important conceptual issues, most of them related to the application of standard quantum theory to the Universe as a whole.
In fact, there is no satisfactory justification on why the initial quantum vacuum state of the perturbations, which is translational and rotational invariant, and which remains so during Schr\"odinger time evolution, results in a non-translational invariant state characteristic of our inhomogeneous Universe. In particular, even when there is suitable decoherence, which suppresses interference between different non-invariant terms into which the quantum state can be decomposed, it is not explained why one of these terms is selected. According to standard quantum theory, a transition to a non-invariant state could only be obtained by collapse. For instance, in a spherically symmetric $s$-wave decay state of an atom,
there are actual photons which are detected in particular directions. This happens because the detection involves a classical apparatus, a separate entity outside the quantum system which is not described by the wave function, and which brings to actual facts the potentialities inscribed in the $s$-wave. Once the system reaches the classical apparatus, one can invoke the collapse postulate, which will yield a definite direction where the photon is detected. In the cosmological context, however, there is something qualitative new: here we are talking about cosmological primordial fluctuations, that is, fluctuations which will give rise to all structures in the Universe. Hence, in the cosmological case, there is no classical apparatus outside the quantum system, the quantum fluctuations should themselves generate all the structures, including inhomogeneous classical systems and possible classical apparata. However, as said above, the wave function of the perturbations is always translational and rotational
invariant, so how these symmetries can be broken without an external agent, like in the $s$-wave example? On the other hand, how can intrinsically inhomogeneous classical apparata exist without prior braking of these symmetries of the wave function?

These issues can easily be overcome in the framework of the de Broglie-Bohm quantum theory.
As we have seen, in this theory the Universe is described by a universal wave function, together with an actual configuration for gravity (e.g.\ an actual three metric) and a configuration for the matter (e.g.\ particle positions or fields). This assumption, namely, the existence of an actual gravity-matter trajectory in configuration space, is the essential point of the dBB theory which breaks the translational and rotational invariance of the quantum system, in spite of the fact that the universal wave function always respects these symmetries, because in the dBB approach the wave function is not enough to describe the quantum physical system. Hence, although the quantum state remains translational and rotational invariant in this framework, the actual dBB field configuration corresponding to the perturbations breaks this symmetry.
For example, in the simpler case of the $s$-wave decay of an atom, one can incorporate the apparatus in the quantum description (it is not anymore a separate classical entity), and a particular photon detection happens because there is a particular atom-photon-apparatus configuration which is actually taking place, guided by the wave function.
In addition, the dBB theory allows for a simple and unambiguous characterization of the classical limit.

In fact, in Ref. \cite{ward} this quantum to classical transition was described with details from the dBB perspective.
We have shown that, assuming a quantum vacuum
state for the perturbations, the bohmian trajectories for each perturbation mode are non-classical while they are smaller than
the Hubble radius in
the beginning of inflation, and then become classical when they get bigger than the Hubble radius afterwards. Hence,
once again, the assumption of the objective reality of quantum trajectories turns out to be very useful in order to
deal with cosmological issues.

\newpage

\section{Conclusions}

In this paper we reviewed the main results of the application of the
dBB quantum theory to quantum cosmology.
It is a consensus that in this domain we cannot
interpret the wave function of the Universe using the Copenhaguen interpretation.
Fortunately there are alternatives, like the MW theory, the CH interpretation,
and the dBB theory. Among these alternatives, the dBB theory has one advantage over the others:
the notion of a probability measure is not essential for obtaining results from the theory.
In fact, it is a dynamical theory of trajectories in the configuration space of
any general system, including the Universe. This dynamical theory is, of course,
not simple (non-local, contextual) and weird (quantum mechanics is weird),
where trajectories in configuration space are determined by a wave function
satisfying the wave equation appropriate to the system under investigation.
Hence, one can use the resulting trajectories to answer physical
questions about the system without needing
any probabilistic notion. This is very important because quantum cosmology
does not furnish an obvious probabilistic measure in general,
which renders the implementation of alternative quantum theories rather obscure
and involved.

In section 5, we have shown explicitly that in de~Broglie-Bohm theory we can
investigate the entire evolution of the Universe,
even without the notion of probability.
It is worth remarking that every quantum model of the Universe based on probabilistic outcomes has to face a non-trivial problem:
we have only one copy of the system under investigation, the Universe, which forbids us to repeat experiments, hence raising doubts about the physical meaning of any kind of probability in this context. Thus, the lack of
an obvious probabilistic measure which usually happens in quantum cosmology should not come as a surprise. On the contrary, this may be viewed as the rule of the game, pushing us to carefully analyze whether one can consistently extract information and predictions from such models without any notion of probability.

We should stress that one can recover probabilistic predictions in quantum cosmology using the de Broglie-Bohm theory in the situations where they are necessary when one implements a more
complex modeling of the Universe, adding new degrees of freedom, like linear perturbations described in section 6. In that case, a probability measure naturally appears in the quantum description of the sub-systems of the Universe (see section 6 for details), and the usual Born rule
can be recovered. In that case, for the sub-systems, the de Broglie-Bohm approach should coincide
with other quantum theories.

We have also shown very precisely, under the perspective of this
quantum theory, how to obtain the cosmological classical
limit, how singularities can be avoided, and how the notion
of time and probabilities can be recovered in the circumstances they are
crucial. In fact, the simple assumption of the objective reality of trajectories
in configuration space turns all these results possible. For instance, the dBB quantum theory
yields simple explanations to old issues, like the quantum to classical
transition of cosmological perturbations in the standard cosmological
model, which are not so easy to obtain in other quantum theories \cite{ward}.

In the case of full superspace, the dBB interpretation of canonical
quantum cosmology yields a quantum geometrodynamical picture where
the bohmian quantum evolution
of three-geometries may form, depending on the wave functional, a consistent
non degenerate
four geometry which must be euclidean (but only for a very special local form
of the quantum potential), or a consistent but degenerate four-geometry,
indicating the presence of special
vector fields and the breaking of the spacetime structure as a single
entity.
Hence, in general, and always when the quantum potential is non-local,
the three-geometries evolved under the influence
of a quantum potential do not in general stick together to form a
non degenerate four-geometry, a single spacetime with the causal structure
of relativity. Among the consistent bohmian evolutions, the more general
structures that are formed are degenerate four-geometries with alternative causal structures.
We obtained these results taking a minimally coupled
scalar field as the matter source of gravitation, but
it can be generalized to any matter source with non-derivative
couplings with the metric, like Yang-Mills fields.
We have also seen that any real solution of the Wheeler-DeWitt equation
yields a structure which is the idealization of the strong gravity limit
of GR. Due to the generality of this picture (it is valid for any
real solution of the Wheeler-DeWitt equation, which is a real equation), it
deserves
further attention. It may well be that these degenerate four-metrics are
the correct quantum geometrodynamical description of the Planckian universe.
We would like to remark that all these results were obtained
without assuming any particular factor ordering and regularization
of the Wheeler-DeWitt equation. Also, we did not use any probabilistic
interpretation
of the solutions of the Wheeler-DeWitt equation. Hence, it is a quite general
result. Of course these conclusions are limited by many strong
assumptions we have tacitly made, as supposing that a continuous three-geometry
exists at the quantum level (quantum effects could also destroy it), or
the validity of quantization of standard GR, forgetting other developments
like string theory. However, even if this approach is not the appropriate one,
it is nice to see how far we can go with the dBB quantum theory,
even in such incomplete stage of canonical quantum gravity.

One may object that the conclusions based on the dBB theory may have
no physical significance, that they are abstractions with no observational
consequences. However, in our section 6 we have shown that the hypothesis
of the objective reality of quantum bohmian trajectories in
configuration space may lead to peculiar observational consequences related to the evolution
of quantum cosmological perturbations in quantum cosmological backgrounds,
which might be tested and which are not known how to be obtained in other
approaches. Note that if one had used other interpretations of quantum mechanics
instead of dBB theory, where the notion of trajectories is not immediate, the
implementation of the calculations for the perturbations we have presented
in section 6 based, e.g., on Eq.~(\ref{equacoes-mukhanov}) could have been much more involved,
if possible.
Also, if the simplifications discussed just before Eq.~(\ref{equacoes-mukhanov})
had not been made, corrections to the Schr\"odinger equation for the perturbations in the
quantum background could have led to a departure from quantum equilibrium during the
bounce, turning possible to find physical systems, which were created and frozen after
the bounce, were the dBB theory could be
tested against usual quantum theory because they would not satisfy the Born rule.

We are presently in a situation in quantum physics where we have a precise
algorithm that furnish correct probabilistic predictions concerning all
known quantum phenomena, but with no underlying consensual theoretical
understanding of this algorithm. There are many proposals, each one with
their own merits and difficulties, which furnish the same known experimental
results. The Universe is a very peculiar physical system,
where quantum concepts are pushed to their limits. We have seen, for instance, that the theoretical
framework of the Copenhagen interpretation simply cannot be applied to quantum cosmology,
and that some conclusions about the existence of cosmological singularities may depend strongly
on the quantum theory we are using.
Hence, quantum cosmology can be viewed
as the arena on which one may select the quantum theory, among the many viable present
possibilities, which is the most suitable to describe all quantum phenomena, including the
Universe itself. If it turns out that one or more of the results obtained in section
6 within the framework of the dBB theory applied to quantum cosmology
are indeed observed and confirmed in the future, with no alternative
classical explanation for them,
then we will have a situation where a particular quantum theory is capable
to predict and describe some physical phenomena which are not known how to
be obtained in other quantum theories.
Hence, we will face the situation where not only quantum theory is helping cosmology
but also cosmology is helping
quantum theory.
These are exciting possibilities that
must be investigated in the future, not only because they are relevant
for cosmology, but also because they are important for quantum physics itself.
\par
{\bf Acknowledgments:} We thank CNPq (Brazil) for partial financial support. J.C.F. thanks also FAPES (Brazil) for partial financial support.


\begin{thebibliography}{}

\bibitem{pinto} J.~Acacio de Barros, N.~Pinto-Neto, and
M.~A.~Sagioro-Leal, Phys. Lett. A {\bf 241}, 229 (1998).

\bibitem{lemos-ini} F.G. Alvarenga and N.A. Lemos, Gen. Rel. Grav. 30, 681 (1998).

\bibitem{brasil2} F.G. Alvarenga, A.B. Batista, J.C. Fabris and S.V.B. Gon\c{c}alves, {\it Anisotropic quantum cosmological models: a discrepancy between many-worlds and dBB interpretations}, gr-qc/0202009.

\bibitem{fabris} F.G. Alvarenga, J.C. Fabris, N.A. Lemos and G.A. Monerat, Gen. Rel. Grav. {\bf 34}, 651 (2002).

\bibitem{tri} J. Ambjorn, A. Goerlich, J. Jurkiewicz and R. Loll,
Phys. Rep. {bf 519}, 127 (2012).

\bibitem{ash4} A. Ashtekar, in {\it Non-Perturbative Canonical Gravity},
IUCAA Lecture Notes (Syracuse, New York, 1990); T. Thiemann,
Lect. Notes Phys. {\bf 721}, 185 (2007).

\bibitem{ash1} A. Ashtekar, Gen. Rel. Grav. {\bf 41}, 707-741, (2009);
arXiv:gr-qc/0812.0177v1.

\bibitem{ash2} A. Ashtekar, A. Corichi, P. Singh, Phys. Rev. {\bf D77}, 024046 (2008);
arXiv:gr-qc/0710.3565.

\bibitem{ash3} A. Ashtekar, T. Pawlowski, P. Singh;
Phys. Rev. {\bf D73}, 124038 (2006).

\bibitem{ash41} A. Ashtekar and P. Singh, Class. Quant. Grav. {\bf 28}, 213001 (2011).

\bibitem{kerner} R. Balbinot, J.C. Fabris and R. Kerner, Phys. Rev. {\bf D42}, 1023 (1990).

\bibitem{ghirardi} A. Bassi and G. Ghirardi, Phys.Lett. {\bf A257},
247 (1999); arXiv:gr-qc/9811050v4.

\bibitem{ghirardi2} A. Bassi and G. Ghirardi,  J. Statist. Phys.
{\bf 98}, 457 (2000); arXiv:quant-ph/9912031v1.

\bibitem{denis} D. Bessada, N. Pinto-Neto, B.B. Siffert and O.D. Miranda,
JCAP {\bf 1211}, 054 (2012).

\bibitem{birrell} N.D. Birrell and P.C.W. Davies, {\it Quantum fields in curved space} (Cambridge University Press, Cambridge, 1982).

\bibitem{dBB} D. Bohm, Phys. Rev. {\bf 85}, 166 (1952); Phys. Rev.
{\bf 85},180 (1952); D. Bohm, B. J. Hiley and P. N. Kaloyerou, Phys.
Rep. {\bf 144}, 349 (1987); D. Bohm and B.J. Hiley, Phys. Rep. 144,
323 (1987).

\bibitem{boh} D. Bohm and B.J. Hiley, {\it The undivided universe:
an ontological interpretation of quantum theory} (Routledge, London, 1993).

\bibitem{bohr} N. Bohr, {\it Atomic Physics and Human Knowledge} (Science
Editions, New York, 1961); N. Bohr; Phys. Rev. {\bf 48}, 696 (1935).

\bibitem{CS} D.A. Craig, P. Singh, Phys. Rev. D {\bf 82}, 123526 (2010).
arXiv:gr-qc/1006.3837.

\bibitem{iso} R. Colistete Jr., N. Pinto-Neto and A. F. Velasco,
Phys. Lett. {\bf A 277}, 194 (2000).

\bibitem{gaussian} R.~Colistete Jr., J.~C.~Fabris and N.~Pinto-Neto,
Phys. Rev. {\bf 62}, 083507 (2000).

\bibitem{dew3} B.S. DeWitt, Phys. Rev. {\bf 160}, 1113 (1967).

\bibitem{mwi} B.S. DeWitt and N. Graham, Editors, {\it The Many-Worlds Interpretation of
Quantum Mechanics}, ed. by  (Princeton University Press, Princeton, 1973).

\bibitem{constr1} P.A.M. Dirac, {\it Lectures on Quantum Mechanics} (Yeshiva University, New York, 1964).

\bibitem{his1} H. F. Dowker and J. J. Halliwell, Phys. Rev. {\bf D46}, 1580 (1992).

\bibitem{eve} H. Everett, Rev. Mod. Phys. {\bf 29}, 454 (1957).

\bibitem{falciano} F.T. Falciano and N. Pinto-Neto, Phys. Rev. {D \bf 79} 023507 (2009).

\bibitem{gg} J. C. Fabris and A. F. Velasco, Class. Quantum Grav. {\bf 16}, 3807 (1999).

\bibitem{har} M. Gell-Mann and J.B. Hartle, in {\it Complexity, Entropy and the Physics of Information}, ed. by W. H. Zurek (Addison Wesley, 1990).

\bibitem{kiefer} D. Giulini and C. Kiefer, Class. Quantum Grav. {\bf 12}, 403 (1995).

\bibitem{rim} G.C. Ghirardi, A. Rimini and T. Weber, Phys. Rev. {\bf D 34} 470 (1986);
G.C. Ghirardi , P. Pearle and A. Rimini, Phys. Rev.{\bf A 42}, 78 (1990).

\bibitem{gle} A. M. Gleason, J. Math. Mechanics {\bf 6}, 895 (1953).

\bibitem{nen1} M. Gleiser, R. Holman and N. Pinto-Neto, Nucl. Phys.
{\bf B294}, 1164 (1987).

\bibitem{halli} J. J. Halliwell and S. W. Hawking, Phys. Rev. {\bf D31},
1777 (1985).

\bibitem{hal0} J. J. Halliwell, in {\it Quantum Cosmology and Baby
Universes}, ed. by S. Coleman, J.B. Hartle, T. Piran and S. Weinberg
(World Scientific, Singapore, 1991).

\bibitem{halli1} J. J. Halliwell and J. Thorwart, Phys. Rev. {\bf
D64},  124018 (2001); arXiv:gr-qc/0106095v2.

\bibitem{halli2} J. J. Halliwell, \textit{Journal of Physics:
Conference Series} {\bf 306}, 012023 (2011).

\bibitem{har2} J. B. Hartle, Found. Phys. {\bf 41}, 982 (2011).

\bibitem{he} S. W. Hawking and G. F. R. Ellis, {\it The large scale structure
of space-time} (Cambridge University Press, Cambridge, 1973).

\bibitem{hei} W. Heisenberg, {\it The Physical Principles of the Quantum Theory} (Dover, New York, 1949).

\bibitem{hen} M. Henneaux, M. Pilati and C. Teitelboim, Phys. Lett. {\bf 110 B},
123 (1982).

\bibitem{constr2} M. Henneaux and C. Teitelboim; {\it Quantization of
Gauge Systems} (Princeton University Press, Princeton, 1992).

\bibitem{kuc2}  S. A. Hojman, K. Kucha$\check{\mbox{r}}$ and C. Teitelboim,
Ann. Phys. {\bf 96}, 88 (1976).

\bibitem{hol} P. R. Holland, {\it The Quantum Theory of Motion: An
Account of the de Broglie-Bohm Interpretation of Quantum
Mechanichs} (Cambridge University Press, Cambridge, 1993).

\bibitem{ter2} A. Hosoya and M. Morikawa, Phys. Rev. {\bf D39}, 1123 (1989).

\bibitem{fou} J.N. Islam, Found. of Phys. {\bf 24}, 593 (1994).

\bibitem{oqed} P N. Kaloyerou, Phys. Rep. {\bf 244}, 287 (1994).

\bibitem{kie1} C. Kiefer, Phys. Rev. {\bf D45}, 2044 (1992).

\bibitem{2kie} C. Kiefer, Class. Quantum Grav. {\bf 18},
379 (1991); D. Giulini, E. Joos, C. Kiefer, J. Kupsch, I. O. Stamatescu
and H. D. Zeh, {\it Decoherence and the Appearance of a Classical
World in Quantum Theory} (Springer-Verlag, Berlin, 1996).

\bibitem{kiefer2} C. Kiefer, Gen. Rel. Grav. {\bf 41}, 877, (2009).

\bibitem{his4} C. Kiefer and H. D. Zeh, Phys. Rev. {\bf D51}, 4145 (1995).

\bibitem{kuc1} K. Kuchar, in {\it Quantum Gravity 2: A Second Oxford
Symposium}, eds. C. J. Isham, R. Penrose and D. W. Sciama
(Clarendon Press, Oxford, 1981).

\bibitem{kucmin} K.V. Kuchar and M.P. Ryan, Phys. Rev. {\bf D40}, 3982 (1989).

\bibitem{rubakov} V.G. Lapchinskii and V.A. Rubakov, Theor. Math. Phys. {\bf 33}, 1076 (1977).

\bibitem{poin} J.M. L\'evy Leblond, {\it Ann. Inst. Henri Poincar\`e} {\bf 3}, 1(1965).

\bibitem{lif}  E.M. Lifshitz and I.M. Khalatnikov, Adv. Phys. {\bf 12}, 185 (1963).

\bibitem{luka1} V.N. Lukash, Sov. Phys. JETP {\bf 52}, 807(1980).

\bibitem{luka2} V.N. Lukash, JETP Lett. {\bf 31}, 596(1980).

\bibitem{bia} R. Maier, S. Pereira, N. Pinto-Neto and B.B. Siffert, Phys. Rev. {\bf D 85}, 023508 (2012).

\bibitem{majumder} B. Majumder and N. Banerjee, Gen. Rel. Grav. {\bf 45}, 1(2013).

\bibitem{mis1} C. W. Misner, Phys. Rev. {\bf 186}, 1319 (1969).

\bibitem{mis2} C.W. Misner, Phys. Rev. {\bf 186}, 1328 (1969).

\bibitem{mtw} C.W. Misner, K.S. Thorne and J.A. Wheeler, {\it Gravitation}
(W. H. Freeman and Company, New York, 1973). 

\bibitem{mukha} V.F. Mukhanov, Sov. Phys. JETP {\bf 67}, 1297(1988).

\bibitem{MFB} V.F.~Mukhanov, H.A.~Feldman, and R.H.~Brandenberger,
Phys. Rep. {\bf 215}, 203 (1992).

\bibitem{muk} V.F. Mukhanov, in {\it Physical Origins of Time Asymmetry}, ed.
by J. J. Halliwell, J. P\'erez-Mercader and W. H. Zurek (Cambridge University
Press, 1994).

\bibitem{sake} W. Nelson and M. Sakellariadou,	Phys. Rev. {\bf D80}, 063521 (2009).

\bibitem{omn} R. Omn\`es, {\it The Interpretation of Quantum Mechanics}
(Princeton University Press, Princeton, 1994).

\bibitem{his5} J.P. Paz and S. Sinha, Phys. Rev. {\bf D44}, 1038 (1991).

\bibitem{his3} J.P. Paz and W. H. Zurek, Phys. Rev. {\bf D48}, 2728 (1993).

\bibitem{pen} R. Penrose, in {\it Quantum Implications: Essays in Honour of David
Bohm}, ed. by B. J. Hiley and F. David Peat (Routledge, London, 1987).

\bibitem{PPN1} P.~Peter, E.J.C.~Pinho, and N.~Pinto-Neto, J. Cosmol. Astropart. Phys. {\bf 07}, 014(2005).

\bibitem{PPN2} P.~Peter, E.J.C.~Pinho and N.~Pinto-Neto, Phys. Rev. {D \bf 73} 104017 (2006).

\bibitem{PPN4} P.~Peter, E.~J.~C.~Pinho, and N.~Pinto-Neto, Phys. Rev. {D \bf 75} 023516 (2007).

\bibitem{PPN3} E.J.C.~Pinho and N.~Pinto-Neto, Phys. Rev. {D \bf 76} 023506 (2007).

\bibitem{pereira} N. Pinto-Neto, F. T. Falciano, R. Pereira and E. Sergio Santini,
Phys. Rev. {\bf D 86}, 063504 (2012).

\bibitem{ward} N. Pinto-Neto, G. Santos and W. Struyve,
Phys. Rev. {\bf D 85}, 083506 (2012).

\bibitem{nen2} N. Pinto-Neto and A. F. Velasco, GRG {\bf 25}, 991 (1993).

\bibitem{euc} N. Pinto-Neto and E. Sergio Santini, {\it Phys. Rev.} D  59,
123517 (1999).

\bibitem{nelson1} N. Pinto-Neto, E. Sergio Santini and F. T. Falciano, Phys. Lett. {\bf A344}, 131 (2005).

\bibitem{supernova} A.G. Riess {\em et al.}, Astron. J. \textbf{116}, 1009 (1998); S.
Perlmutter {\em et al.}, Astrophys. J. \textbf{517}, 565 (1999).

\bibitem{rya} M. P. Ryan, lecture notes from the {\it $6^{th}$ Brazillian
School on Cosmology and Gravitation} (Rio de Janeiro, 1989).

\bibitem{sasa}  M. Sasaki, Prog. Theor. Phys. {\bf 76}, 1036(1986).

\bibitem{schutz1} B.F. Schutz, Phys. Rev. {\bf D2}, 2762 (1970).

\bibitem{schutz2} B.F. Schutz, Phys. Rev.{\bf D4}, 3559 (1971).

\bibitem{ter1} A. Strominger, Phys. Rev. Lett {\bf 52}, 1733 (1984).

\bibitem{str} W. Struyve, J. Phys. Conf. Ser. {\bf 306}, 012047 (2011).

\bibitem{sun} K. Sundermeyer, {\it Constrained Dynamics}
(Springer-Verlag, Berlin, 1982).

\bibitem{tei1} C. Teitelboim, {\it Ann. Phys.}  80, 542 (1973).

\bibitem{tei2} C. Teitelboim, {\it Phys. Rev.}   D 25  3159 (1982).

\bibitem{tipler} F. Tipler, Phys. Rep. {\bf 137}, 231 (1986).

\bibitem{valentini} A. Valentini, Phys. Lett. {\bf A156}, 5 (1991).

\bibitem{valentini2} A. Valentini, ``Inflationary Cosmology as a Probe of Primordial Quantum Mechanics", arXiv:quant-ph/0805.0163.

\bibitem{dBBbis} A. Valentini,  J. Phys. A: Math. Theory, {\bf 40}, 3285 (2007); M. Kenmoku, T. Matsuyama, R. Sato and S. Uchida, Class. Quant. Grav. {\bf 19}, 799 (2002); M. Kenmoku, Hiroto Kubotani, Eiichi Takasugi and Yuki Yamazaki, Prog. Theor. Phys. {\bf 105}, 897 (2001); M. Kenmoku, H. Kubotani, E. Takasugi and Y. Yamazaki, Phys. Rev. {\bf D57} 4925, (1998).

\bibitem{von} J. von Neumann, {\it Mathematical Foundations of Quantum
Mechanics} (Princeton University Press, Princeton, 1955).

\bibitem{ed1} E. Witten, Nucl. Phys. {\bf B311}, 46 (1988).

\bibitem{deco} H. D. Zeh, Found. Phys. {\bf 1}, 69 (1970); E. Joos and
H. D. Zeh, Z. Phys. {\bf B59}, 223 (1985); W. H. Zurek, Phys. Rev. {\bf D26},
1862 (1982);  W. H. Zurek, Phys. Today  {\bf 44}, 36 (1991).

\bibitem{zeh} H. D. Zeh, in {\it Decoherence and the Appearance of a Classical
World in Quantum Theory} (Springer-Verlag, Berlin, 1996).

\bibitem{zel} Y.B. Zeldovich, Zh. Eksp. Teor. Fiz. {\bf 41}, 1609 (1961); Y.B. Zeldovich, Sov. Phys. JETP {\bf 14}, 1143 (1962).

\end{thebibliography}
\end{document}